\documentclass[reprint,onecolumn,secnumarabic,amssymb, nobibnotes, aps, prfluids]{revtex4-2}

\usepackage{psfrag}

\usepackage{xcolor}
\usepackage{bbm}

\usepackage{lmodern}
\usepackage[T1]{fontenc}
\usepackage[utf8]{inputenc}
\usepackage{indentfirst}
\usepackage{color}
\usepackage{microtype}
\usepackage{lipsum}	
\usepackage{parskip}
\usepackage{setspace}

\usepackage{graphicx}
\usepackage{subcaption}
\usepackage{epstopdf}
\usepackage{epsfig}
\usepackage{rotating}
\usepackage{morefloats}

\usepackage{amsmath}
\usepackage{amssymb}
\usepackage{bm}
\usepackage{wasysym}
\usepackage{icomma}
\usepackage{multirow}
\usepackage{placeins}
\usepackage{mathrsfs}
\usepackage{mathtools}
\usepackage{subeqnarray}

\usepackage[english]{babel}
\usepackage{bibentry}
\usepackage{hyperref}
\usepackage{cleveref}
\usepackage{natbib}

\usepackage{etoolbox}
\makeatletter
\patchcmd\affils@present@group
 {\frontmatter@affiliationfont}
 {\frontmatter@affiliationfont\begingroup\lccode`~=`,\lowercase{\endgroup\let~}\active@comma}
 {}{}
\makeatother

\usepackage{makecell}

\newcommand{\difp}[3][2]{\frac{\partial #2}{\partial #3}}
\newcommand{\hq}{{\hat{q}}}
\newcommand{\hu}{{\hat{u}}}
\newcommand{\hv}{{\hat{v}}}
\newcommand{\hw}{{\hat{w}}}

\begin{document}

%%%%%%%%%%%%%%%%%%%%%%%%%%%%%%%%%%%%%%%%%%%%%%
\title{Wavepackets in subsonic elliptical jets: coherent-structure eduction and linear modelling}
%%%%%%%%%%%%%%%%%%%%%%%%%%%%%%%%%%%%%%%%%%%%%%

\author{Filipe R. do Amaral}
\thanks{framaral@ita.br}
\affiliation{Divisão de Mecânica, Departamento de Energia, Instituto Tecnológico de Aeronáutica (ITA, IEM-E), 12228-900 São José dos Campos, Brazil}
\affiliation{Département Fluides, Thermique, Combustion (DFTC-2AT), Institut Pprime, CNRS--Université de Poitiers--ENSMA, 86000 Poitiers, France}
\author{Mateus P. Avanci}
\author{Jean-Christophe Robinet}
\affiliation{Laboratoire de dynamique des fluides (DynFluid), École Nationale Supérieure d'Arts et Métiers (ENSAM), 75013 Paris, France}
\author{Peter Jordan}
\affiliation{Département Fluides, Thermique, Combustion (DFTC-2AT), Institut Pprime, CNRS--Université de Poitiers--ENSMA, 86000 Poitiers, France}
\author{Jérôme Huber}
\affiliation{Airbus Operations SAS, 31300 Toulouse, France}
\author{Grégoire Pont}
\affiliation{Airbus, 31300 Toulouse, France}

\date{\today}
\maketitle

%%%%%%%%%%%%%%%%%%%
% ABSTRACT
%%%%%%%%%%%%%%%%%%%

We study coherent structures extracted from time-resolved particle image velocimetry (TR-PIV) measurements of isothermal elliptical jets with aspect ratio of 1:2, at Mach numbers 0.4 and 0.7. 
The velocity snapshots are decomposed according to the D$_2$ symmetry.
Spectral proper orthogonal decomposition (SPOD) is then used to identify coherent structures.
Modelling is performed using three-dimensional parabolised stability equations (3D-PSE) based on linearisation about the time-averaged mean flows. 
A comparison between SPOD and 3D-PSE results shows alignment of SPOD and single-mode PSE structures of up to 70\%. 
A linear combination of PSE solutions, obtained from the family of initially unstable local modes, leads to SPOD-PSE alignment of up to 90\%. 
Limitations are observed at low frequencies and near the nozzle and are attributed to the existence of non-modal amplification mechanisms.
The study shows that coherent structures in turbulent elliptical jets can be educed using the D$_2$-SPOD approach, and modelled using 3D PSE.

%%%%%%%%%%%%%%%%%%%%%%%%%%%%%%%%%%%%%%%%%%%%%%
\section{Introduction}
\label{sec:introduction}
%%%%%%%%%%%%%%%%%%%%%%%%%%%%%%%%%%%%%%%%%%%%%%

% Coherent structures in round jets
Coherent structures drive sound generation in round jets~\citep{jordan2013wave, cavalieri2019wave}, and are expected to play a similar role in turbulent elliptical jets, which have not yet been extensively characterized.
In the case of round turbulent jets, axisymmetry allows for Fourier-series decomposition in the azimuthal direction, and further post-processing with techniques such as spectral proper orthogonal decomposition (SPOD)~\citep{picard2000pressure, towne2018spectral, schmidt2020guide} is typically performed for each azimuthal mode (and frequency) to extract coherent structures~\citep{cavalieri2013wavepackets, rodriguez2013inlet, sinha2014wavepacket, sasaki2017high, schmidt2017wavepackets, schmidt2018spectral, lesshafft2019resolvent, nogueira2019large, pickering2020lift, maia2023effect, maia2024effect, eichberger2024localized}.
To understand the underlying physical mechanisms, these structures can be compared with those obtained through modeling approaches such as locally parallel linear stability theory (LST)~\citep{rodriguez2015study, schmidt2017wavepackets, maia2023effect}, parabolized stability equations (PSE)~\citep{gudmundsson2011instability, rodriguez2013inlet, sinha2014wavepacket, rodriguez2015study, cavalieri2013wavepackets, sasaki2017high, padilla2026plane}, one-way Navier--Stokes (OWNS)~\citep{towne2015one, zhu2023recursive, nogueira2023prediction, nogueira2024wavepackets, amaral2025tabbed}, and resolvent analysis~\citep{towne2018spectral, schmidt2018spectral, nogueira2019large, lesshafft2019resolvent, pickering2020lift, maia2024effect, eichberger2024localized}.
This body of work has shown how the coherent structures that arise in round turbulent jets can be attributed to amplification mechanisms, that draw fluctuation energy from the mean in a manner analogous to that by which  modal and non-modal instability mechanisms draw fluctuation energy from laminar flows.

Regarding round turbulent jets,~\citet{cavalieri2013wavepackets} investigated Kelvin--Helmholtz (K--H) wavepackets in subsonic jets using particle image velocimetry (PIV), hot-wire anemometry, and linear PSE for modelling.
Good agreement between experimental results and PSE predictions was found for both streamwise and radial velocity components, up to the end of the jet potential core and for $\mathit{St} < 1$.
Their study was extended by~\citet{sasaki2017high}, who analysed a $M_j = 0.9$ jet using LES and modelled the wavepackets with linear PSE.
Close agreement between empirical structures educed from data and linear PSE predictions was observed for high Strouhal numbers near the nozzle exit and for lower Strouhal numbers further downstream.

The prevalence of such coherent structures in round turbulent jets has been shown in the study of~\citet{pickering2020lift}.
Using data from a validated large-eddy simulation (LES)~\citep{bres2019modelling}, they show that three kinds of coherent structure account for the fluctuation energy across a broad range of scales and frequencies: K--H wavepackets, lift-up/streaks, and Orr structures.
K--H wavepackets, which dominate over a range of intermediate frequencies, undergo modal, spatial amplification upstream of the end of the potential core and are primarily forced by Orr structures concentrated in or near the nozzle~\citep{lesshafft2019resolvent, kaplan2021nozzle, bugeat2024acoustic}.
The lift-up mechanism~\citep{ellingsen1975stability, landahl1980note}, which leads to low-frequency streaks, acts throughout the jet shear layer.
The Orr-type wavepacket involves low-frequency axisymmetric structures that are extensively forced over the first ten jet diameters: forcing structures inclined against the mean shear produce a response structure that rotates and gradually aligns with the shear as it evolves downstream under the action of the critical layer~\citep{tissot2017wave, schmidt2018spectral}.
These three types of structures have also been observed in round jets subjected to a flight stream~\citep{maia2023effect, maia2024effect}.

% Studies on elliptical jet structures
In elliptical jets, the Fourier symmetry in the azimuthal direction is lost: this raises questions regarding the kinds of coherent structures that arise, the underlying physical mechanisms, and our ability to both extract them from data and model them using linearised methods based on the mean flow.

Early experimental investigations of non-circular jets focused on the observation of their complex three-dimensional development, including axis switching in both rectangular~\citep{zaman1996axis, gutmark1999flow} and elliptical jets~\citep{hussain1989elliptic, husain1991elliptic, husain1993elliptic}.
These subsonic studies analysed jets in the range 9.5 m/s $\leq U_j \leq$ 30 m/s (corresponding to $10^4 \leq \mathit{Re}_{D_h} \leq 10^5$, based on the nozzle hydraulic diameter $D_h$ and jet exit velocity $U_j$).
Axis switching has also been observed in supersonic elliptical jets~\citep{edgington-mitchell2013near, beekman2024acoustic, mazharmanesh2024multimodal, suzuki2026coherent}, whereas other measurements in the range 0.4 $\leq M_j \leq$ 0.9 did not confirm the presence of such a phenomenon~\citep{amaral2023elliptical, nogueira2023prediction}.
Additionally,~\citet{suzuki2026coherent} conducted direct numerical simulations (DNS) of an incompressible elliptical jet at $\mathit{Re}_{D_h} = 400$, with a nozzle aspect ratio $\mathit{AR} = 1{:}2$, and identified the axis switching phenomenon.

The studies by~\citet{morris1988instability} and~\citet{morris1995spatial} investigated the stability properties of elliptical jets using analytical mean-flow profiles.
Both studies formulated the problem in elliptical-cylindrical coordinates and used modified Mathieu functions~\citep{mclachlan1947mathieu} to classify eigenmodes.
This approach allows for the extraction of even and odd modes, symmetric or antisymmetric with respect to the ellipse axes.
The combination of elliptical-cylindrical coordinates and Mathieu functions yields analogues to the axisymmetric ($m = 0$) varicose mode in circular jets, as well as other modes such as flapping, wagging, double-flapping-sidewards, and double-flapping-diagonally~\citep{kinzie1997azimuthal}.
\citet{morris1988instability} found no single dominant mode in the initial mixing region.
In contrast,~\citet{morris1995spatial} studied compressible, inviscid elliptical jets and identified the flapping mode—odd about the ellipse's major axis and even about the minor axis (analogous to the $m = 1$ mode in circular jets) as most unstable, exhibiting preferential directivity in the jet’s minor-axis plane.
This result was also confirmed by experiments on supersonic elliptical jets at $M_j = 1.5$~\citep{kinzie1997azimuthal, kinzie1999aeroacoustic}.

% Modeling of elliptical jets
More recent modelling has been undertaken by~\citet{nogueira2023prediction} in an elliptical jet at $M_j = 0.6$.
They solved the one-way Navier--Stokes (OWNS) equations for jet profiles obtained from both an analytical expression and a fit to PIV measurements.
They showed that wavepacket peak position and phase speed depend on mode symmetry with respect to the ellipse axes.
Following the notation of~\citet{morris1988instability} and~\citet{morris1995spatial}, they identified varicose, flapping, and wagging modes.
In a subsequent study,~\citet{suzuki2026coherent} investigated DNS databases of low-Reynolds-number ($Re_{D_h} = 400$), incompressible elliptical jets under different forcing amplitudes and in the presence of axis switching.
The authors decomposed the flow into the four D$_2$ dihedral symmetry components and performed SPOD analysis, observing that the wagging mode dominates in the pre-axis-switching region, whereas the flapping mode becomes dominant downstream of the axis-switching location, particularly at low frequencies.

% Acoustic aspects of elliptical jets
Experimental measurements of subsonic elliptical jets have shown that their acoustic fields closely resemble those of equivalent axisymmetric jets~\citep{powell1958experimental, bridges1987roles, tesson2005experiments, proenca2022far, amaral2023elliptical, viswanathan2023elliptic, beekman2024acoustic}.
This contrasts with the behavior of supersonic jets, for which elliptical nozzles tend to produce lower noise levels compared to round ones~\citep{kinzie1997azimuthal, kinzie1999aeroacoustic}.

% Objective and paper organization
In the present study, the focus is placed on the identification of coherent structures in elliptical jets and their linear modelling via PSE.
The flows considered are subsonic, isothermal elliptical jets, of aspect ratio $\mathit{AR} = 1{:}2$, which are probed using time-resolved particle image velocimetry (TR-PIV) in cross-flow planes.
Jets at $M_j = 0.4$ and 0.7 are considered.
Given the loss of Fourier symmetry, a D$_2$ decomposition is used, yielding four symmetry families with respect to the ellipse’s minor and major axes: symmetric–symmetric (SS), symmetric–antisymmetric (SA), antisymmetric–symmetric (AS), and antisymmetric–antisymmetric (AA).
Coherent structures associated with each symmetry, extracted from the TR-PIV database using SPOD, are compared with solutions obtained by solution of the three-dimensional parabolised stability equations (3D-PSE)~\citep{avanci2022application}.

The remainder of the paper is organized as follows.
Section~\ref{sec:methodology} describes the experimental methods employed, including details of the setup and data post-processing procedures.
Section~\ref{sec:theoretical_approach} presents the theoretical framework and stability codes.
The results are presented in \S~\ref{sec:results}, and the main conclusions and perspectives are summarized in \S~\ref{sec:conclusions}.

%%%%%%%%%%%%%%%%%%%%%%%%%%%%%%%%%%%%%%%%%%%%%%%
\section{Experiments}
\label{sec:methodology}
%%%%%%%%%%%%%%%%%%%%%%%%%%%%%%%%%%%%%%%%%%%%%%

%%%%%%%%%%%%%%%%%%%%%%%%%%%%%%%%%%%%%%%%%%%%%%
\subsection{Nozzle configuration}
\label{sec:nozzle}

The nozzle has an aspect ratio $\mathit{AR} = 1{:}2$.
The hydraulic diameter $D_h = 50$ mm is defined as $D_h = 2\sqrt{ab}$, where $a$ and $b$ are the semi-major and semi-minor axes of the ellipse, respectively.
The nozzle contains a convergent section with a contraction ratio of $1{:}49$, which smoothly reduces the diameter from 350 mm over a length of 350 mm.
To ensure a turbulent boundary layer at the nozzle exit, a carborundum strip was installed $2.7 D_h$ upstream of the exit~\citep{cavalieri2013wavepackets}.

\begin{figure}
    \centering
        \includegraphics[width=\textwidth]{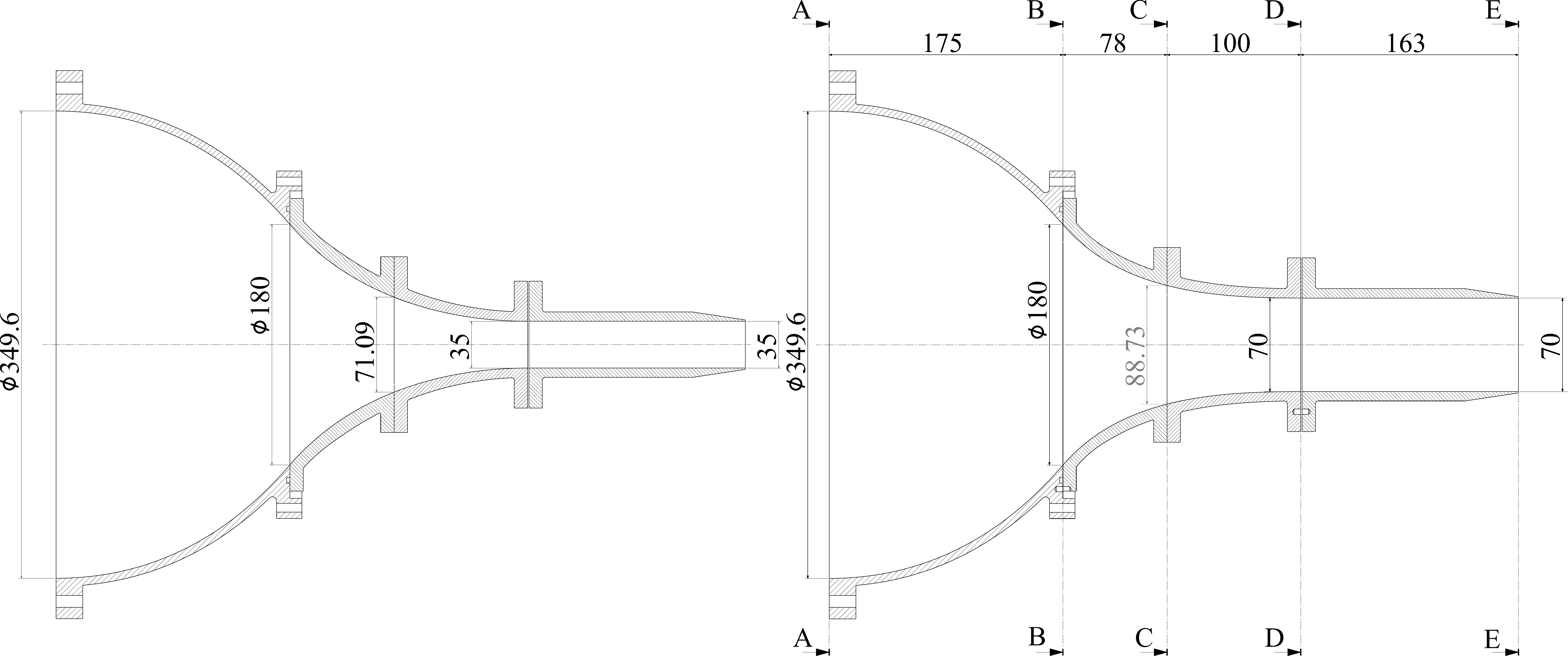}
        \caption{CAD views of the convergent section and the elliptical nozzle with $\mathit{AR} = 1{:}2$. Left frame: minor axis. Right frame: major axis. Dimensions in millimeters.}
        \label{fig:nozzle_axial}
\end{figure}

Figure~\ref{fig:nozzle_axial} shows CAD views of the nozzle in the axial direction, with the left frame showing the minor axis and the right frame showing the major axis.
The assembly is divided into four parts: the first (section AA--BB) reduces the diameter of the circular section from approximately 350~mm to 180~mm; the second and third parts (sections BB--CC and CC--DD, respectively) provide the transition from the circular to the elliptical cross-section; and the fourth part (section DD--EE) constitutes the elliptical nozzle itself.
This configuration allows the nozzle aspect ratio to be varied by replacing the second, third, and fourth parts.

\begin{figure}
    \centering
        \includegraphics[width=\textwidth]{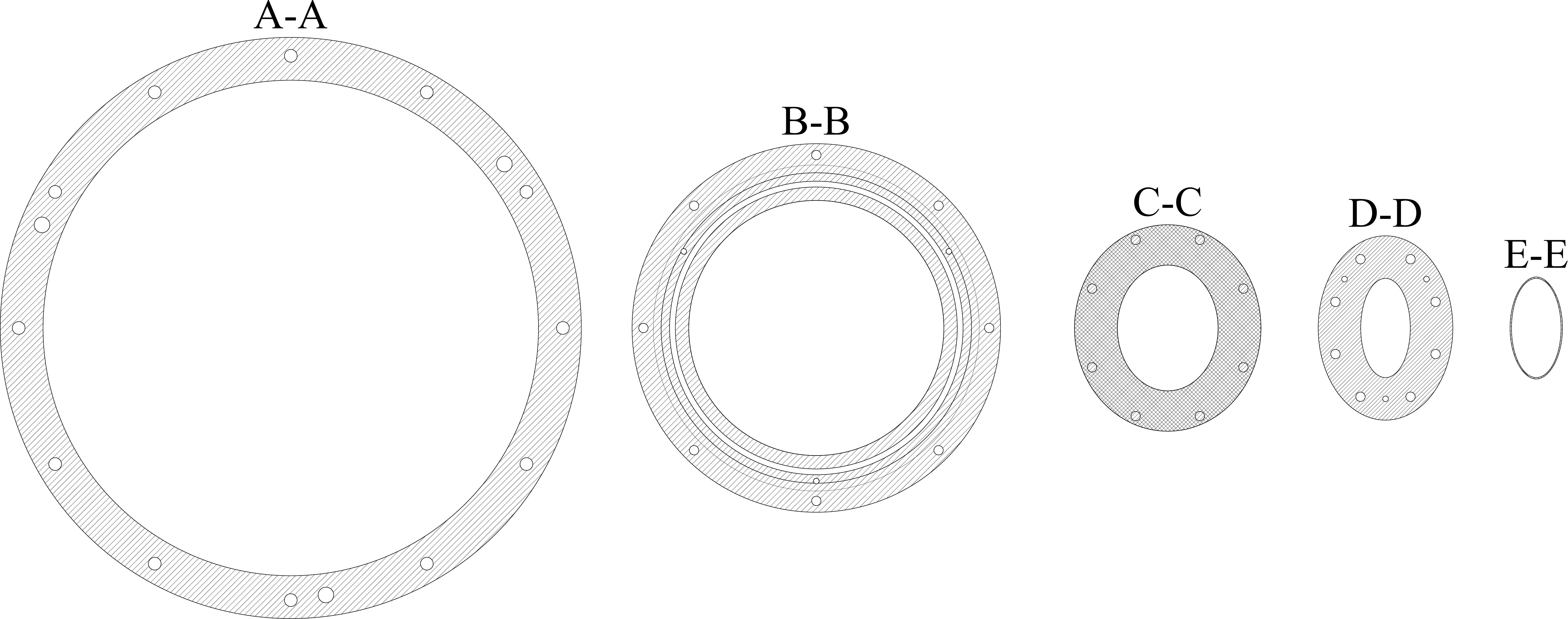}
        \caption{Cross-sectional views of the nozzle. Sections AA to EE indicated in Figure~\ref{fig:nozzle_axial}.}
        \label{fig:nozzle_cross}
\end{figure}

Figure~\ref{fig:nozzle_cross} shows cross-sectional views of the nozzle corresponding to the sections indicated in Figure~\ref{fig:nozzle_axial}.
The nozzle is seen to transition smoothly from a circular to an elliptical cross-section.

%%%%%%%%%%%%%%%%%%%%%%%%%%%%%%%%%%%%%%%%%%%%%%
\subsection{TR-PIV}
\label{sec:experimental_database}

Time-resolved particle image velocimetry (TR-PIV) measurements, employing a two-dimensional–three-component (2D–3C) configuration, were conducted at the \emph{Bruit \& Vent} facility of the Pprime Institute.
The facility comprises an anechoic chamber and a closed-circuit wind tunnel.
Cross-flow planes of the jet were acquired using LaVision hardware and software.

\begin{figure}
    \centering
    \begin{subfigure}{0.48\textwidth}
        \begin{center}
            \includegraphics[width=\textwidth]{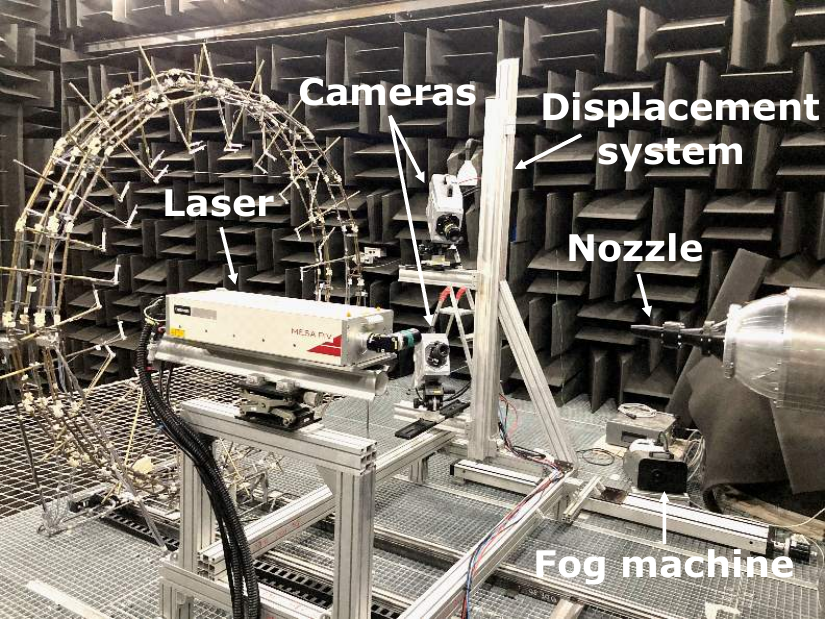}
        \end{center}
        \caption{Side view}
        \label{fig:picture_piv_lateral}
    \end{subfigure}
    \begin{subfigure}{0.48\textwidth}
        \begin{center}
            \includegraphics[width=\textwidth]{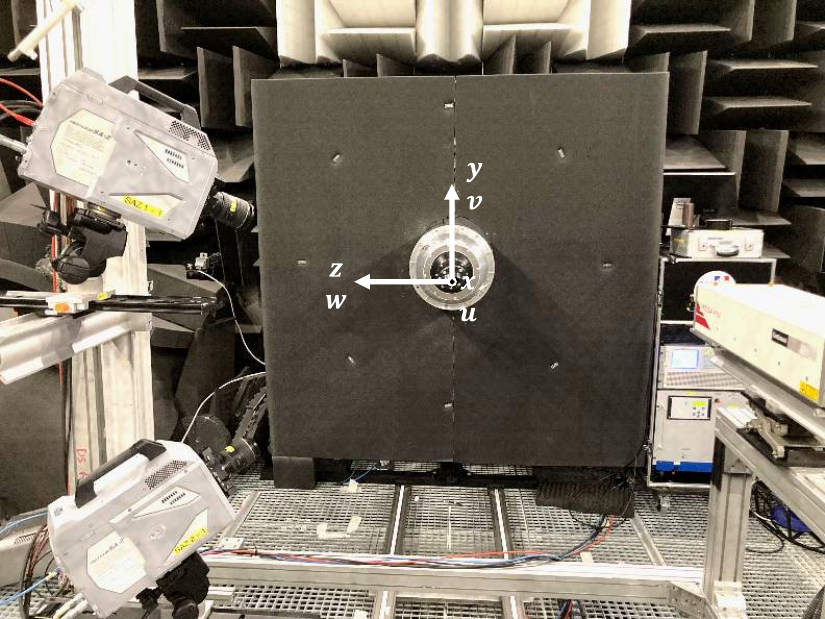}
        \end{center}
        \caption{Front view}
        \label{fig:picture_piv_front}
    \end{subfigure}
    \caption{Photographs of the stereo PIV setup inside the anechoic chamber, showing the cameras, laser, elliptical nozzle, displacement system, fog machine, and other equipment used in the measurements, along with the adopted coordinate system.}
    \label{fig:picture_piv}
\end{figure}

Figure~\ref{fig:picture_piv} shows the TR-PIV setup.
The laser and cameras are mounted on a displacement system that allows both the cameras and the laser to move together in the streamwise direction, such that each cross-plane is measured independently.
 procedures ensure the three velocity components are correctly measured and the cameras and laser are aligned.
The Cartesian coordinate system used is shown in the figure: where $x$ represents the streamwise direction (with velocity component $u$), while $y$ and $z$ denote the cross-stream directions (with velocity components $v$ and $w$, respectively).

Measurements were performed for isothermal, elliptical jets at Mach numbers of $M_j = 0.4$ and 0.7.
The three components of the velocity field were measured at $N_x = 11$ and 32 streamwise positions (cross-flow planes) over the streamwise range $0.03 \leq x/D_h \leq 15$.
Originally, 32 cross-flow planes were measured for both Mach numbers~\citep{amaral2023elliptical,amaral2024elliptical}, with an acquisition time of approximately $T \approx$ 2.16 s, corresponding to $N_s = 27{,}000$ snapshots.
These data-acquisition parameters were based on best-practice for round turbulent jets extensively studied in the past.
In the case of the round jet, the Fourier series decomposition is an effective filter of ``incoherent'' fluctuations, which facilitates statistical convergence of SPOD. 
Having processed the initial elliptical-jet data on the basis of best-practice for round jets, it was noted that the SPOD, based now on D$_2$-symmetry decomposition was not fully converged.
This underlines the challenge of coherent-structure identification in 3D flows where spatial Fourier decomposition can no longer be applied.
A second experimental campaign was therefore undertaken for the $M_j = 0.4$ case, where three additional acquisitions of 50,512 snapshots each (corresponding to roughly 4.04 s) were performed.
These were concatenated to form a time series comprising $N_s = 151{,}536$ snapshots (total duration $T \approx 12.12$ s).
A subset of 11 streamwise positions was selected for the long-duration measurements at $M_j = 0.4$ up to $x/D_h \leq 8$, while the original 32-plane dataset was thus retained for the $M_j = 0.7$ case (see Table~\ref{tab:streamwise_stations}).
In terms of storage, the second experimental campaign produced approximately 2 terabytes of raw TR-PIV image data.
After velocity field extraction and post-processing, the total data volume reaches approximately 8 terabytes.
We mention this because it highlights the data-acquisition, storage and processing challenges that are faced for 3D flows.

\begin{table}
    \begin{center}
		\begin{tabular}{c c c c }
			$x$ [mm]	& $x/D_h$	& $T = 2.16$ s	& $T = 12.12$ s	\\
			\hline
			1.5   		& 0.03  	& $\checkmark$ 	&   						\\
			5     		& 0.1   	& $\checkmark$ 	& $\checkmark$ 	\\
			10    		& 0.2   	& $\checkmark$ 	& $\checkmark$ 	\\
			15    		& 0.3   	& $\checkmark$ 	&   						\\
			20    		& 0.4   	& $\checkmark$ 	&   						\\
			25    		& 0.5   	& $\checkmark$ 	& $\checkmark$ 	\\
			30    		& 0.6   	& $\checkmark$ 	&   						\\
			35    		& 0.7   	& $\checkmark$ 	&   						\\
			40    		& 0.8   	& $\checkmark$ 	&   						\\
			50    		& 1     	& $\checkmark$ 	& $\checkmark$ 	\\
			60    		& 1.2   	& $\checkmark$ 	&   						\\
			70    		& 1.4   	& $\checkmark$ 	&   						\\
			75    		& 1.5   	& $\checkmark$ 	& $\checkmark$ 	\\
			80    		& 1.6   	& $\checkmark$ 	&   						\\
			100   		& 2     	& $\checkmark$ 	& $\checkmark$ 	\\
			125   		& 2.5   	& $\checkmark$ 	&   						\\
			150   		& 3     	& $\checkmark$ 	& $\checkmark$ 	\\
			175   		& 3.5   	&$\checkmark$ 	&   						\\
			200   		& 4     	& $\checkmark$ 	& $\checkmark$ 	\\
			225   		& 4.5   	& $\checkmark$ 	&   						\\
			250   		& 5     	& $\checkmark$ 	& $\checkmark$ 	\\
			275   		& 5.5   	& $\checkmark$ 	&   						\\
			300   		& 6     	& $\checkmark$ 	& $\checkmark$ 	\\
			350   		& 7     	& $\checkmark$ 	&   						\\
			400   		& 8     	& $\checkmark$ 	& $\checkmark$ 	\\
			450   		& 9     	& $\checkmark$ 	&   						\\
			500   		& 10    	& $\checkmark$ 	&   						\\
			550   		& 11    	& $\checkmark$ 	&   						\\
			600   		& 12    	& $\checkmark$ 	&   						\\
			650   		& 13    	& $\checkmark$ 	&   						\\
			700   		& 14    	& $\checkmark$ 	&   						\\
			750   		& 15    	& $\checkmark$ 	&   						\\
		\end{tabular}
		\caption{Streamwise stations $x$ corresponding to the $T = 2.16$~s and $T = 12.12$~s acquisitions. Stations marked with a $\checkmark$ indicate where measurements were taken.}
		\label{tab:streamwise_stations}
	\end{center}
\end{table}

Table~\ref{tab:Mach_Re_U} summarizes the main parameters of the two jet flows.
The Reynolds number is computed as
\begin{equation}
    \mathit{Re}_{D_h} = \frac{U_j D_h}{\nu} \mbox{,}
    \label{eq:Re}
\end{equation}
\noindent where $\nu$ is the kinematic viscosity and $U_j$ is the jet exit velocity.
The number of convective time units is given by
\begin{equation}
    N_c = \frac{T U_j}{D_h} \mbox{.}
    \label{eq:Nc}
\end{equation}

\begin{table}
    \begin{center}
        \def~{\hphantom{0}}
        \begin{tabular}{c c c c c c c c c}
            $M_j$   & $U_j$ [m/s] & $\mathit{Re}_{D_h}$   & $N_s$     & $\Delta t_e$ [$\mu$s] & $T$ [s]   & $St_\mathit{Nyquist}$ & $N_c$ 	& $N_x$	\\
            \hline
            0.4     & 137         & $4.5 \times 10^5$     & 151,536   & 4                     & 12.12     & 2.28                  & 33,216	& 11   	\\
                    &               &           & (27,000)                &                   & (2.16)        &           &   (5,918) &   \\
            0.7     & 240         & $7.8 \times 10^5$     & 27,000    & 2                     & 2.16      & 1.30                  & 10,368 	& 32   	\\
        \end{tabular}
        \caption{Summary of experimental parameters: jet Mach number ($M_j$), jet exit velocity ($U_j$), Reynolds number ($\mathit{Re}_{D_h}$), number of snapshots ($N_s$), time between laser pulses ($\Delta t_e$), total acquisition time ($T$), Nyquist Strouhal number ($St_\mathit{Nyquist}$), number of convective time units ($N_c$), and number of streamwise cross-flow planes ($N_x$).}
        \label{tab:Mach_Re_U}
    \end{center}
\end{table}

From the raw PIV data, the average image over all acquisitions was subtracted from each frame in order to enhance the signal-to-noise ratio by reducing background noise.
A perspective correction was then applied to account for optical distortions.
The initial interrogation pass employed windows of $64 \times 64$ pixels with 50\% overlap, while the final pass used $16 \times 16$ pixel windows, also with 50\% overlap.
During the multi-pass post-processing, additional procedures were implemented, including the removal of vectors with correlation coefficients below 0.3 and the identification and replacement of outliers, among other standard quality-control steps.

The resulting velocity fields were obtained in Cartesian coordinates with $N_z = 194$ and $N_y = 142$ grid points in the spanwise and vertical directions, respectively.
The field of view (FoV) spanned approximately $(L_y,~L_z) \approx (4 D_h,~5 D_h)$ and the mesh resolution is of approximately $0.025 D_h$ (or $1.25$ mm) in both direction.
Further details on the experimental setup and processing procedures are available in~\citet{amaral2023installed}, \citet{amaral2023elliptical} and~\citet{amaral2024elliptical}.

%%%%%%%%%%%%%%%%%%%%%%%%%%%%%%%%%%%%%%%%%%%%%%
\subsection{Data post-processing}
\label{sec:experimental_processing}

%%%%%%%%%%%%%%%%%%%%%%%%%%%%%%%%%%%%%%%%%%%%%%
\subsubsection{D$_2$-symmetry}
\label{sec:D2}

As mentioned, given that Fourier symmetries are not available, we performed decomposition based on the dihedral group D$_2$.
This approach is commonly employed in flows that exhibit symmetry with respect to two orthogonal axes, such as twin jets~\citep{rodriguez2018wavepacket, yeung2022three, rodriguez2023preferred, padillamontero2024pressure} and rectangular jets~\citep{rodriguez2021near, marshall2024prediction, yeung2025spectral}.

The first step in the D$2$ decomposition involves identifying the jet centreline position $(\bar{y}, \bar{z})$, since in the experiments the jet axis may not be perfectly aligned with the geometric centre of the nozzle exit.
The centre is estimated using the mean streamwise velocity component $\bar{u}$ at each streamwise plane as
\begin{subeqnarray}
    \bar{y} &=& \frac{\sum_{i}{\bar{u}_i y_i}}{\sum_{i}{\bar{u}_i}} \mbox{,} \\
    \bar{z} &=& \frac{\sum_{i}{\bar{u}_i z_i}}{\sum_{i}{\bar{u}_i}} \mbox{,}
    \label{eq:center}
\end{subeqnarray}
\noindent where $i$ denotes the grid point index.
A similar approach is adopted by~\citet{gudmundsson2011instability}.

The second step involves decomposing the perturbation field into four symmetry components with respect to the minor and major axes: symmetric-symmetric (${\boldsymbol{q}_\mathit{SS}}^{\prime}$, SS), antisymmetric-symmetric (${\boldsymbol{q}_\mathit{AS}}^{\prime}$, AS), symmetric-antisymmetric (${\boldsymbol{q}_\mathit{SA}}^{\prime}$, SA), and antisymmetric-antisymmetric (${\boldsymbol{q}_\mathit{AA}}^{\prime}$, AA).
The mean flow is used as the reference state for the SPOD and stability analysis and is approximated by its symmetric-symmetric component, $\overline{\boldsymbol{q}}_\mathit{SS}$.
This choice is consistent with the symmetry of the ideal nozzle geometry and the symmetry-based formulation adopted for the perturbation analysis (dihedral group D$_2$).
However, the geometric symmetry does not necessarily imply that the experimentally measured time-averaged flow is perfectly symmetric, since coherent fluctuations can contribute to asymmetries in the mean field.
Accordingly, the use of $\overline{\boldsymbol{q}}_\mathit{SS}$ should be understood as a modelling assumption rather than a direct demonstration that the measured mean flow contains no antisymmetric components.

The total velocity field is then expressed as
\begin{equation}
    \boldsymbol{q} = {\boldsymbol{q}_\mathit{SS}}^{\prime} + {\boldsymbol{q}_\mathit{AS}}^{\prime} + {\boldsymbol{q}_\mathit{SA}}^{\prime} + {\boldsymbol{q}_\mathit{AA}}^{\prime} + \overline{\boldsymbol{q}}_\mathit{SS} \mbox{.}
    \label{eq:D2_decomposition}
\end{equation}
In equation~\eqref{eq:D2_decomposition}, $\boldsymbol{q}$ represents the velocity field, with $\boldsymbol{q} = [\boldsymbol{u}~\boldsymbol{v}~\boldsymbol{w}]$, where $\boldsymbol{u}$ is the streamwise velocity component and $\boldsymbol{v}$ and $\boldsymbol{w}$ are the cross-flow components.
The primes and overbars denote fluctuating and mean quantities, respectively, and the subscripts refer to the associated symmetry class.

\begin{figure}
    \centering
    \includegraphics[width=\textwidth]{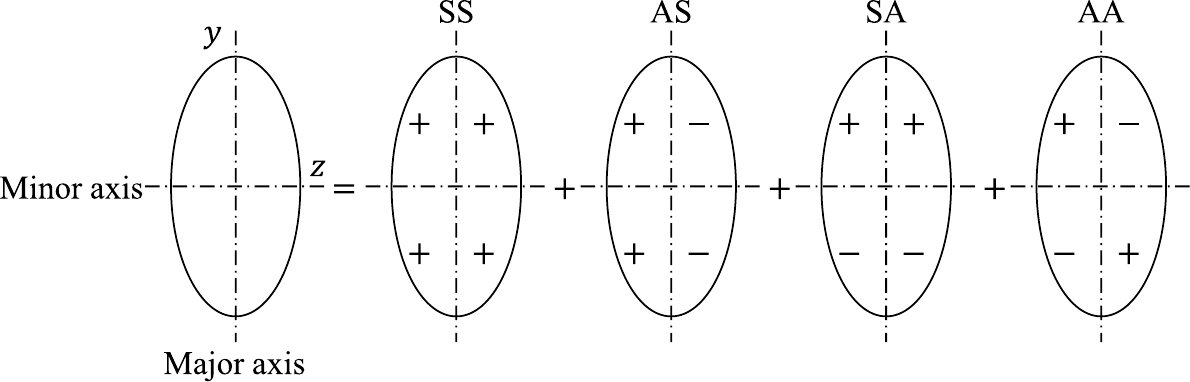}
    \caption{Dihedral group D$_2$ decomposition sketch.}
    \label{fig:D2groups}
\end{figure}

Figure~\ref{fig:D2groups} exhibits a sketch of the symmetry groups employed in~\eqref{eq:D2_decomposition} and a detailed methodology for extracting these symmetry components is provided in~\citet{yeung2025spectral}, with key equations presented in appendix~\ref{app:D2-decomposition}.

%%%%%%%%%%%%%%%%%%%%%%%%%%%%%%%%%%%%%%%%%%%%%%
\subsubsection{SPOD}
\label{sec:SPOD_method}

After extracting the D$_2$-symmetry components, the data are further processed using spectral proper orthogonal decomposition (SPOD)~\citep{picard2000pressure, towne2018spectral, schmidt2020guide}.
For each frequency, SPOD modes are computed via the eigenvalue decomposition of the cross-spectral density (CSD) matrix of the velocity field, following the D$_2$ decomposition and an appropriate inner-product norm $\boldsymbol{W}$, as
\begin{equation}
    \boldsymbol{C} \boldsymbol{W} \boldsymbol{\Phi} = \boldsymbol{\Phi} \boldsymbol{\Lambda} \mbox{,}
    \label{eq:spod}
\end{equation}
\noindent where $\boldsymbol{C}$ is the CSD matrix, $\boldsymbol{W}$ incorporates both integration weights and numerical quadrature over the discrete grid (i.e., the norm or inner product), $\boldsymbol{\Phi}$ denotes the SPOD eigenfunctions, and $\boldsymbol{\Lambda}$ is the diagonal matrix of eigenvalues.

We employ the kinetic-energy norm for the three velocity components~\citep{schmidt2020guide}.
The energy contribution of each SPOD mode is associated with the square of the eigenvalue, ranked in descending order of magnitude: $\lambda_1 \geq \lambda_2 \geq \cdots \geq \lambda_n$.
We used an in-house implementation of the snapshot-based SPOD algorithm~\citep{towne2018spectral, schmidt2020guide} for post-processing.

The CSD matrices for each D$_2$-symmetry component were computed using Welch’s method~\citep{welch1967fft}. 
Specifically, a Hanning window was applied, with block size $N_\mathit{fft} = 128$ and 50\% overlap between blocks.
The remaining Welch parameters for both jet Mach numbers are provided in Table~\ref{tab:Welch_parameters}.

\begin{table}
    \begin{center}
        \def~{\hphantom{0}}
        \begin{tabular}{c c c c c}
            $M_j$   & $N_\mathit{fft}$  & Overlap [\%]  & $N_b$ & $\Delta\mathit{St}$  \\
            \hline
            0.4     & 128               & 50            & 2367  & 0.036                 \\
                    &                   &               & (421) &                       \\
            0.7     & 128               & 50            & 421   & 0.020                 \\
        \end{tabular}
        \caption{Welch parameters used for SPOD: block size ($N_\mathit{fft}$), block overlap percentage, number of blocks ($N_b$), and Strouhal number discretization ($\Delta\mathit{St}$) for each jet Mach number ($M_j$).}
        \label{tab:Welch_parameters}
    \end{center}
\end{table}

Appendix~\ref{app:SPOD_convergence} presents a SPOD convergence study for the $M_j = 0.4$ databases, comparing acquisition times of $T = 2.16$ seconds and $T = 12.12$ seconds (see \S~\ref{sec:experimental_database}).
Overall, the $T = 12.12$ seconds dataset is much better converged.
Nevertheless, as discussed in Appendix~\ref{app:time_series_length}, the shorter database was still able to provide useful comparisons, due likely to the filtering property of the alignment projection used: the PSE mode filters the non-converged details of the SPOD mode.

%%%%%%%%%%%%%%%%%%%%%%%%%%%%%%%%%%%%%%%%%%%%%%
\section{Modelling}
\label{sec:theoretical_approach}
%%%%%%%%%%%%%%%%%%%%%%%%%%%%%%%%%%%%%%%%%%%%%%

%%%%%%%%%%%%%%%%%%%%%%%%%%%%%%%%%%%%%%%%%%%%%%
\subsection{3D-PSE}
\label{sec:pse_codes}

The non-uniformity in elliptical jets is due to the differing characteristics along the major and minor axes.  
Although it is theoretically possible to solve the full eigenvalue problem arising from discretization in all three spatial directions, such computations are expensive.  
An efficient alternative is to use a parabolised form of the stability equations (PSE) in three dimensions (3D-PSE), which is applicable when the mean flow varies slowly along one spatial coordinate.
In this framework, the three-dimensional eigenvalue problem is reformulated as an initial-value problem, which can be solved by marching spatially along the slowly varying direction~\citep{herbert1987stability,herbert1991boundary,chang1991compressible,bertolotti1991analysis}.

We analyse PIV mean flows measured at $M_j = 0.4$ and $M_j = 0.7$, with Reynolds numbers $\mathit{Re}_{D_h}$ high enough to justify the inviscid approximation.
The analysis begins with the compressible Euler equations,
\begin{subequations}
    \begin{equation}
        \rho \frac{D \mathbf{u}}{Dt} = - \frac{1}{\gamma M_j^2} \nabla (\rho T) \mbox{,}
    \end{equation}
    \begin{equation}
        \rho \frac{DT}{Dt} = - \frac{(\gamma - 1)}{\gamma} \rho T \nabla \cdot \mathbf{u} \mbox{,}
    \end{equation}
    \begin{equation}
        \frac{\partial \rho}{\partial t} + \nabla \cdot (\rho \mathbf{u}) = 0 \mbox{.}
    \end{equation}
    \label{eq:ns_comp}
\end{subequations}

The mean flow vector is defined as
\begin{equation}
    \bar{\boldsymbol{q}}(x,y,z) = \left[\bar{\boldsymbol{u}}(x,y,z)~\bar{\rho}(x,y,z)~\bar{T}(x,y,z)\right]^T \mbox{,}
    \label{eq:q}
\end{equation}
\noindent where $\bar{\boldsymbol{u}} = [\bar{U}~\bar{V}~\bar{W}]$ denotes the mean velocity field, $\bar{\rho}$ and $\bar{T}$ are obtained via the Crocco--Busemann relation, and $T$ denotes the transpose operator.

The quasi-parallel assumption decomposes the flow field $\boldsymbol{q}(x,y,z,t)$ into a base state $\bar{\boldsymbol{q}}(x,y,z)$ and a small-amplitude perturbation $\varepsilon \boldsymbol{q}^{\prime}(x,y,z,t)$ as
\begin{equation}
    \boldsymbol{q}(x,y,z,t) = \bar{\boldsymbol{q}}(x,y,z) + \varepsilon \boldsymbol{q}^{\prime}(x,y,z,t) \mbox{,}
    \label{eq:decomq}
\end{equation}
\noindent where $\boldsymbol{q}^{\prime}$ represents a rapidly oscillating wave with a slowly varying amplitude, expressed as
\begin{equation}
    \boldsymbol{q}^{\prime}(x,y,z,t) = \boldsymbol{\hat{q}}(x,y,z) \, e^{\mathrm{i}\left(\int_{x_0}^x \alpha(\xi) \, \mathrm{d}\xi - \omega t\right)} + \mathrm{c.c.} \mbox{,}
    \label{eq:modnormal_PSE}
\end{equation}
\noindent where $\mathrm{c.c.}$ denotes the complex conjugate, $\mathrm{i} = \sqrt{-1}$, $t$ is time, $\omega$ is the real frequency, $\alpha$ is the complex streamwise wavenumber, and the hat symbol indicates the frequency-domain representation.

Substituting \eqref{eq:modnormal_PSE} into the Euler equations \eqref{eq:ns_comp}, and neglecting terms of order $\mathcal{O}(\varepsilon^2)$, as well as $\partial \alpha / \partial x$, and higher-order derivatives of $\alpha$ and $\hat{\boldsymbol{q}}$ with respect to $x$ (where the hat denotes frequency-domain quantities), we obtain linearised, compressible, and inviscid 3D-PSE system as
\begin{subequations}
    \begin{equation}
        \begin{aligned}
            \bar{\rho} \bar{U} \frac{\partial \hat{w}}{\partial x} &= 
            i\omega \bar{\rho} \hat{w} 
            - \bar{\rho} \bar{W} \frac{\partial \hat{w}}{\partial z} 
            - \bar{\rho} \hat{w} \frac{\partial \bar{W}}{\partial z} 
            - \bar{\rho} \bar{V} \frac{\partial \hat{w}}{\partial y} 
            - \bar{\rho} \hat{v} \frac{\partial \bar{W}}{\partial y} \\
            &- i\alpha \bar{\rho} \bar{U} \hat{w} 
            - \bar{\rho} \hat{u} \frac{\partial \bar{W}}{\partial x} 
            - \rho' \left( \bar{W}\frac{\partial \bar{W}}{\partial z} + \bar{V}\frac{\partial \bar{W}}{\partial y} + 
            \bar{U}\frac{\partial \bar{W}}{\partial x} \right) \\
            &- \frac{1}{\gamma Ma^2} \left( \bar{T} \frac{\partial \rho'}{\partial z} + \tau' \frac{\partial \bar{\rho}}{\partial z} + \bar{\rho} \frac{\partial \tau'}{\partial z} + \rho' \frac{\partial \bar{T}}{\partial z} \right) \mbox{,}
        \end{aligned}
    \end{equation}
    
    \begin{equation}
        \begin{aligned}
        \bar{\rho} \bar{U} \frac{\partial \hat{v}}{\partial x} &=
        i\omega \bar{\rho} \hat{v} 
        - \bar{\rho} \bar{W} \frac{\partial \hat{v}}{\partial z} 
        - \bar{\rho} \hat{w} \frac{\partial \bar{V}}{\partial z} 
        - \bar{\rho} \bar{V} \frac{\partial \hat{v}}{\partial y} 
        - \bar{\rho} \hat{v} \frac{\partial \bar{V}}{\partial y} \\ 
        &- \bar{\rho} \bar{U} \hat{v} i\alpha 
        - \bar{\rho} \hat{u} \frac{\partial \bar{V}}{\partial x}
        -\rho' \left( \bar{W}\frac{\partial \bar{V}}{\partial z} + \bar{V}\frac{\partial \bar{V}}{\partial y} + 
        \bar{U}\frac{\partial \bar{V}}{\partial x} \right) \\
        &-\frac{1}{\gamma Ma^2} \left( 
        \bar{T} \frac{\partial \hat{\rho}}{\partial y} 
        + \hat{\tau} \frac{\partial \bar{\rho}}{\partial y} 
        + \bar{\rho} \frac{\partial \hat{\tau}}{\partial y} 
        + \hat{\rho} \frac{\partial \bar{T}}{\partial y} \right) \mbox{,}
        \end{aligned}
    \end{equation}

    \begin{equation}
        \begin{aligned}
        \bar{\rho} \bar{U} \frac{\partial \hat{u}}{\partial x} + \frac{1}{\gamma Ma^2} \bar{\rho} \frac{\partial \hat{\tau}}{\partial x} &+ \frac{1}{\gamma Ma^2} \bar{T} \frac{\partial \hat{\rho}}{\partial x} = 
        i\omega \bar{\rho} \hat{u} 
        - \bar{\rho} \bar{W} \frac{\partial \hat{u}}{\partial z} 
        - \bar{\rho} \hat{w} \frac{\partial \bar{U}}{\partial z} \\
        &- \bar{\rho} \bar{V} \frac{\partial \hat{u}}{\partial y}  
        - \bar{\rho} \hat{v} \frac{\partial \bar{U}}{\partial y} 
        - \bar{\rho} \bar{U} \hat{u} i\alpha 
        - \bar{\rho} \hat{u} \frac{\partial \bar{U}}{\partial x} \\
        &-\rho' \left( \bar{W}\frac{\partial \bar{U}}{\partial z} + \bar{V}\frac{\partial \bar{U}}{\partial y} + 
        \bar{U}\frac{\partial \bar{U}}{\partial x} \right) \\
        &-\frac{1}{\gamma Ma^2} \left( 
        i\alpha \bar{T} \hat{\rho} 
        + \hat{\tau} \frac{\partial \bar{\rho}}{\partial x} 
        + i\alpha \bar{\rho} \hat{\tau} 
        + \hat{\rho} \frac{\partial \bar{T}}{\partial x} \right) \mbox{,}
        \end{aligned}
    \end{equation}

    \begin{equation}
        \begin{aligned}
        \bar{\rho} \bar{U} \frac{\partial \hat{\tau}}{\partial x} &+ \frac{(\gamma - 1)}{\gamma} \bar{\rho} \bar{T} \frac{\partial \hat{u}}{\partial x} =  
        \bar{\rho} i \omega \hat{\tau} - \bar{\rho} \bar{W} \frac{\partial \hat{\tau}}{\partial z} - \bar{\rho} \hat{w} \frac{\partial \bar{T}}{\partial z} \\
        &- \bar{\rho} \bar{V} \frac{\partial \hat{\tau}}{\partial y} - \bar{\rho} \hat{v} \frac{\partial \bar{T}}{\partial y} - \bar{\rho} i \alpha \bar{U} \hat{\tau} - \bar{\rho} \hat{u} \frac{\partial \bar{T}}{\partial x} \\
        &- \hat{\rho} \bar{W} \frac{\partial \bar{T}}{\partial z} -\hat{\rho} \bar{V} \frac{\partial \bar{T}}{\partial y} -\hat{\rho} \bar{U} \frac{\partial \bar{T}}{\partial x} -\frac{(\gamma - 1)}{\gamma} \Bigg[ 
        \hat{\rho} \bar{T} \frac{\partial \bar{W}}{\partial z} 
        + \hat{\rho} \bar{T} \frac{\partial \bar{V}}{\partial y} \\
        &+ \hat{\rho} \bar{T} \frac{\partial \bar{U}}{\partial x} 
        + \bar{\rho} \bar{T} \frac{\partial \hat{w}}{\partial z} 
        + \bar{\rho} \bar{T} \frac{\partial \hat{v}}{\partial y} 
        + i \alpha \bar{\rho} \bar{T} \hat{u} \\
        &+\tau \bar{\rho} \frac{\partial \bar{W}}{\partial z} 
        + \tau \bar{\rho} \frac{\partial \bar{V}}{\partial y} 
        + \tau \bar{\rho} \frac{\partial \bar{U}}{\partial x} \Bigg] \mbox{,}
        \end{aligned}
    \end{equation}

    \begin{equation}
        \begin{aligned}
        \bar{\rho} \frac{\partial \hat{u}}{\partial x} + \bar{U} \frac{\partial \hat{\rho}}{\partial x} &= 
        - \hat{\rho} \left( -i\omega + \frac{\partial \bar{W}}{\partial z} + \frac{\partial \bar{V}}{\partial y} + \frac{\partial \bar{U}}{\partial x} + i\alpha \bar{U} \right) \\
        &- \hat{w} \frac{\partial \bar{\rho}}{\partial z}
        - \hat{v} \frac{\partial \bar{\rho}}{\partial y}
        - \hat{u} \left( \frac{\partial \bar{\rho}}{\partial x} + i\alpha \bar{\rho} \right) \\
        &- \bar{\rho} \left( \frac{\partial \hat{w}}{\partial z} + \frac{\partial \hat{v}}{\partial y} \right)
        - \frac{\partial \hat{\rho}}{\partial z} \bar{W}
        - \frac{\partial \hat{\rho}}{\partial y} \bar{V} \mbox{.}
        \end{aligned}
    \end{equation}
    \label{eq:nspse3D}
\end{subequations}

In matrix form, the operators $\mathcal{M}$ and $\mathcal{N}$ are defined as
\begin{equation}
    \mathcal{M} =  
   \begin{pmatrix}
    %%%%%%%%%%%%%%%%%%%%%%%%% M1 %%%%%%%%%%%%%%%%%%%%%%%%%
    \mathcal{M}_{11} & 0 & 0 & 0 & 0 \\
    %%%%%%%%%%%%%%%%%%%%%%%%% M2 %%%%%%%%%%%%%%%%%%%%%%%%%
    0 & \mathcal{M}_{22} & 0 & 0 & 0 \\
    %%%%%%%%%%%%%%%%%%%%%%%%% M3 %%%%%%%%%%%%%%%%%%%%%%%%%
    0 & 0 & \mathcal{M}_{33} & \mathcal{M}_{34} & \mathcal{M}_{35} \\
    %%%%%%%%%%%%%%%%%%%%%%%%% M4 %%%%%%%%%%%%%%%%%%%%%%%%%
    0 & 0 & \mathcal{M}_{43} & \mathcal{M}_{44} & 0 \\ 
    %%%%%%%%%%%%%%%%%%%%%%%%% M5 %%%%%%%%%%%%%%%%%%%%%%%%%
    0 & 0 & \mathcal{M}_{53} & 0 & \mathcal{M}_{55} \\ 
    \end{pmatrix}
    \mbox{,}
\end{equation}
\noindent and
\begin{equation}
    \mathcal{N} =  
  \begin{pmatrix}
    %%%%%%%%%%%%%%%%%%%%%%%%% N1 %%%%%%%%%%%%%%%%%%%%%%%%%
    \mathcal{N}_{11} & \mathcal{N}_{12} & \mathcal{N}_{13} & \mathcal{N}_{14} & \mathcal{N}_{15} \\
    %%%%%%%%%%%%%%%%%%%%%%%%% N2 %%%%%%%%%%%%%%%%%%%%%%%%%
    \mathcal{N}_{21} & \mathcal{N}_{22} & \mathcal{N}_{23} & \mathcal{N}_{24} & \mathcal{N}_{25} \\
    %%%%%%%%%%%%%%%%%%%%%%%%% N3 %%%%%%%%%%%%%%%%%%%%%%%%%
    \mathcal{N}_{31} & \mathcal{N}_{32} & \mathcal{N}_{33} & \mathcal{N}_{34} & \mathcal{N}_{35} \\
    %%%%%%%%%%%%%%%%%%%%%%%%% N4 %%%%%%%%%%%%%%%%%%%%%%%%%
    \mathcal{N}_{41} & \mathcal{N}_{42} & \mathcal{N}_{43} & \mathcal{N}_{44} & 0 \\
    %%%%%%%%%%%%%%%%%%%%%%%%% N5 %%%%%%%%%%%%%%%%%%%%%%%%%
    \mathcal{N}_{51} & \mathcal{N}_{52} & \mathcal{N}_{53} & \mathcal{N}_{54} & \mathcal{N}_{55} 
    \end{pmatrix}
    \mbox{.}
\end{equation}
Operators $\mathcal{M}_{ij}$ and $\mathcal{N}_{ij}$ are defined in Appendix~\ref{app:3dpse_operators}.

In compact matrix form, defining $\hat{\boldsymbol{q}}=[\hu~\hv~\hw~\hat{\rho}~\hat{\tau}]^T$,  equation~\eqref{eq:nspse3D} becomes
\begin{equation}
    \mathcal{M}\difp{\boldsymbol{\hq}}{x} = \mathcal{N}\boldsymbol{\hq} \mbox{.}
    \label{eq:pseprob}
\end{equation}
\noindent To solve equation~\eqref{eq:pseprob} numerically, a first-order implicit Euler scheme is employed, yielding the semi-discrete form
\begin{equation}
    (\mathcal{M} - \Delta x \mathcal{N}) \hat{\boldsymbol{q}}_{j+1} = \mathcal{M}\hat{\boldsymbol{q}}_{j} \mbox{,}
    \label{eq:pseprocess}
\end{equation}
\noindent where $j$ denotes the streamwise index position.

At each streamwise station $x$, the complex wavenumber $\alpha$ is iteratively updated—where $n$ denotes the iteration index—until a normalization condition is satisfied. This condition ensures that the assumption of weak streamwise variation remains valid.
The normalization condition is given by
\begin{equation}
    \alpha_{n+1}^{j+1} = \alpha_{n}^{j+1} - \frac{i}{\Delta x}\frac{ \underset{\Omega}{\int\int}(\hat{\boldsymbol{q}}_n^{j+1})^*(\hat{\boldsymbol{q}}_n^{j+1} - \hat{\boldsymbol{q}}^{j}) dz dy   }{\underset{\Omega}{\int\int}\mid \hat{\boldsymbol{q}}_n^{j+1} \mid^2 dz dy } \mbox{,}
    \label{eq:psenormalization}
\end{equation}
\noindent where $\Delta x$ is the streamwise step size, and $\Omega$ denotes the $yz$-plane domain.

%%%%%%%%%%%%%%%%%%%%%%%%%%%%%%%%%%%%%%%%%%%%%%
\subsection{2D Local instability}
\label{sec:2dlst}

An initial solution is required to start the initial-value problem defined by equations~\eqref{eq:pseprocess} and \eqref{eq:psenormalization}. 
This is typically obtained by performing a two-dimensional (2D) local stability analysis at a given streamwise location $x_0$, often chosen near the nozzle, for a prescribed frequency $\omega$.

In this approach, the non-parallel components of the mean flow, $\bar{V}$ and $\bar{W}$, are neglected, and the streamwise variation of the mean flow is assumed to be weak.
Accordingly, it is assumed that $d\alpha/dx \approx 0$ and $\partial \boldsymbol{q}/\partial x = i\alpha\boldsymbol{q}$.
Under these assumptions, equation~\eqref{eq:nspse3D} reduces to the following eigenvalue problem (EVP)
\begin{equation}
    \mathcal{A}\cdot\tilde{\boldsymbol{q}} = \alpha\mathcal{B}\cdot\tilde{\boldsymbol{q}} \mbox{,}
    \label{eq:egv}
\end{equation}
\noindent where the operators $\mathcal{A}$ and $\mathcal{B}$ collect the relevant terms from the simplified linearized Euler equations.
The resulting system reads
\begin{subequations}
    \begin{equation}
        \begin{aligned}
        &\bar{\rho} \left( -i \omega \hat{w} + i \alpha \bar{U} \hat{w} \right) = 
        - \frac{1}{\gamma Ma^2} \left( \bar{T} \frac{\partial \hat{\rho}}{\partial z} + \hat{\tau} \frac{\partial \bar{\rho}}{\partial z} + \bar{\rho} \frac{\partial \hat{\tau}}{\partial z} + \hat{\rho} \frac{\partial \bar{T}}{\partial z} \right) \mbox{,}
        \end{aligned}
    \end{equation}

    \begin{equation}
        \begin{aligned}
        &\bar{\rho} \left( -i \omega \hat{v} + i \alpha \bar{U} \hat{v} \right) = 
        - \frac{1}{\gamma Ma^2} \left( \bar{T} \frac{\partial \hat{\rho}}{\partial y} + \hat{\tau} \frac{\partial \bar{\rho}}{\partial y} + \bar{\rho} \frac{\partial \hat{\tau}}{\partial y} + \hat{\rho} \frac{\partial \bar{T}}{\partial y} \right) \mbox{,}
        \end{aligned}
    \end{equation}

    \begin{equation}
        \begin{aligned}
        &\bar{\rho} \left( -i\omega \hat{u} + \hat{w} \frac{\partial \bar{U}}{\partial z} + \hat{v} \frac{\partial \bar{U}}{\partial y} + \bar{U} i\alpha \hat{u} \right) =
        - \frac{1}{\gamma Ma^2} \left( i \alpha \bar{T} \hat{\rho} + i \alpha \bar{\rho} \hat{\tau} \right) \mbox{,}
        \end{aligned}
    \end{equation}
    
    \begin{equation}
        \begin{aligned}
        &\bar{\rho} \left( -i \omega \hat{\tau} + \hat{w} \frac{\partial \bar{T}}{\partial z} + \hat{v} \frac{\partial \bar{T}}{\partial y} + \bar{U} i \alpha \hat{\tau} \right) = - \frac{(\gamma - 1)}{\gamma} \bar{\rho} \bar{T} \left( \frac{\partial \hat{w}}{\partial z} + \frac{\partial \hat{v}}{\partial y} + i \alpha \hat{u} \right) \mbox{,}
        \end{aligned}
    \end{equation}
    
    \begin{equation}
        \begin{aligned}
        -i\omega \hat{\rho} + \bar{\rho} \frac{\partial \hat{w}}{\partial z} + \hat{w} \frac{\partial \bar{\rho}}{\partial z} 
        + \bar{\rho} \frac{\partial \hat{v}}{\partial y} + \hat{v} \frac{\partial \bar{\rho}}{\partial y} 
        + i\alpha \bar{\rho} \hat{u} + i\alpha \bar{U} \hat{\rho} = 0 \mbox{.}
        \end{aligned}
    \end{equation}
\end{subequations}

In matrix notation, the operators $\mathcal{A}$ and $\mathcal{B}$ that define the eigenvalue problem in equation~\eqref{eq:egv} can be written as
\begin{equation}
    \mathcal{A} =  
    \begin{pmatrix}
    %%%%%%%%%%%%%%%%%%%%%%%%% A1 %%%%%%%%%%%%%%%%%%%%%%%%%
    \mathcal{A}_{11} & 0 & 0 & \mathcal{A}_{14} & \mathcal{A}_{15} \\
    %%%%%%%%%%%%%%%%%%%%%%%%% A2 %%%%%%%%%%%%%%%%%%%%%%%%%
    0 & \mathcal{A}_{22} & 0 & \mathcal{A}_{24} & \mathcal{A}_{25} \\
    %%%%%%%%%%%%%%%%%%%%%%%%% A3 %%%%%%%%%%%%%%%%%%%%%%%%%
    \mathcal{A}_{31} & \mathcal{A}_{32} & \mathcal{A}_{33} & 0 & 0 \\
    %%%%%%%%%%%%%%%%%%%%%%%%% A4 %%%%%%%%%%%%%%%%%%%%%%%%%
    \mathcal{A}_{41} & \mathcal{A}_{42} & 0 & \mathcal{A}_{44} & 0 \\ 
    %%%%%%%%%%%%%%%%%%%%%%%%% A5 %%%%%%%%%%%%%%%%%%%%%%%%%
    \mathcal{A}_{51} & \mathcal{A}_{52} & 0 & 0 & \mathcal{A}_{55} \\
    \end{pmatrix} \mbox{,}
\end{equation}
\noindent and
\begin{equation}
    \mathcal{B} =  
    \begin{pmatrix}
    %%%%%%%%%%%%%%%%%%%%%%%%% B1 %%%%%%%%%%%%%%%%%%%%%%%%%
    \mathcal{B}_{11} & 0 & 0 & 0 & 0 \\
    %%%%%%%%%%%%%%%%%%%%%%%%% B2 %%%%%%%%%%%%%%%%%%%%%%%%%
    0 & \mathcal{B}_{22} & 0 & 0 & 0 \\
    %%%%%%%%%%%%%%%%%%%%%%%%% B3 %%%%%%%%%%%%%%%%%%%%%%%%%
    \mathcal{B}_{31} & \mathcal{B}_{32} & \mathcal{B}_{33} & \mathcal{B}_{34} & \mathcal{B}_{35} \\
    %%%%%%%%%%%%%%%%%%%%%%%%% AB4 %%%%%%%%%%%%%%%%%%%%%%%%%
    0 & 0 & \mathcal{B}_{43} & \mathcal{B}_{44} & 0 \\ 
    %%%%%%%%%%%%%%%%%%%%%%%%% B5 %%%%%%%%%%%%%%%%%%%%%%%%%
    0 & 0 & \mathcal{B}_{53} & 0 & \mathcal{B}_{55} \\
    \end{pmatrix} \mbox{.}
\end{equation}
Operators $\mathcal{A}_{ij}$ and $\mathcal{B}_{ij}$ are defined in Appendix~\ref{app:2dlocal_operators}.

Solution of the eigenvalue problem~\eqref{eq:egv} provides the initial pair $(\alpha_0,~\tilde{\boldsymbol{q}}_0)$ required to initialize the PSE marching procedure.

%%%%%%%%%%%%%%%%%%%%%%%%%%%%%%%%%%%%%%%%%
\subsection{Numerical methods}
\label{sec:numerical_methods}

The numerical solution of both the PSE linear system~\eqref{eq:pseprocess} and the eigenvalue problem (EVP)~\eqref{eq:egv} requires spatial discretization over the $zy$-plane.
For elliptic jets, this is accomplished via a coordinate transformation to generalized coordinates.
In this approach, the Cartesian domain is mapped onto an elliptic domain, which can be interpreted as a stretched cylindrical coordinate system.

This transformation involves expressing the differential operators $\mathcal{D}_y$ and $\mathcal{D}_z$ in terms of generalized coordinates~\citep{visbal2002use,merle2009resolution,paredes2014advances}.
The mapping from Cartesian to elliptic coordinates is defined by
\begin{equation}
    \begin{aligned}
        z &= \mathit{AR} \, r\cos{\theta} \mbox{,} \\
        y &= r\sin{\theta} \mbox{,}
    \end{aligned}
\end{equation}
\noindent where $\mathit{AR}$ is the aspect ratio of the mesh, acting as a stretching factor, and is set to 1.8.
This value should be distinguished from the geometrical aspect ratio of the jet: here, $\mathit{AR}$ defines the computational coordinate mapping and controls the distribution of grid points across the cross-stream direction.
Note the nozzle has an aspect ratio of $\mathit{AR} = 1{:}2$.
However, as the jet develops downstream, this ellipticity gradually decreases, as will be addressed in \S\ref{sec:mean_flow}.
In other words, sufficiently far from the nozzle, the jet tends to recover the axisymmetry.
Since the PSE analysis is started at $x/D_h=0.1$, the flow has already slightly evolved from the nozzle exit.
The value $\mathit{AR}=1.8$ was therefore selected based on the mean-flow distribution at this location, as it provides a suitable representation of the shear-layer region and a satisfactory numerical convergence of the PSE calculation.
It is not intended to exactly match the geometrical aspect ratio of the jet at each streamwise location, which evolves downstream.

The resulting mesh in the $zy$-plane consists of $N_r \times N_\theta$ grid points.
In the radial direction $r$, $N_r = 140$ points are distributed following a Chebyshev–Gauss–Lobatto (CGL) distribution over $\xi \in [-1,~1]$.
Following~\citep{malik1990numerical}, a mapping transforms the computational domain to the physical interval $[0,~r_\infty]$, given by
\begin{equation}
    r = l \frac{(1-\xi)}{(1+s+\xi)} \mbox{,}
    \quad
    s = 2l/r_\infty \mbox{,}
    \quad
    l = \frac{r_\infty r_c}{r_\infty - 2r_c} \mbox{,}
    \label{eq:rcgl}
\end{equation}
\noindent where half of the grid points are distributed within the region $0 \leq r \leq r_c$ and $r_\infty$ the location where the calculation domain is truncated.
In the present calculations, the outer boundary is located at $r_\infty = 20D_h$.
The mean flow obtained from the PIV measurements is interpolated from the experimental $(y,z)$ grid onto the elliptical PSE grid.
The PIV field covers a region of approximately $(4 \times 5)D_h$ in the $yz$-plane, whereas the PSE domain extends to $r_\infty=20D_h$.
The portion of the PSE domain lying outside the measured PIV region is filled with the ambient state.
At these radial locations, the mean velocity and velocity gradients are already very small, so this outer extension is not expected to significantly affect the shear-layer dynamics governing the instability.
Consequently, part of the PSE domain lies outside the experimentally measured region.
In this outer region, the measured mean flow is extended toward the ambient state, where the velocity gradients and the mean-flow velocity are already very small.
Since the instability dynamics considered here are primarily governed by the shear layers close to the jet, this outer extension is not expected to significantly affect the PSE solutions.

In the azimuthal direction $\theta$, a uniform grid with $N_\theta = 41$ points is employed over a quarter domain $0 \leq \theta \leq \pi/2$, leveraging the symmetry of the elliptic jet about its major and minor axes.

Spatial derivatives are approximated using the Dispersion-Relation-Preserving (DRP) finite-difference scheme~\citep{tam1993dispersion,lele1992compact}, which is superior to classical finite-difference methods in preserving wave propagation characteristics, as it targets the numerical dispersion relation rather than solely minimizing truncation errors.

Dirichlet boundary conditions are imposed at the outer boundary $r \to r_\infty$, while symmetry conditions at the centerline $r \to 0$ are enforced using a combination of Neumann and Dirichlet boundary conditions, depending on the parity of each perturbation quantity with respect to the jet's symmetry axes.

The EVP~\eqref{eq:egv} is solved using the Arnoldi method.
Studies by~\citet{natarajan1992arnoldi} and~\citet{nayar1993computation} demonstrated the effectiveness of this algorithm for linear stability analyses, particularly when combined with the QZ method~\citep{golub2013matrix}.
Subsequent applications to multidimensional and large-scale problems were performed by~\citet{ehrenstein1996linear} and~\citet{theofilis1997verification}, respectively.

The PSE system~\eqref{eq:pseprocess} is integrated using sparse linear algebra routines from the \texttt{SciPy} library~\citep{virtanen2020fundamental}, which offer efficient algorithms for solving large-scale sparse systems.

%%%%%%%%%%%%%%%%%%%%%%%%%%%%%%%%%%%%%%%%%%%%%%
\subsubsection{D$_2$ decomposition}
\label{sec:azimuthal_decomposition}

Solving the eigenvalue problem associated with local parallel instability, followed by downstream integration using the three-dimensional PSE, is computationally demanding.
A common strategy to reduce this cost is to exploit the symmetries of the flow.
In the case of elliptical jets, symmetric or anti-symmetric boundary conditions can be enforced along the major axis ($z = 0$) and minor axis ($y = 0$) of the ellipse.
This approach enables decomposition of the full problem into the four distinct mode families used in decomposition of the experimental data: SS, AS, SA, and AA.

This symmetry-based classification not only reduces computational expense but also aids in physical interpretation and comparison with data, which, as shown above, is similarly decomposed.
The strategy has been successfully employed in the analysis of rectangular and twin jets, as shown by~\citet{rodriguez2018wavepacket, rodriguez2021near, yeung2025spectral, rodriguez2023preferred, padillamontero2024pressure, marshall2024prediction}.
Note that, because symmetry is explicitly enforced through the D$_2$ decomposition, the stability analysis is less sensitive to possible three-dimensional asymmetries or flow drift.

Due to the azimuthal dependence of the eigenfunctions governed by their symmetry properties, these functions can be expanded in sine and cosine series.
This leads to a natural analogy with the classical azimuthal mode decomposition in axisymmetric jets.
Table~\ref{tab:mode-symmetry} summarizes the symmetry classification of the mode families and shows their approximate correspondence with the azimuthal wavenumbers m in the axisymmetric case, following the same convention adopted by \citet{rodriguez2021near} for rectangular jets.

\begin{table}
    \centering
    \begin{tabular}{c c c c}
        $(x,z)$-plane  & $(x,y)$-plane      & Nomenclature  & Equivalent $m$        \\
        \hline
        Symmetric       & Symmetric         & SS            & $m = 0, 2, 4, \dots$  \\
        Anti-Symmetric  & Symmetric         & AS            & $m = 1, 3, 5, \dots$  \\
        Symmetric       & Anti-Symmetric    & SA            & $m = 1, 3, 5, \dots$  \\
        Anti-Symmetric  & Anti-Symmetric    & AA            & $m = 2, 4, 6, \dots$  \\
    \end{tabular}
    \caption{Classification of mode families based on symmetry in the $(x,z)$ and $(x,y)$ planes, along with their approximate correspondence to azimuthal wavenumber $m$ in axisymmetric jets.}
    \label{tab:mode-symmetry}
\end{table}

%%%%%%%%%%%%%%%%%%%%%%%%%%%%%%%%%%%%%%%%%%%%%%
\subsection{Comparison with SPOD}
\label{sec:alignment}

To compare experimentally extracted flow structures and theoretical predictions, the alignment between the leading D$_2$-SPOD mode and the corresponding 3D-PSE solution is assessed for each symmetry family.
This comparison is quantified using an alignment metric defined as~\citep{cavalieri2013wavepackets}
\begin{equation}
    \beta(\mathit{St},x) = \frac{\left\lvert\ \langle \boldsymbol{\hat{q}}_\mathit{PSE}(\mathit{St},x,y,z),~{\boldsymbol{\hat{q}}_\mathit{SPOD}}(\mathit{St},x,y,z) \rangle \right\rvert}{ \| \boldsymbol{\hat{q}}_\mathit{PSE}(\mathit{St},x,y,z) \| \hspace{2pt} \| {\boldsymbol{\hat{q}}_\mathit{SPOD}}(\mathit{St},x,y,z) \|} \mbox{,}
    \label{eq:beta}
\end{equation}
\noindent where $\beta$ is the alignment coefficient, with $\vert\beta\vert \in [0,~1]$.
Here, $\langle \cdot, \cdot \rangle$ and $\|\cdot\|$ denote the $L_2$ inner product and norm, respectively, and the numerator includes the absolute value to account for global phase differences between modes.

A value of $\vert\beta\vert = 1$ indicates perfect alignment between the SPOD and PSE modes, while $\vert\beta\vert = 0$ denotes orthogonality.
The computation includes all three components of the velocity field, and both the real and imaginary parts of the complex-valued mode shapes.

Note that one cannot know \emph{a priori} whether a given PSE solution will match an SPOD mode.
The SPOD modes contains the true physics associated with the signature of the full nonlinear turbulent dynamics associated with the symmetry under consideration.
When agreement with PSE is observed, however, the latter provides a physical interpretation of the SPOD mode, which would otherwise remain a purely empirical object.
Indeed, the PSE solutions describe the spatial evolution, in a weakly non-parallel and non-axisymmetric flow, of instability-wave mechanisms initiated from the locally parallel stability analysis and associated with Kelvin--Helmholtz-type amplification.
Therefore, when a linear model is able to reproduce the experimentally extracted structures, one may infer that the amplification mechanisms embedded in the linear operator are relevant to the dynamics of the turbulent flow.
At the same time, it is important to emphasize that there is no fundamental reason why a linearized mean-flow model should accurately reproduce the dynamics of a fully turbulent flow.
The validity of the model can only be established \emph{a posteriori}, through comparison with experimental observations, from which the corresponding physical interpretation then follows.

%%%%%%%%%%%%%%%%%%%%%%%%%%%%%%%%%%%%%%%%%%%%%%
\section{Results}
\label{sec:results}
%%%%%%%%%%%%%%%%%%%%%%%%%%%%%%%%%%%%%%%%%%%%%%

In this section, the majority of results focus on the $M_j = 0.4$ jet flow case, while the $M_j = 0.7$ case is briefly discussed for completeness and to illustrate a Mach-number effect on coherent structures in elliptical jets.

%%%%%%%%%%%%%%%%%%%%%%%%%%%%%%%%%%%%%%%%%%%%%%
\subsection{Mean flow}
\label{sec:mean_flow}

\begin{figure}
    \centering
    \includegraphics[width=\textwidth]{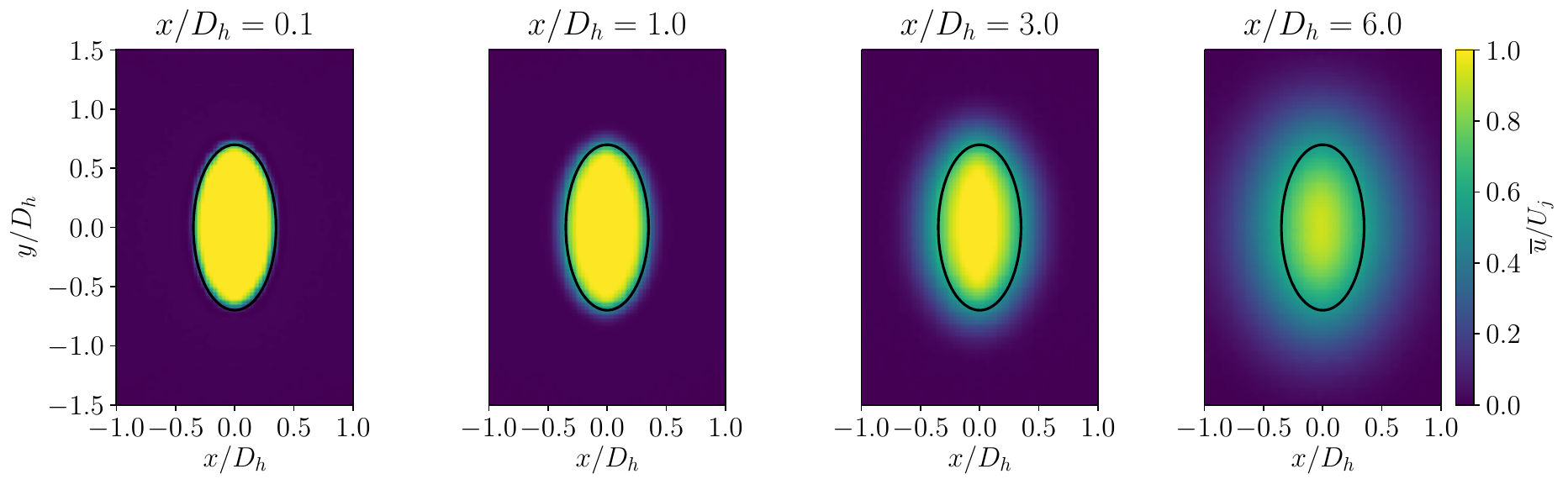}
    \caption{Symmetric-symmetric streamwise velocity component mean flow for the $M_j = 0.4$ case, showing the three velocity components at streamwise locations $x/D_h = 0.1$, $1$, $3$, and $6$ (left to right). All quantities are normalized by the jet exit speed $U_j$. The elliptical curve indicates the nozzle lip.}
    \label{fig:u_mean_Mj04}
\end{figure}

\begin{figure}
    \centering
    \includegraphics[width=\textwidth]{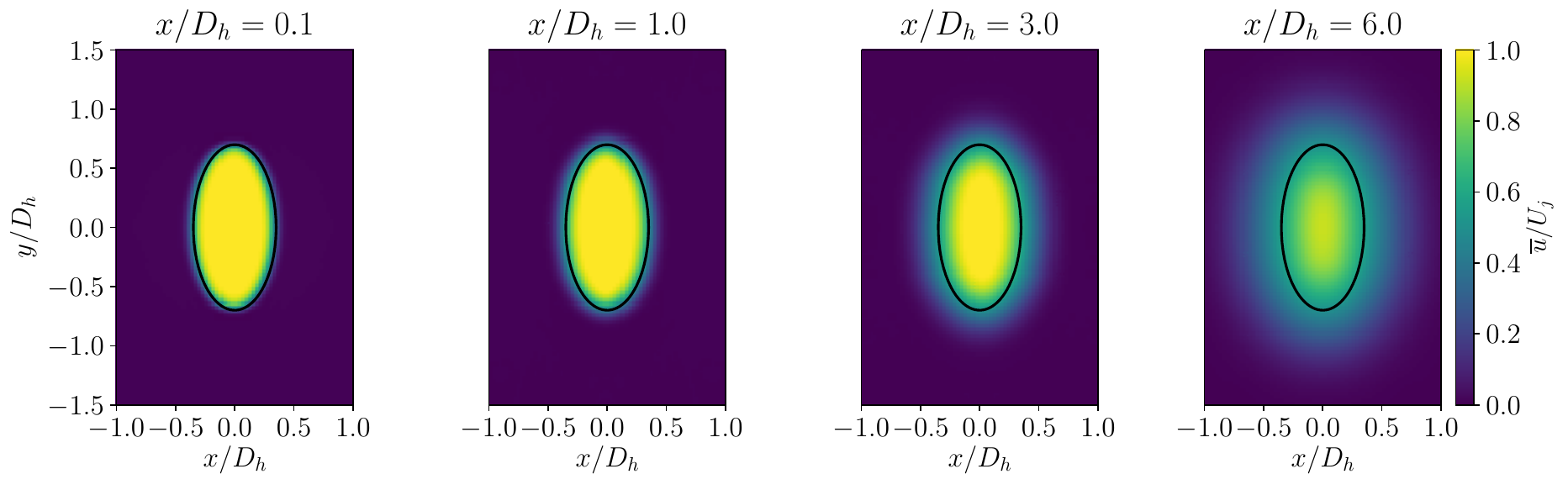}
    \caption{Symmetric-symmetric streamwise velocity component mean flow for the $M_j = 0.7$ case. See caption of Figure~\ref{fig:u_mean_Mj04} for details.}
    \label{fig:u_mean_Mj07}
\end{figure}

Figures~\ref{fig:u_mean_Mj04} and \ref{fig:u_mean_Mj07} show the symmetric-symmetric (SS) streamwise velocity component mean flow for $M_j = 0.4$ and $M_j = 0.7$, respectively.
The velocity components are displayed at selected streamwise stations ($x/D_h$), with all quantities normalized by the jet exit velocity $U_j$.
These representations provide a direct visualization of the mean-flow development for the two Mach number and serve as a reference for the subsequent analysis of the flow topology.

\begin{figure}
    \centering
    \includegraphics[width=0.7\textwidth]{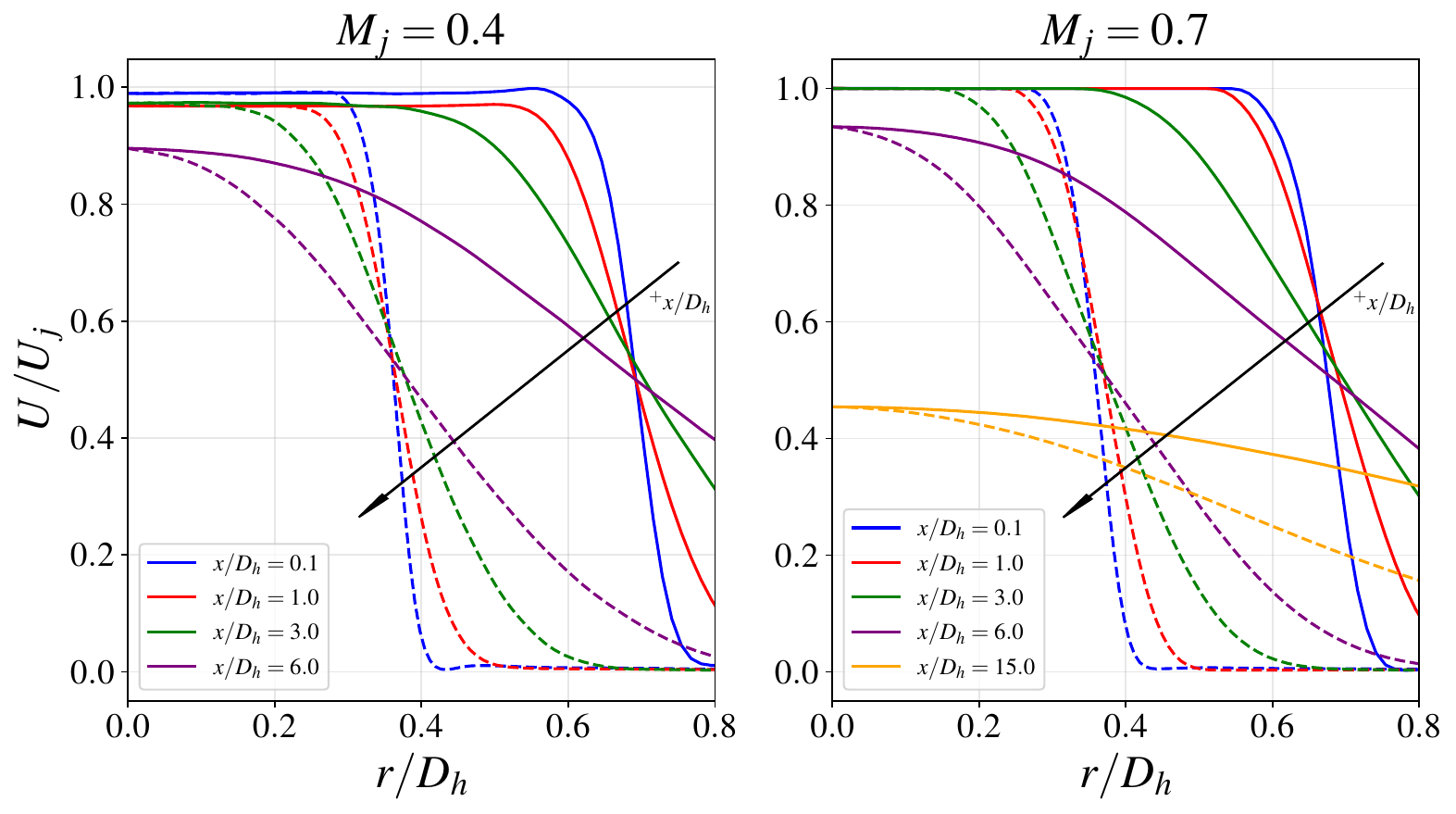}
    \caption{Symmetric-symmetric streamwise velocity component mean flow profiles for $M_j = 0.4$ (left) and 0.7 (right) and selected streamwise stations. Continuous curves denote the jet major axis, whereas dotted curves correspond to the jet minor axis.}
    \label{fig:u_mean_axis}
\end{figure}

Figure~\ref{fig:u_mean_axis} shows the jet profiles for the same $x/D_h$ as in the previous figures, with the addition of $x/D_h = 15$ for $M_j = 0.7$, since this is the most downstream station measured for such jet (see Table~\ref{tab:streamwise_stations}).
Continuous and dashed lines denote profiles measured at the jet major and minor axis, respectively.

\begin{figure}
    \centering
    \includegraphics[width=0.5\textwidth]{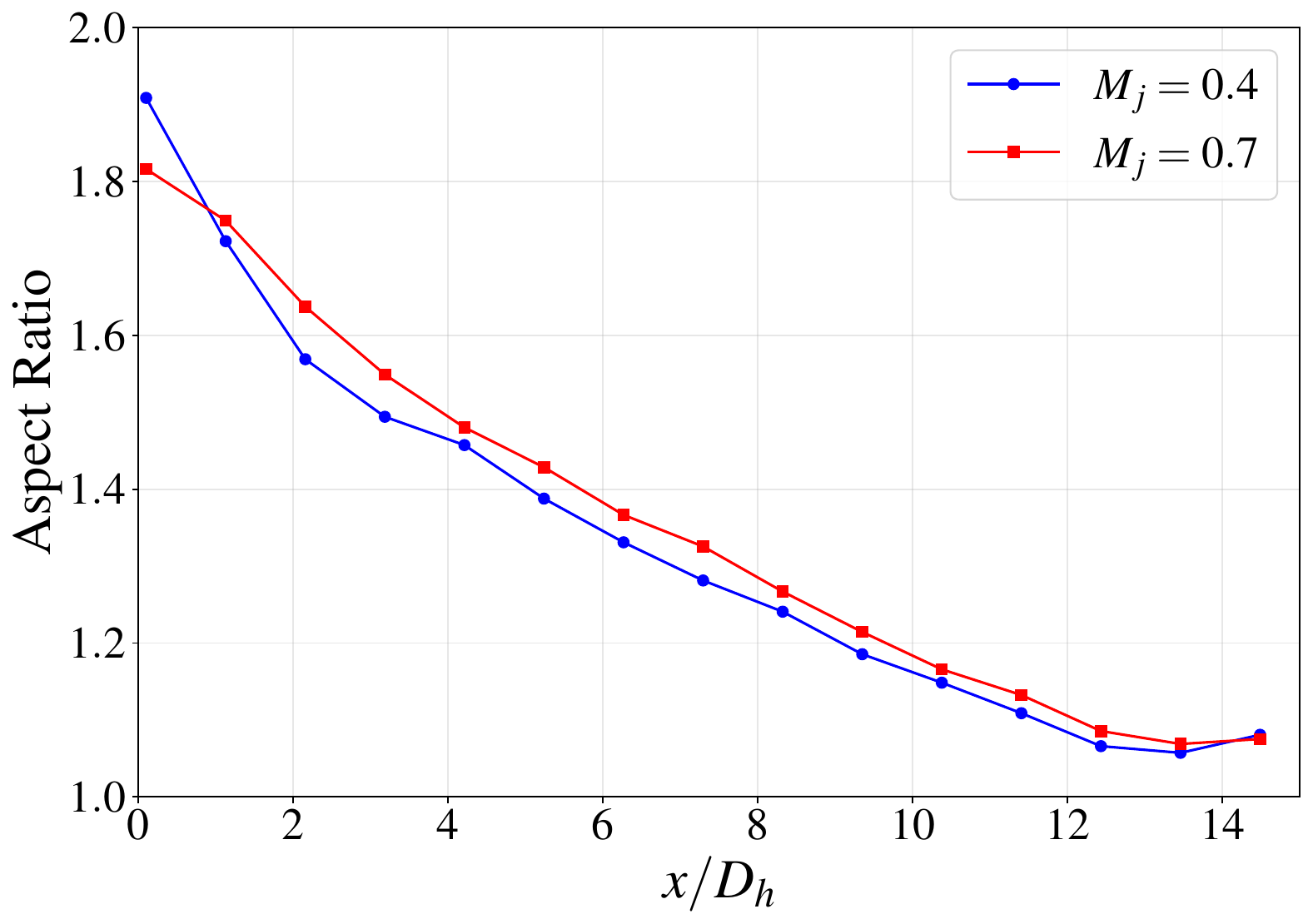}
    \caption{Evolution of the aspect ratio of the elliptical jet shape along the streamwise direction for ($M_j = 0.4$) (blue circles) and ($M_j = 0.7$) (red squares).}
    \label{fig:ar_evolution}
\end{figure}

The jet remains elliptical throughout the entire streamwise measurement range, with variations in the aspect ratio between the major and minor axes at different ($x/D_h$) locations, and no evidence of axis switching is observed, most likely due to the thick turbulent boundary layers present at the nozzle exit plane. This behaviour is illustrated in Figure~\ref{fig:ar_evolution}. Note that the jet progressively evolves toward a more axisymmetric shape downstream, and the stability calculations employ the measured mean flow (after D$_2$ decomposition) at each streamwise station, thereby naturally accounting for the gradual decay of ellipticity.

\begin{figure}
    \centering
    \includegraphics[width=0.5\textwidth]{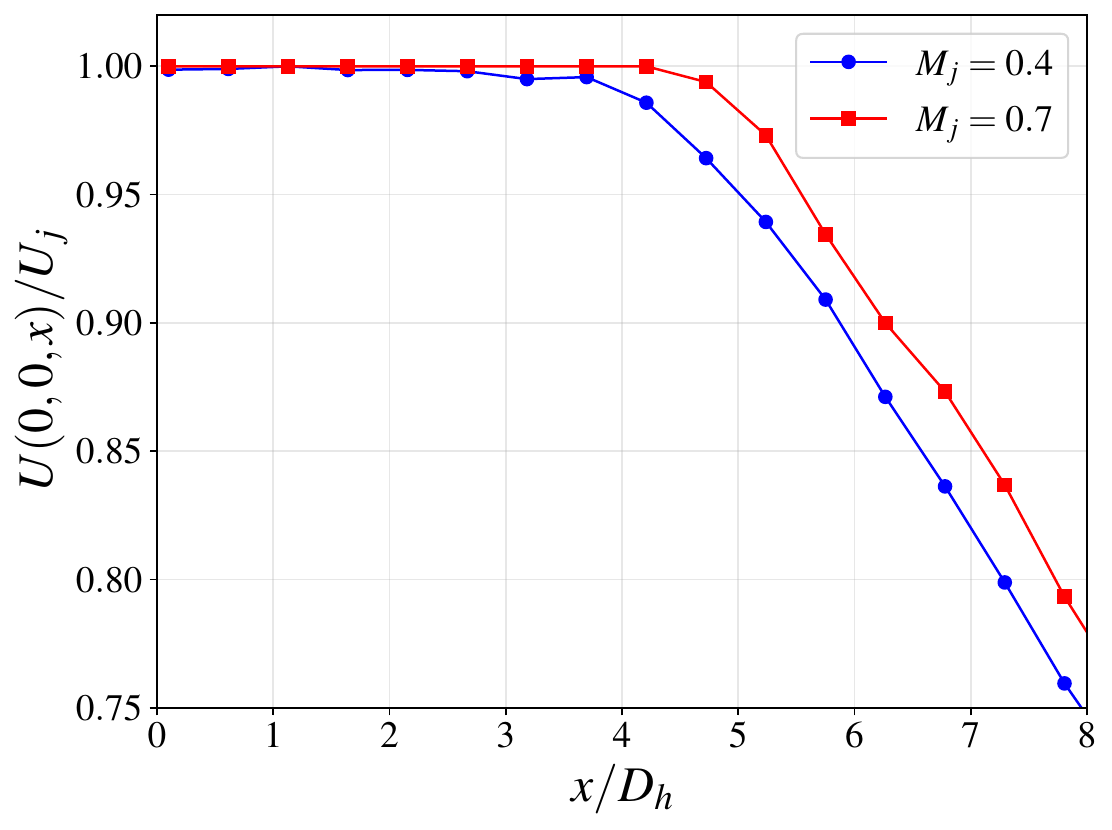}
    \caption{Centreline velocity for the symmetric-symmetric streamwise velocity component mean flow for $M_j = 0.4$ (blue bullets) and 0.7 (red squares).}
    \label{fig:u_centreline}
\end{figure}

The jet centreline velocity is exhibited in Figure~\ref{fig:u_centreline}, normalized by the jet exit velocity $U_j$.
The potential core closes approximately at $x/D_h = 3.5$ for $M_j = 0.4$ and at $x/D_h = 4.5$ for $M_j = 0.7$.

%%%%%%%%%%%%%%%%%%%%%%%%%%%%%%%%%%%%%%%%%%%%%%
\subsection{SPOD structures}
\label{sec:spod_structures}

\begin{figure}
    \centering
    \includegraphics[width=\textwidth]{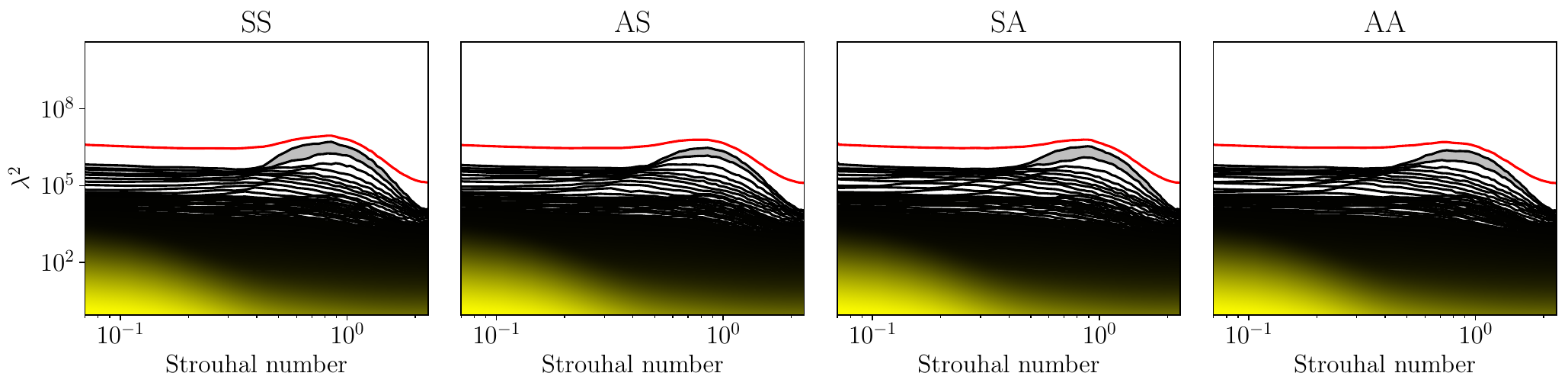}
    \caption{SPOD eigenvalue spectra for $M_j = 0.4$ at $x/D_h = 1$. Modes are colour-coded from optimal (black) to least suboptimal (yellow). The red curve denotes total energy, and the shaded gray region highlights the rank separation between the first and second modes. Frames from left to right: SS, SA, AS, and AA symmetry components.}
    \label{fig:SPODeigenvalues_zD1}
\end{figure}

\begin{figure}
    \centering
    \includegraphics[width=\textwidth]{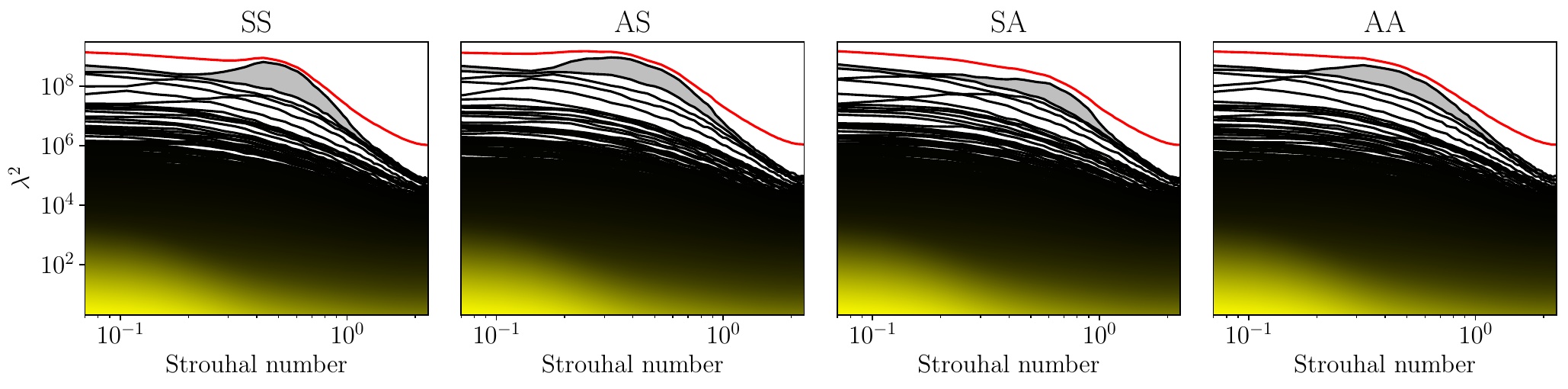}
    \caption{SPOD eigenvalue spectra for $M_j = 0.4$ at $x/D_h = 3$. See caption of Figure~\ref{fig:SPODeigenvalues_zD1} for details.}
    \label{fig:SPODeigenvalues_zD3}
\end{figure}

\begin{figure}
    \centering
    \includegraphics[width=\textwidth]{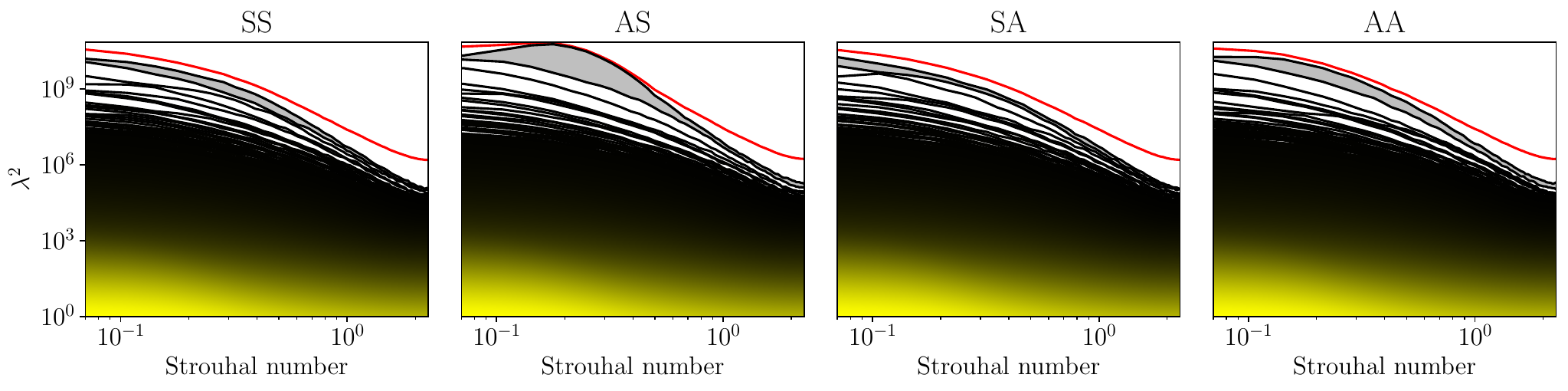}
    \caption{SPOD eigenvalue spectra for $M_j = 0.4$ at $x/D_h = 6$. See caption of Figure~\ref{fig:SPODeigenvalues_zD1} for details.}
    \label{fig:SPODeigenvalues_zD6}
\end{figure}

The SPOD eigenvalue spectra for selected streamwise/axial positions, i.e. near the nozzle exit ($x/D_h = 1$), at $x/D_h = 3$ and just downstream of the end of the potential core ($x/D_h = 6$), for the four D$_2$ symmetries are shown in Figures~\ref{fig:SPODeigenvalues_zD1},~\ref{fig:SPODeigenvalues_zD3} and~\ref{fig:SPODeigenvalues_zD6}, respectively, for the four symmetry components at $M_j = 0.4$.
SPOD spectra for $M_j = 0.7$ are not presented in this section for brevity, since the differences relative to the $M_j = 0.4$ case are primarily quantitative rather than qualitative.
The modes are colour-coded from black (optimal mode) to yellow (least suboptimal mode).
The red curve indicates the total energy, i.e. the sum of all eigenvalues, while the shaded gray region highlights the energy separation between the leading (first) and second modes.
Note that better alignment between SPOD modes and PSE solutions is expected as the SPOD eigenvalue spectra become increasingly low-rank, i.e. when larger energy separation between the first and second modes is attained.

\begin{figure}
    \centering
    \includegraphics[width=\textwidth]{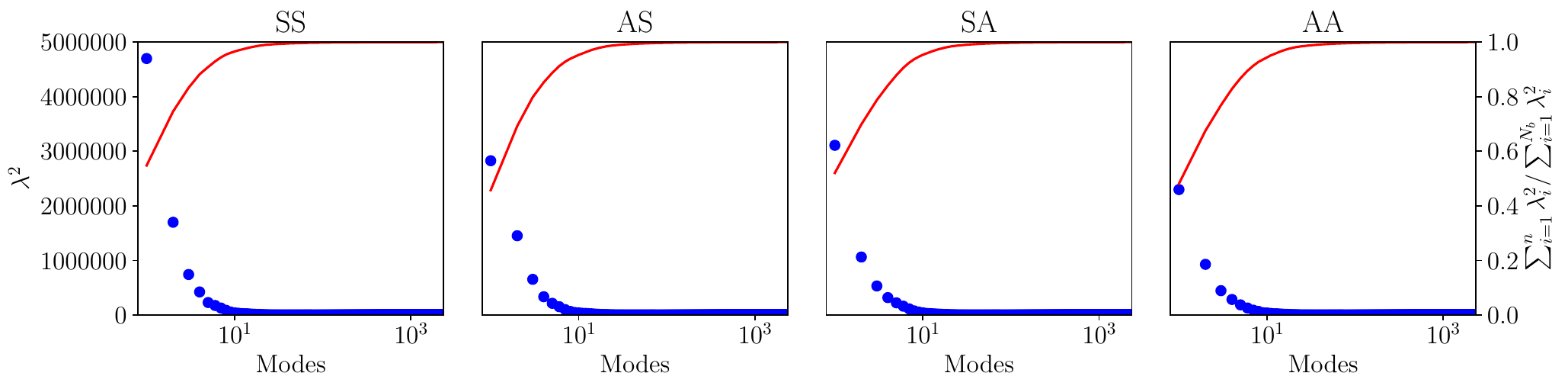}
    \caption{SPOD eigenvalues for $M_j = 0.4$ at $(\mathit{St},~x/D_h) \approx (0.8,~1)$. Blue bullets represent the eigenvalues (left y-axis), and the red curve indicates the accumulated energy (right y-axis). Frames from left to right: varicose (SS), wagging (SA), flapping (AS), and wobbling (AA) symmetry components.}
    \label{fig:SPODeigenvalues_zD1_St075}
\end{figure}

Near the nozzle exit (Figure~\ref{fig:SPODeigenvalues_zD1}), the first mode accounts for up to 50\% of the total energy at $\mathit{St} \approx 0.8$, i.e. at the spectral energy peak.
At this frequency, all four symmetry classes contribute to the flow, but the SS component exhibits the most energetic leading mode and the highest energy separation.
Figure~\ref{fig:SPODeigenvalues_zD1_St075} presents the eigenvalues (blue bullets) and the accumulated energy (red curve) at $(\mathit{St},~x/D_h) \approx (0.8,~1)$, i.e. the frequency at which the highest rank separation is observed after inspection of the spectra shown in Figure~\ref{fig:SPODeigenvalues_zD1}.
The varicose mode (SS) is the most energetic and has the greatest energy separation, which indicates a greater degree of organisation.
The wagging (SA), flapping (AS) and wobbling (AA) modes present comparable energy levels and energy separation.
Thus, the dominance of the SS component at this frequency does not imply that the other symmetry classes are absent; rather, it indicates that the SS component exhibits the strongest modal organisation among the four symmetry families.

\begin{figure}
    \centering
    \includegraphics[width=\textwidth]{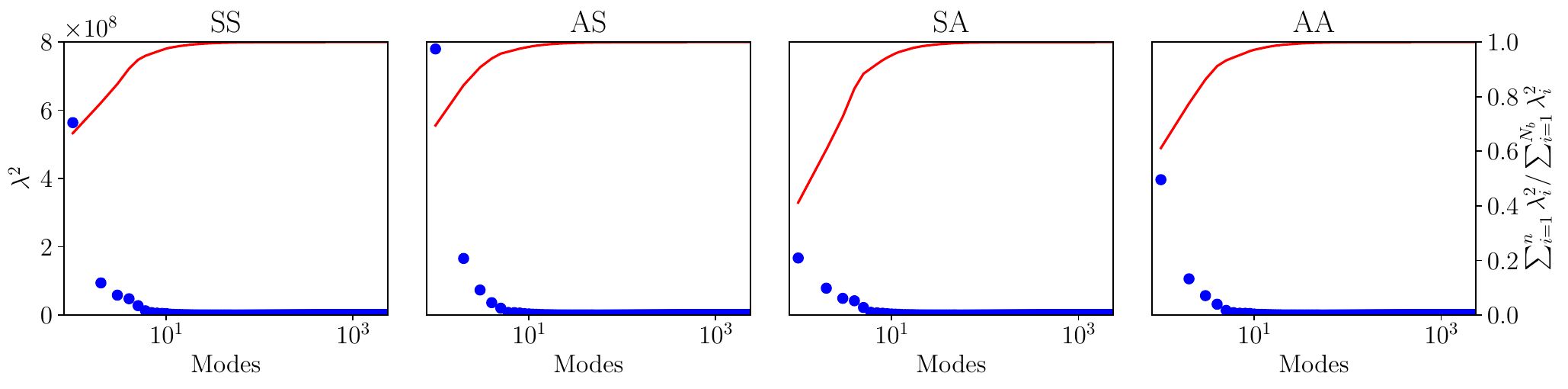}
    \caption{SPOD eigenvalues for $M_j = 0.4$ at $(\mathit{St},~x/D_h) \approx (0.4,~3)$. See caption of Figure~\ref{fig:SPODeigenvalues_zD1_St075} for details.}
    \label{fig:SPODeigenvalues_zD3_St05}
\end{figure}

At $x/D_h = 3$ (Figure~\ref{fig:SPODeigenvalues_zD3}), the flapping (AS) mode begins to emerge, especially around $\mathit{St} \approx 0.4$, again, the frequency exhibiting the highest rank separation in the spectra shown in Figure~\ref{fig:SPODeigenvalues_zD3}.
This is further illustrated in Figure~\ref{fig:SPODeigenvalues_zD3_St05}, which shows the eigenvalues and accumulated energy at $(\mathit{St},~x/D_h) \approx (0.4,~3)$, in the same manner as Figure~\ref{fig:SPODeigenvalues_zD1_St075}.
The AS mode becomes dominant, closely followed by the varicose (SS) mode, while the wobbling (AA) and wagging (SA) modes exhibit lower energy levels.
\citet{morris1995spatial} studied the stability of compressible elliptical jets with analytical mean-flow profiles and found that the mode flapping about the ellipse’s major axis was the most unstable within their parametric space of jet Mach number and aspect ratio.

The instabilities considered here are driven by the velocity gradients across the shear layer.
Since the mean flow has an elliptical shape, the shear-layer thickness is not uniform along the azimuthal direction, being slightly thicker along the major axis and thinner along the minor axis.
As a consequence, the antisymmetric flapping mode (AS), which oscillates about the major axis, is more strongly influenced by the shear layer aligned with the minor axis, where the velocity gradients are sharper.

\begin{figure}
    \centering
    \includegraphics[width=\textwidth]{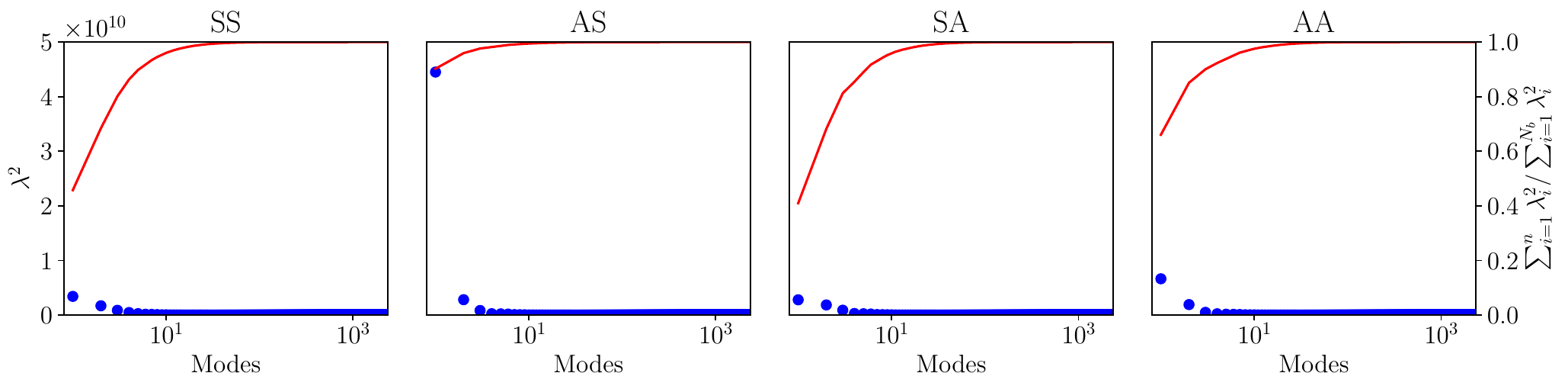}
    \caption{SPOD eigenvalues for $M_j = 0.4$ at $(\mathit{St},~x/D_h) \approx (0.2,~6)$. See caption of Figure~\ref{fig:SPODeigenvalues_zD1_St075} for details.}
    \label{fig:SPODeigenvalues_zD6_St02}
\end{figure}

Farther downstream (Figure~\ref{fig:SPODeigenvalues_zD6}), the maximum rank separation is now observed for $\mathit{St} \approx 0.2$ in the AS symmetry component, where the first mode captures up to 90\% of the total energy (see Figure~\ref{fig:SPODeigenvalues_zD6_St02}).
The only really organised structure is the flapping (AS) mode.
After the flapping (AS), the wobbling (AA) component shows the next highest rank separation, while the wagging (SA) and varicose (SS) components exhibit similar spectra.

In order to avoid repetition, the mode shapes will be addressed in \S~\ref{sec:psestructures} together with the PSE modes, allowing for a direct comparison between data-educed and modelled structures.

%%%%%%%%%%%%%%%%%%%%%%%%%%%%%%%%%%%%%%%%%%%%%%
\subsection{3D-PSE vs. SPOD}
\label{sec:psestructures}

The characterization of coherent structures through plane-marching 3D-PSE begins with a locally parallel analysis at a fixed frequency $\mathit{St}$ and streamwise position $x/D_h$.
The solution $\alpha_0,~\tilde{\mathbf{q}}_0$ of the eigenvalue problem~\eqref{eq:egv} at the specified frequency and $x/D_h$ position is used to initialize the iterative process~\eqref{eq:pseprocess}.
Subsequently, solutions at downstream positions $x_n$ are computed by marching the PSE equations~\eqref{eq:pseprob}.

\begin{figure}
    \centering
    \includegraphics[width=\textwidth]{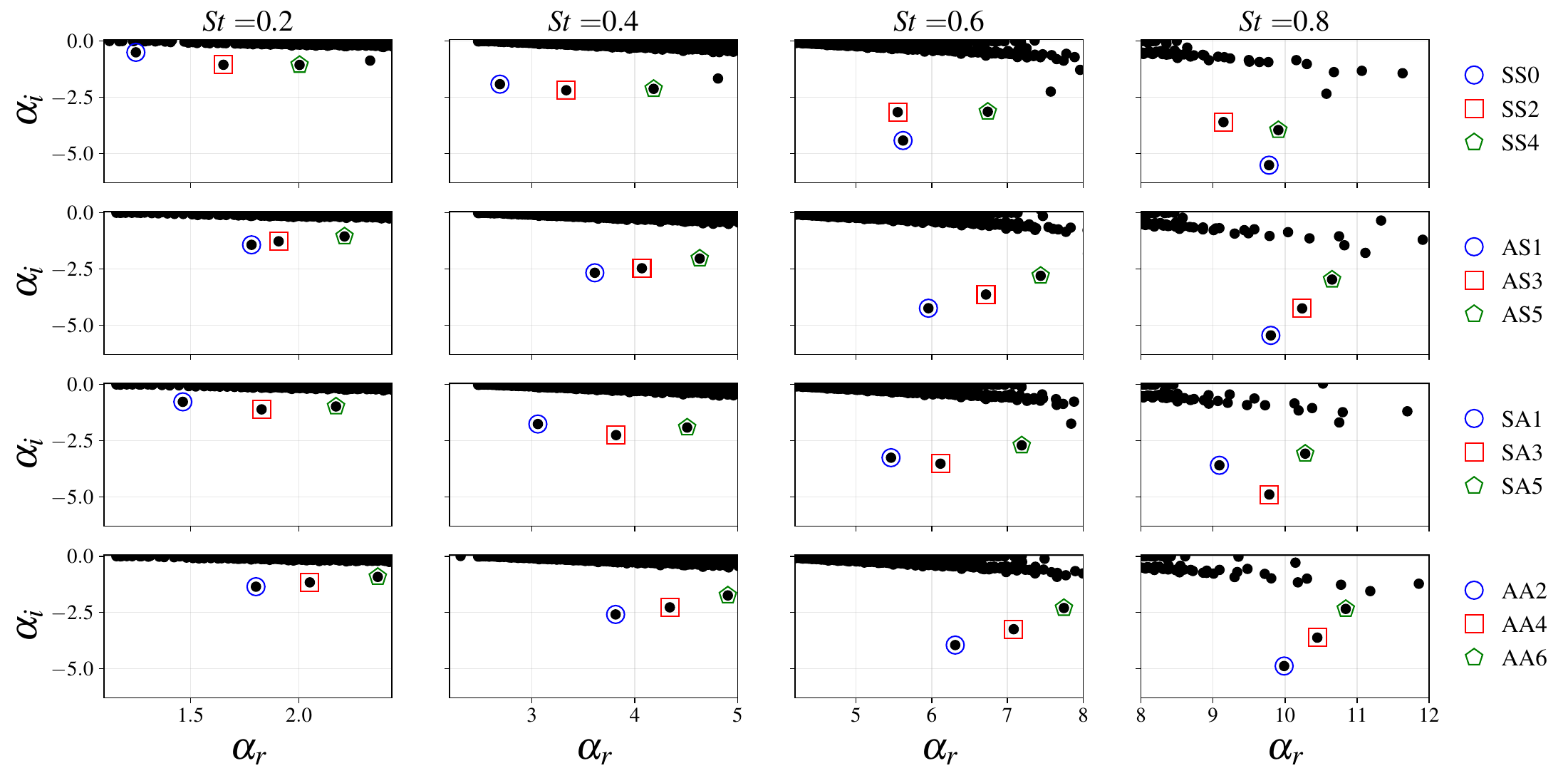}
    \caption{Eigenspectra at $x/D_h = 0.1$ and $M_j = 0.4$. Each line represents a symmetry family and each column corresponds to a Strouhal number. The first three converged modes of each family are highlighted by coloured markers.}
    \label{fig:eigenspectrum}
\end{figure}

Figure~\ref{fig:eigenspectrum} shows the eigenspectrum for several Strouhal numbers (columns) and symmetry boundary conditions (rows) at a location near the nozzle, $x/D_h = 0.1$ which are used to initialize PSE.
The eigenspectra are shown only for the $M_j = 0.4$ jet, since the results for $M_j = 0.7$ are qualitatively similar.
Each eigenspectra corresponds to a specific mode family, with the first three converged unstable modes highlighted by open markers.
Although additional modes appear in the eigenspectra, they lack physical significance (they do not converge even with increased computational resolution), while others that are physically meaningful require a finer mesh for convergence, which considerably increases the computational cost.
In this work, we focus only on the first three converged modes, and the rationale for this choice will be discussed later.
The mode labels follow the classification given in Table~\ref{tab:mode-symmetry}.
Modes associated with low azimuthal wavenumbers $m$ are generally easier to converge.
Consequently, the SS0 mode from the SS family demands the least computational effort, as does the AS1 mode from the AS family, and so forth.
The PSE are solved by spatial marching from this initial position.

To evaluate the PSE as a model for the structures educed from the experimental data, we compare the PSE solution with the leading SPOD mode.
To the authors' knowledge, this paper presents the first attempt to model wavepackets in elliptical jets using 3D-PSE based on an experimental database and to compare the resulting wavepacket structures with data-driven structures obtained through SPOD.
To quantify similarity, we use the alignment coefficient $\beta$, defined in \eqref{eq:beta}.
It should be emphasized that the objective is not to validate the PSE calculations themselves, but rather to assess whether they can provide a meaningful physical interpretation of the SPOD modes.
Where good agreement is observed, the comparison suggests that the experimentally extracted structures are associated with the instability-wave mechanisms captured by the linear model.
Conversely, regions of poor agreement may indicate the presence of physical mechanisms not represented within the PSE framework, such as non-modal amplification, strong non-normal effects, or nonlinear turbulent interactions.

The present analysis focuses on the leading SPOD mode of each symmetry family. This choice is motivated by the pronounced low-rank behaviour of the SPOD spectra at the frequencies of interest: the leading mode accounts for at least 50\% of the spectral energy and, in some cases, up to approximately 90\% of the total energy (see Figures~\ref{fig:SPODeigenvalues_zD1_St075}--\ref{fig:SPODeigenvalues_zD6_St02}).
Moreover, the statistical convergence of SPOD modes progressively deteriorates with increasing mode order, as higher-order modes are associated with smaller eigenvalues and are consequently more sensitive to the finite number of independent realisations used in the spectral estimation (see Appendix~\ref{app:SPOD_convergence}).
Including suboptimal modes would therefore provide a more complete representation of the turbulent field, but at the cost of introducing progressively less statistically robust structures and potentially obscuring the physical interpretation of their comparison with the deterministic 3D-PSE solution.
Since the objective of the present analysis is to assess the ability of 3D-PSE to describe the dominant coherent wavepacket dynamics, we restrict the comparison to the leading SPOD mode, following previous studies of turbulent jets and wavepacket dynamics \citep{gudmundsson2011instability,cavalieri2013wavepackets,rodriguez2015study,sasaki2017high,lesshafft2019resolvent}.

\begin{figure}
    \centering
    \includegraphics[width=\textwidth]{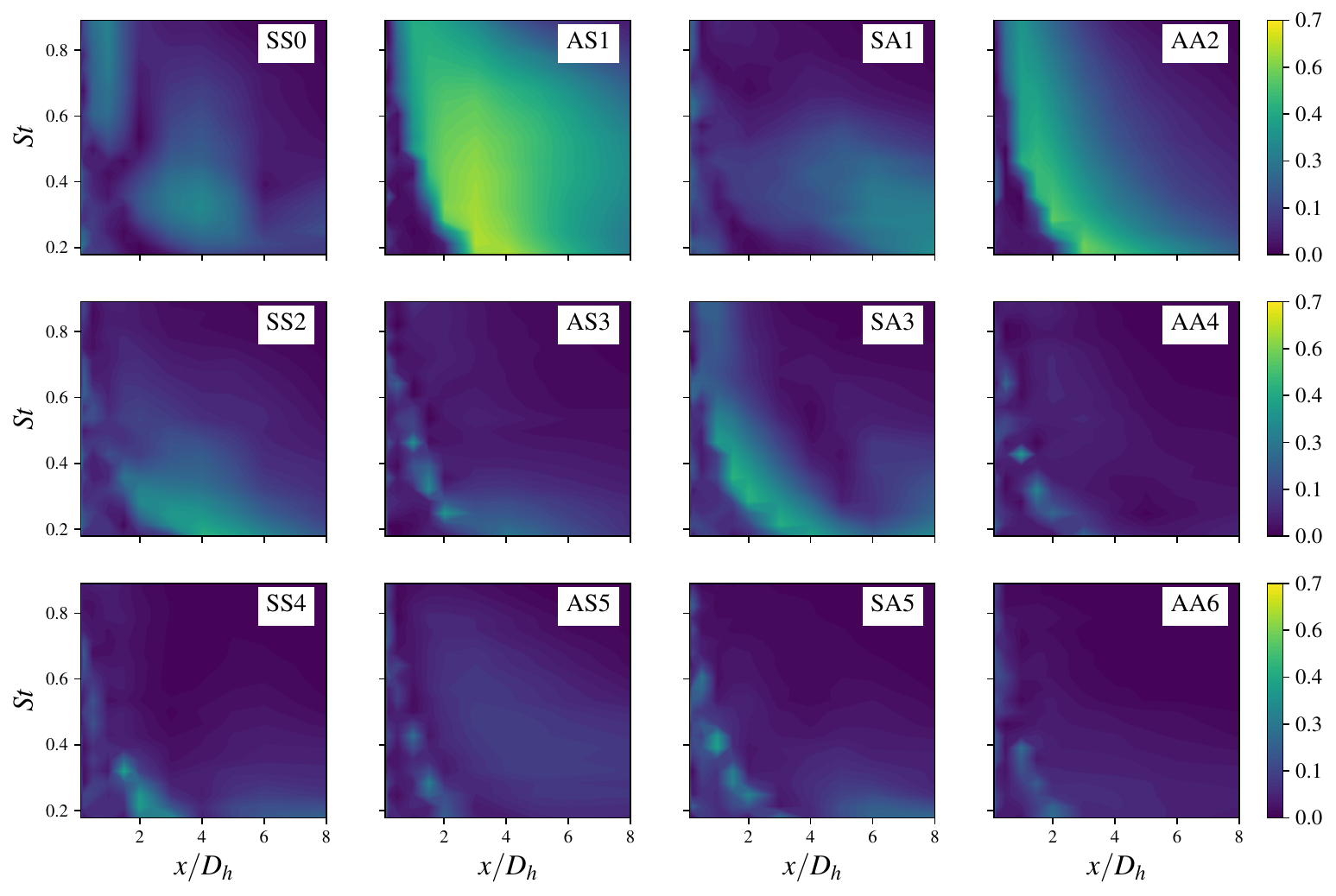}
    \caption{Maps of the alignment coefficient $\vert\beta\vert$ between the various 3D-PSE modes and the most energetic SPOD mode for $M_j = 0.4$.} 
    \label{fig:all_beta_maps_M04}
\end{figure}

\begin{figure}
    \centering
    \includegraphics[width=\textwidth]{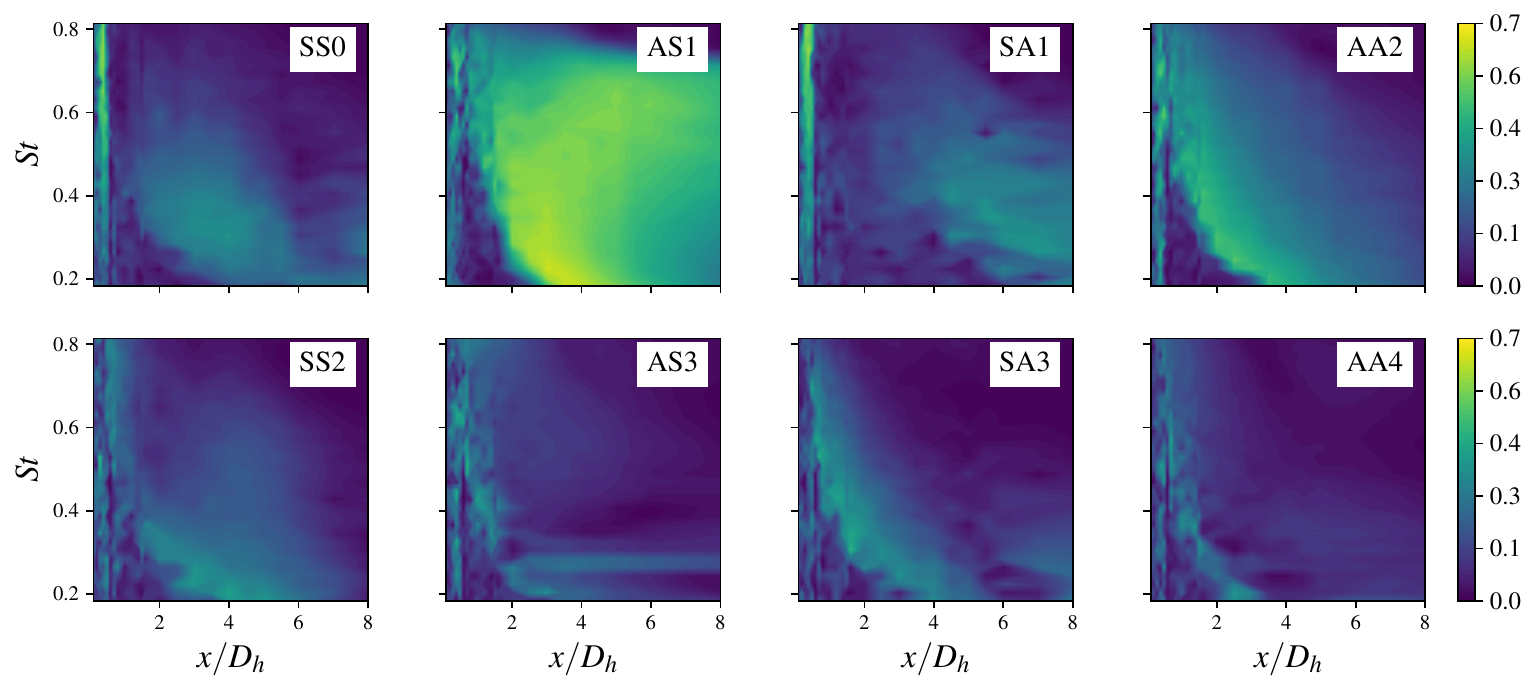}
    \caption{Maps of the alignment coefficient $\vert\beta\vert$ between the various 3D-PSE modes and the most energetic SPOD mode for $M_j = 0.7$.} 
    \label{fig:all_beta_maps_M07}
\end{figure}

Figures~\ref{fig:all_beta_maps_M04} and~\ref{fig:all_beta_maps_M07} present contour maps of $\vert\beta\vert$ as a function of $(\mathit{St},~x/D_h)$ for $M_j = 0.4$ and 0.7, respectively.
The results show that alignment depends on both Strouhal number and streamwise position, with good agreement observed for the flapping mode AS1 at streamwise positions associated with the peak in mode amplitude.
Aside from this mode, the match between SPOD and PSE is globally poor.

The AS modes exhibit the highest energy separation in (\S~\ref{sec:spod_structures}), which may partly explain their better alignment with the PSE solutions shown in Figures~\ref{fig:all_beta_maps_M04} and~\ref{fig:all_beta_maps_M07}.
Indeed, the alignment tends to be best at $(\mathit{St},~x/D_h)$ points where the SPOD spectra exhibit a low-rank behaviour.

\begin{figure}
    \centering
    \includegraphics[width=.7\textwidth]{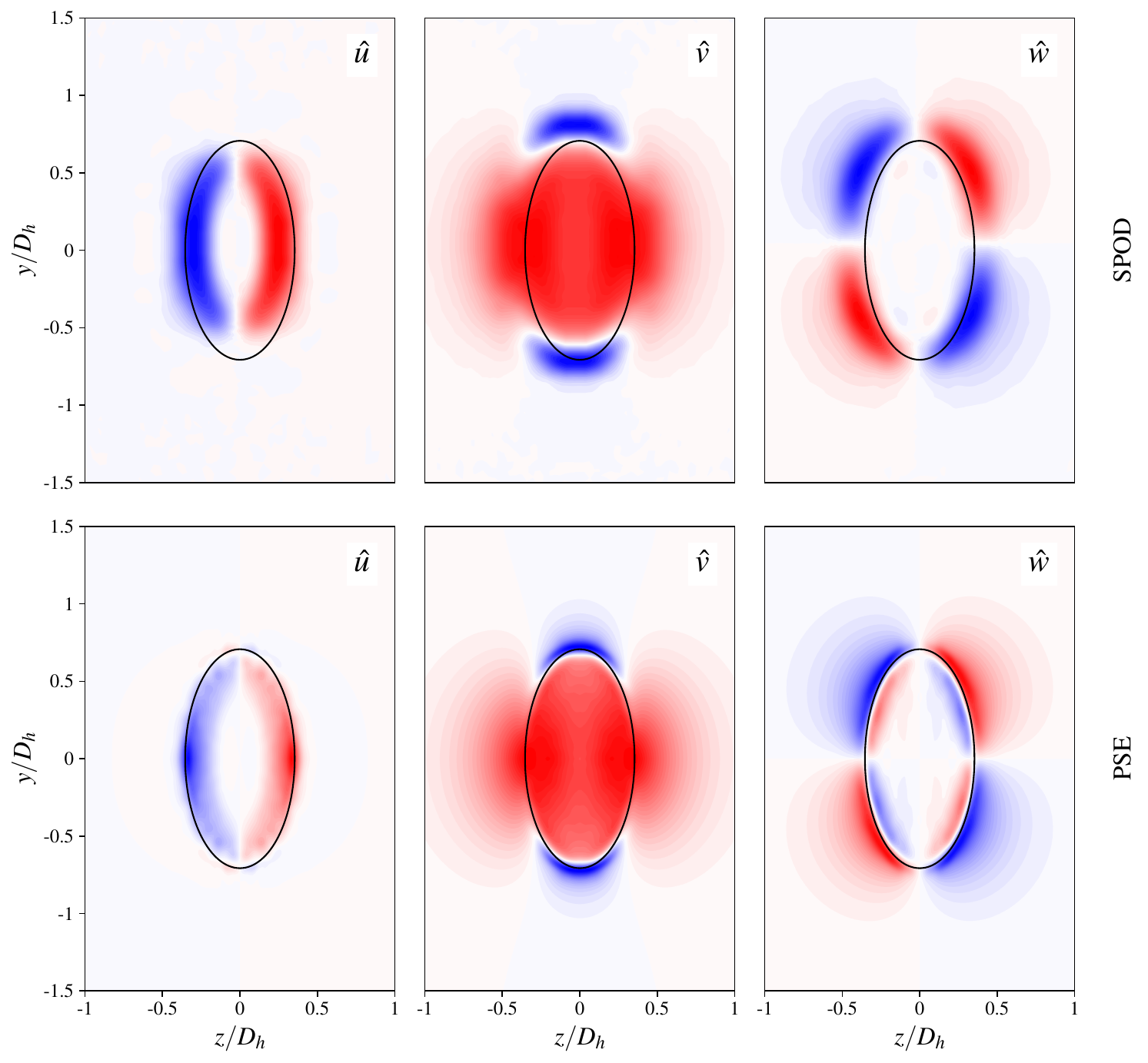}
\caption{Comparison between the real parts of the velocity components obtained using SPOD and 3D-PSE for the mode AS1, $(\mathit{St},~x/D_h) \approx (0.4,~3)$, and $M_j = 0.4$. Alignment coefficient $\vert \beta \vert = 0.59$ (see Figure~\ref{fig:all_beta_maps_M04}).}
    \label{fig:as1_best_beta}
\end{figure}

Figure~\ref{fig:as1_best_beta} shows a comparison of the spatial structures of the three velocity components between SPOD and the AS1 mode obtained from PSE.
These structures correspond to the position $(\mathit{St},~x/D_h) \approx (0.4,~3)$ on the map, with an alignment coefficient of $\vert \beta \vert = 0.59$.
This indicates that, in regions of good alignment, the method is indeed able to reproduce the experimentally observed structures with good accuracy.
Results for $M_j = 0.7$ are similar and not shown in the interest of compactness.

Comparison of an SPOD mode with one PSE solution is problematic, as the SPOD mode is associated with a given symmetry, with which numerous PSE solutions are associated.
The SPOD mode is better thought of as comprising a linear combination of PSE solutions with the same symmetry as the SPOD mode. 
We therefore consider linear combinations of PSE mode, with each mode weighted by a complex coefficient.
As an example, for the SS family at each streamwise station $x_n$, one has 
\begin{equation}
    A_{\mathit{SS},n}^{\mathit{SPOD}} = k_{0,n}B^{\mathit{PSE}}_{\mathit{SS0}} + k_{2,n}C^{\mathit{PSE}}_{\mathit{SS2}} + k_{4,n}D^{\mathit{PSE}}_{SS4} + \dots \mbox{,}
    \label{eq:linear_combination_PSE}
\end{equation}
\noindent where $B^{\mathit{PSE}}_{\mathit{SS0}}$, $C^{\mathit{PSE}}_{\mathit{SS2}}$, and $D^{\mathit{PSE}}_{\mathit{SS4}}$ denote the velocity fields associated with each PSE mode, and $k_{0,n}$, $k_{2,n}$, and $k_{4,n}$ are complex weighting coefficients.

The coefficients are obtained through a least-squares procedure that minimizes the difference between the linear combination and the target SPOD mode.
In this sense, the method can be interpreted as a projection procedure that expresses the SPOD structure as a combination of a limited number of PSE solutions.
The resulting coefficients should therefore be interpreted as the complex weights required for the local projection of the SPOD structure onto the PSE basis, rather than as a measure of the relative contribution or energy content of each PSE solution.
Since both the amplitude and phase of the SPOD modes are available at each streamwise position $x_n$, it is possible to determine a corresponding complex coefficient $k_{m,n} = k_r + j k_i$, where $j=\sqrt{-1}$, $m$ denotes the PSE solution and $n$ the streamwise station.
The least-squares procedure is therefore performed using the full complex representation of the SPOD modes rather than amplitude information alone.
The coefficients in Equation~\eqref{eq:linear_combination_PSE} are determined independently at each streamwise station and are therefore used only to quantify the local representation of the SPOD structure by the PSE basis.
They should not be interpreted as amplitudes that are propagated downstream according to the PSE equations.
Accordingly, the procedure does not constitute a predictive PSE framework, but rather a local reconstruction of the experimentally extracted SPOD structures using a basis formed by PSE solutions.
The resulting coefficients quantify how the selected PSE solutions are combined to locally reconstruct the SPOD structure, but they do not provide a direct measure of the individual modal energy or physical contribution of each PSE solution.
Their values depend on the streamwise position and on the mutual linear dependence of the PSE solutions, and should therefore be interpreted only within the context of the local reconstruction.
Once the local projection weights are determined, the spatial structure of the PSE solution is reconstructed and the alignment coefficient is recalculated.

\begin{figure}
    \centering
    \includegraphics[width=\textwidth]{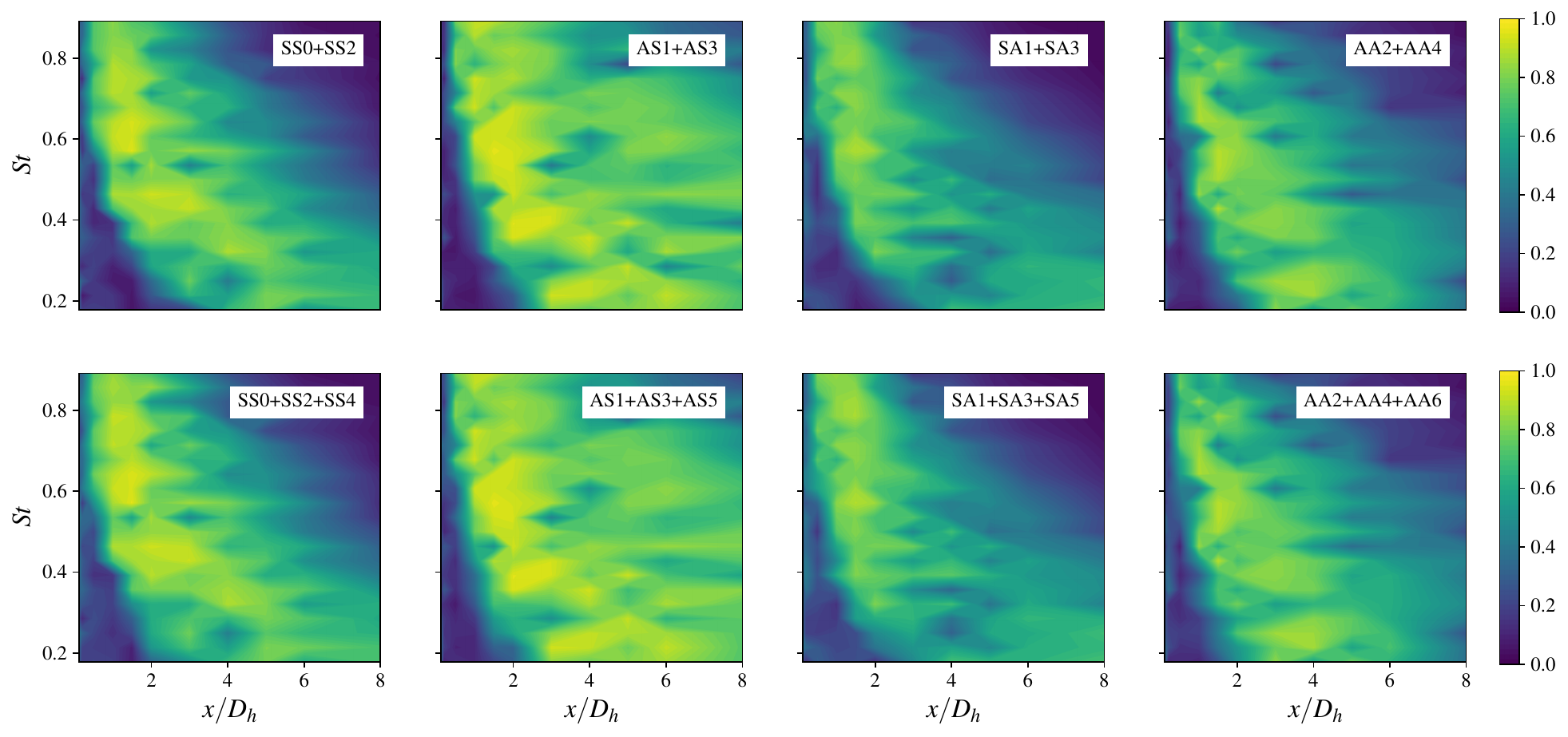}
    \caption{Maps of the alignment coefficient $\beta$ obtained from the linear combination of PSE modes for $M_j = 0.4$.}
    \label{fig:beta_comb}
\end{figure}

\begin{figure}
    \centering
    \includegraphics[width=\textwidth]{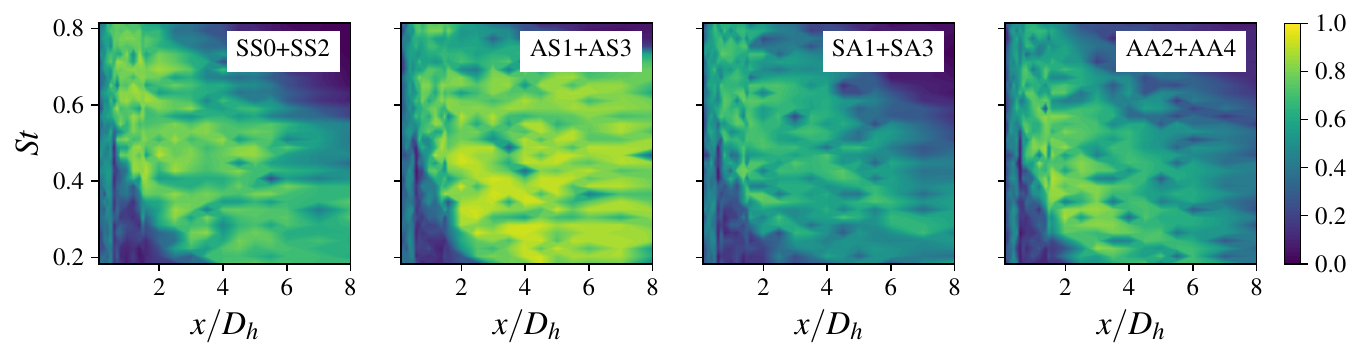}
    \caption{Maps of the alignment coefficient $\beta$ obtained from the linear combination of PSE modes for $M_j = 0.7$.}
    \label{fig:beta_comb_M07}
\end{figure}

\begin{figure}
    \centering
    \includegraphics[width=.7\textwidth]{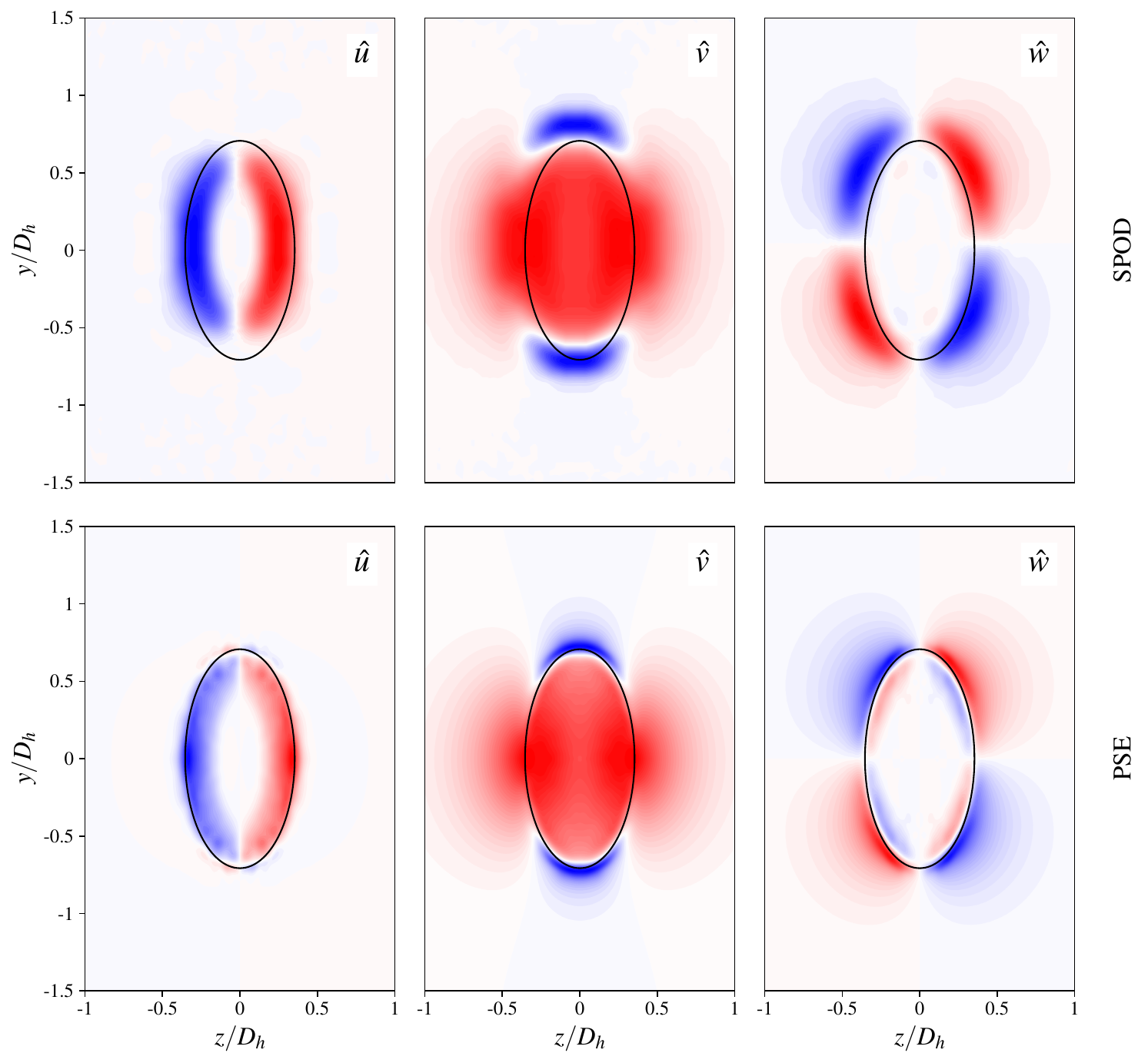}
    \caption{Comparison between the real parts of the velocity components obtained using SPOD and 3D-PSE for combinations of modes AS1 and AS3, $(\mathit{St},~x/D_h) \approx (0.4,~3)$ and $M_j = 0.4$. Alignment coefficients $\vert \beta \vert = 0.93$ (see Figure~\ref{fig:beta_comb}).}
    \label{fig:ss_comb}
\end{figure}

Figure~\ref{fig:ss_comb} shows the real part of the velocity components associated with the combined AS1 and AS3 modes, compared with those of the most energetic SPOD mode at $(\mathit{St},~x/D_h) \approx (0.4,~3.0)$, same frequency and streamwise position pair shown in figure \ref{fig:as1_best_beta}. An improvement in the representation of the modes is noted, mainly for the $\hat{u}$ component, where regions near the nozzle lip now exhibit higher amplitudes. By combining the modes with their respective relative weights, PSE provides a better description of the structures observed experimentally.
For conciseness, results for $M_j = 0.7$ are not presented here, as they display similar trends.

Figure~\ref{fig:beta_comb} also shows that including a third PSE mode produces only a marginal additional improvement in the alignment coefficient.
Indeed, the maps obtained from combinations of the first two modes are nearly indistinguishable from those obtained when the third mode is included.
This indicates that the experimentally observed structures can be effectively represented as combinations of the two lowest-order PSE modes of each symmetry family.
Despite the overall improvement in the alignment coefficient, certain regions in the maps of Figures~\ref{fig:beta_comb} and~\ref{fig:beta_comb_M07} still show poor correlation.
At low frequencies near the nozzle, this behaviour may be associated with non-modal amplification mechanisms that cannot be captured by PSE and require resolvent analysis, as in by~\citet{lesshafft2019resolvent},~\citet{nogueira2019large},~\citet{pickering2020lift},~\citet{maia2024effect} and~\citet{bugeat2024acoustic}.

At high frequencies further downstream, poor correlation is also observed.
This is consistent with the growth and decay of coherent structures as shown, for instance, in~\citet{cavalieri2013wavepackets}. 
High frequency wavepackets peak upstream (e.g., below $x/D_h \approx 3$), low-frequency structures peak farther downstream (e.g., above $x/D_h \approx 3$) and have globally larger streamwise spatial support.
The maps of Figure~\ref{fig:beta_comb}, are qualitatively similar to those of~\citet{sasaki2017high} in round jets, showing that PSE is an appropriate tool for modelling convectively unstable coherent structures in both elliptical and round turbulent jets.

\begin{figure}
    \centering
    \includegraphics[width=\textwidth]{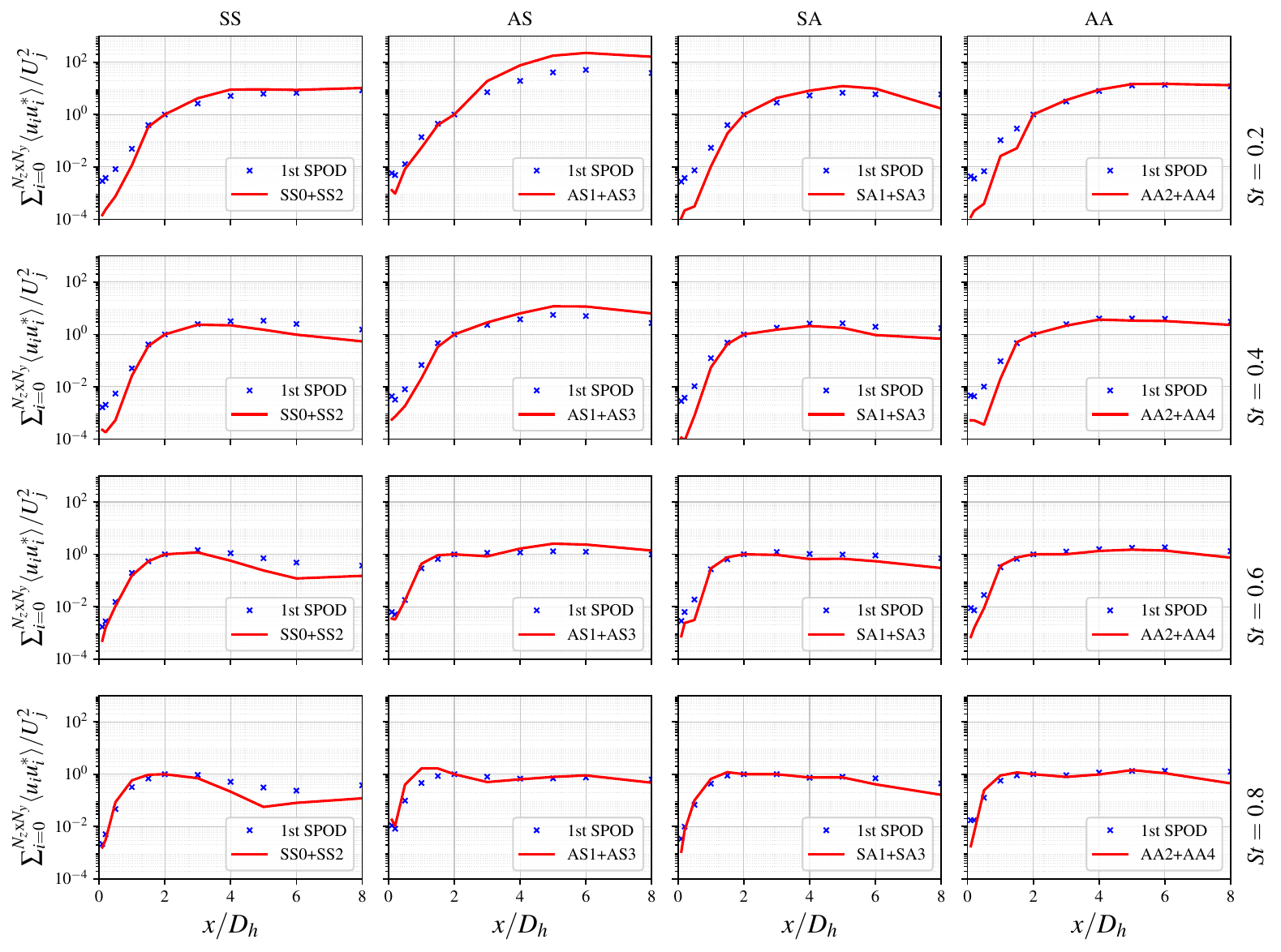}
    \caption{Comparison of the PSD computed from the most energetic SPOD mode (markers) and PSE mode combination (lines) for $M_j = 0.4$.}
    \label{fig:psd_Mj04}
\end{figure}

\begin{figure}
    \centering
    \includegraphics[width=\textwidth]{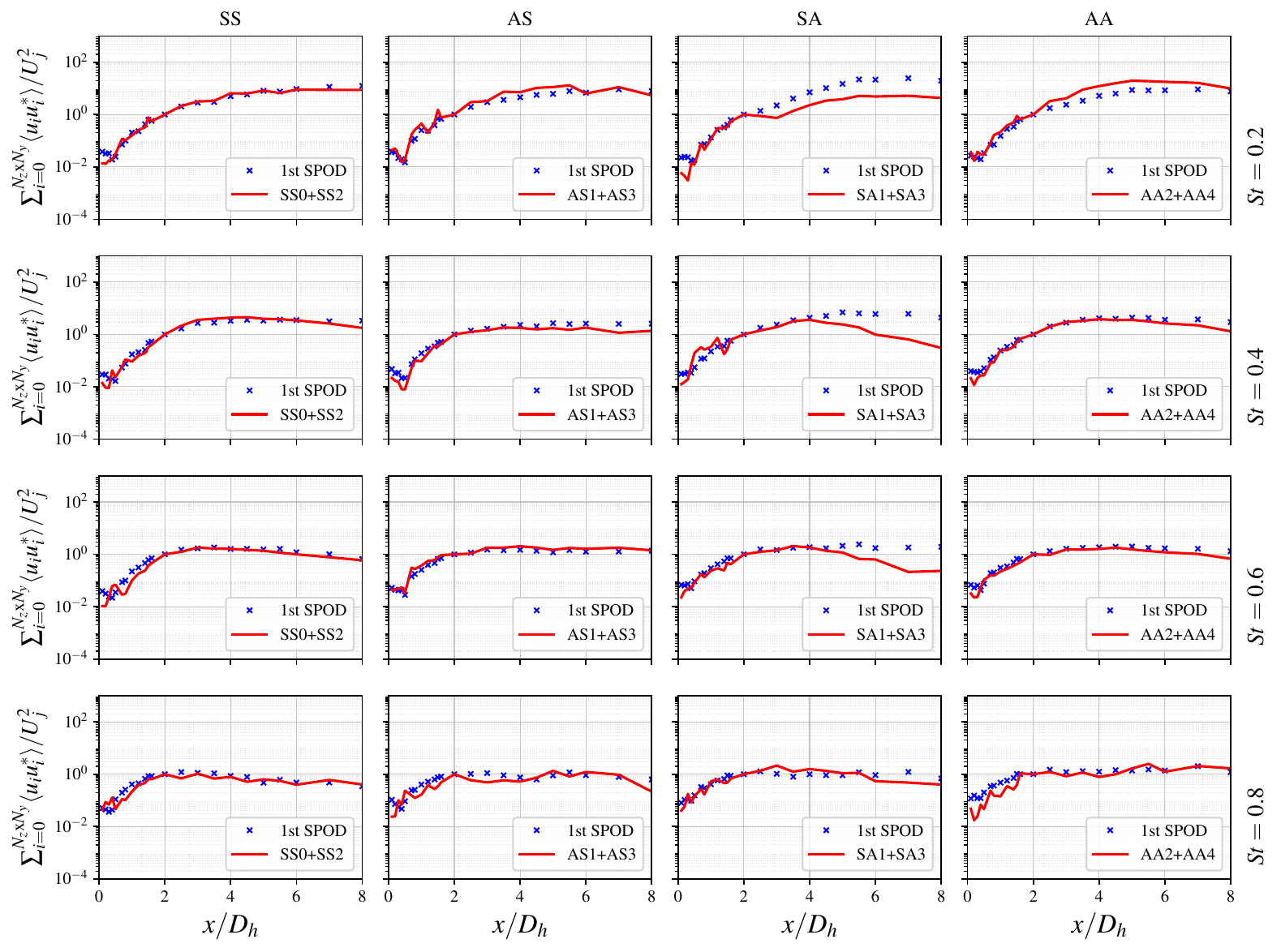}
    \caption{Comparison of the PSD computed from the most energetic SPOD mode (markers) and PSE mode combination (lines) for $M_j = 0.7$.}
    \label{fig:psd_Mj07}
\end{figure}

Figures~\ref{fig:psd_Mj04} and \ref{fig:psd_Mj07} compare the wavepacket envelope (power spectral density, PSD) of the streamwise velocity component ($\hat{u}$) from the first SPOD mode with those obtained from PSE after mode combination (see~\eqref{eq:linear_combination_PSE}) for $M_j = 0.4$ and 0.7, respectively.
For each streamwise location ($x/D_h$), the energy is computed as the sum of the squared magnitudes of the $\hat{u}$ component over the cross-section ($yz$-plane).
The columns correspond to different mode families, while the rows indicate Strouhal numbers in the 0.2 to 0.8 range.
All curves are normalized such that they coincide at $x/D_h = 2$, facilitating a direct comparison of the streamwise energy distribution.

At the lower frequencies, $\mathit{St} \approx 0.2$ and $0.4$, PSE fails to reproduce the PSD distribution for $x/D_h \leq 2$, especially for $M_j = 0.4$.
As seen in comparing the results of~\citet{cavalieri2013wavepackets} and in~\citet{lesshafft2019resolvent}, this mismatch can be attributed to non-modal mechanisms that dominate low-frequency growth and that cannot be captured by PSE.
Nevertheless, resolvent analysis~\citep{lesshafft2019resolvent,chevalier2024resolvent,maia2024effect,marshall2024prediction}, OWNS~\citep{nogueira2023prediction,nogueira2024wavepackets,amaral2025tabbed}, or other frameworks capable of capturing non-modal amplification mechanisms could be applied to the present database to further investigate these structures and potentially improve the agreement with the data-driven (SPOD) modes.
Such an analysis, however, lies beyond the scope of the present study.
In contrast, for $\mathit{St} \geq 0.4$, dominated by modal growth mechanisms, PSE provides provides a good match with the most energetic SPOD mode.
The results for $M_j = 0.7$ are globally similar to those of $M_j = 0.4$, although for $x/D_h \geq 3$ and mode SA, the SPOD levels are higher than those obtained with PSE.

\begin{figure}
    \centering
    \includegraphics[width=\textwidth]{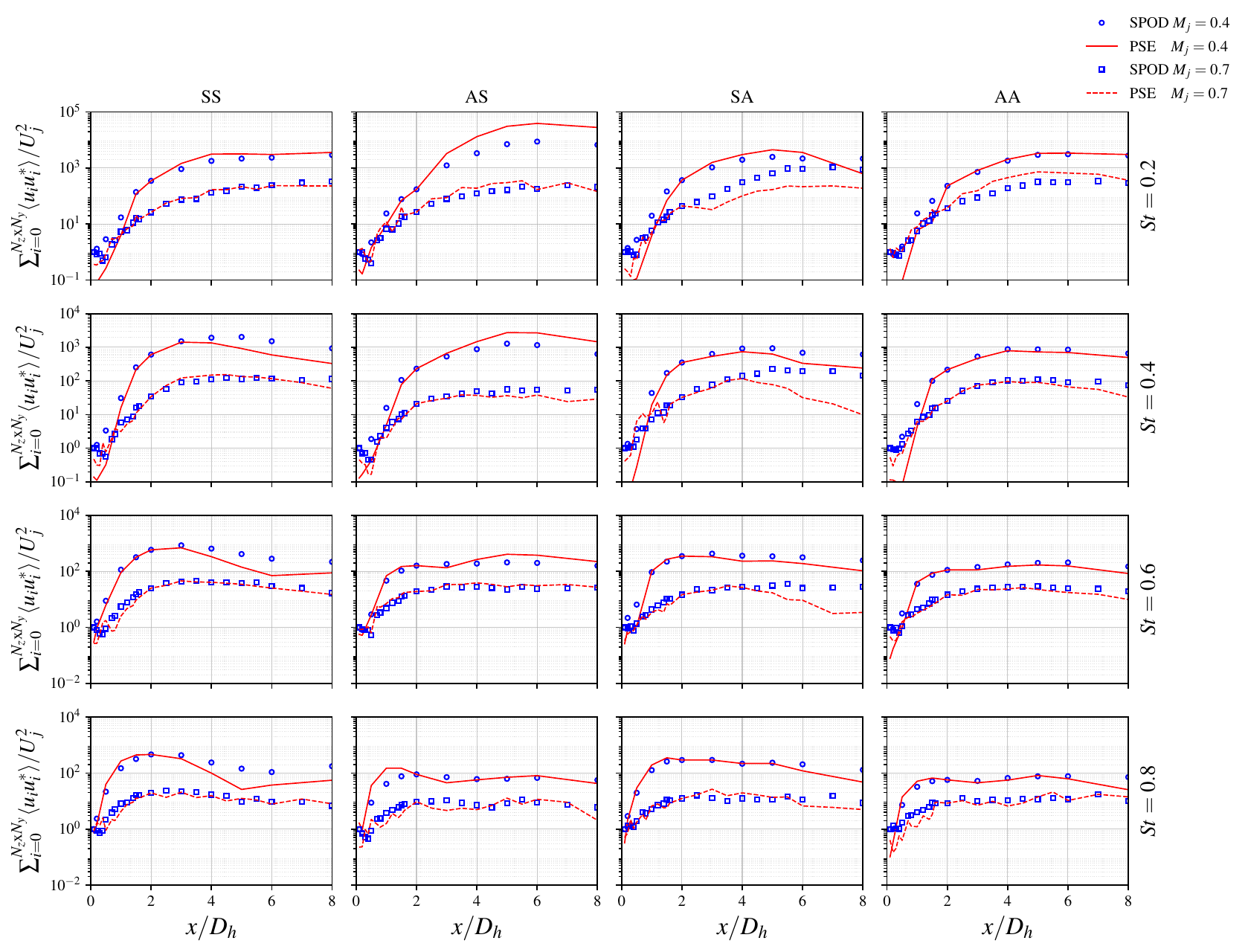}
    \caption{Compressibility effect in PSD computed from the most energetic SPOD mode (markers) and PSE mode combination (lines).}
    \label{fig:psd_Mj_effect}
\end{figure}

It is known that in axisymmetric jets compressibility effects affect the growth rate and wavepacket shape~\citep{cavalieri2012axisymmetric,cavalieri2013wavepackets}.
Therefore, Figure~\ref{fig:psd_Mj_effect} shows the PSE wavepacket envelopes ($M_j = $ 0.4 and 0.7) normalized at $x/D_h=2$ (red curves), whereas the SPOD (blue markers) are further normalized with respect to their initial axial position 
$x/D_h = 0.03$, such that each curve has a value of 1 at its first streamwise point.
After such procedure, it becomes possible to verify the elliptical jets possess the same trends observed in round jets~\citep{cavalieri2012axisymmetric,cavalieri2013wavepackets}.

Overall, it is observed that PSE can describe the flapping (AS) SPOD mode when only the most unstable solution is used to initialize the marching process.
But a linear combination of PSE modes improves the alignment with SPOD, with all symmetry families (SS, AS, SA, and AA) exhibiting satisfactory levels of agreement.
This shows that the dominant coherent structures in elliptical jets can be represented by the PSE.

%%%%%%%%%%%%%%%%%%%%%%%%%%%%%%%%%%%%%%%%%%%%%%
\section{Conclusions}
\label{sec:conclusions}
%%%%%%%%%%%%%%%%%%%%%%%%%%%%%%%%%%%%%%%%%%%%%%

Time-resolved, stereoscopic PIV measurements were used to characterise coherent structures in subsonic elliptical jets, at Mach numbers of 0.4 and 0.7, through D$_2$ symmetry decomposition and SPOD.
Three-dimensional parabolised stability equatoins (3D-PSE) calculations were performed using the PIV-based mean flow and tested as a simplified model to describe the observed flow structures.
Four important points are highlighted by the study.

Firstly, the loss of Fourier symmetries, which facilitate coherent-structure eduction and modelling in round jets, makes these tasks more challenging.
Fourier symmetries are replaced by D$_2$ dihedral-group symmetries, which do not filter out small-scale structures as effectively.
The convergence of subsequent spectral proper orthogonal decomposition (SPOD) is strongly affected: D$_2$-SPOD convergence in the elliptical jet required time series six times longer than what is required for Fourier-SPOD convergence in the round jet.

Secondly, regarding PSE as a model for coherent structures in elliptical jets: it was found that solutions obtained following initialisation by a single (most-unstable) upstream eigenmode only describe the dominant flapping mode observed in the data; but if the leading SPOD modes are interpreted as linear combinations of PSE solutions—constructed from initially unstable eigenmodes sharing the same symmetry as the SPOD mode—the PSE framework can closely reproduce the empirical coherent structures.
In several cases, linear combinations of PSE solutions associated with the two lowest-order azimuthal wavenumbers for a given symmetry group, led to alignment levels with SPOD that exceed 90\%, confirming that experimentally observed structures can be interpreted as a combination of the two lowest-order PSE modes.

Thirdly, by considering jets at Mach 0.4 and Mach 0.7, the compressibility effect observed by~\citet{cavalieri2013wavepackets} for round jets is present in elliptical jets and captured by PSE.

Finally, poorer performance of the PSE at low-frequency is consistent with what is found in round jets and suggests that non-modal mechanisms are here important and resolvent-analysis is required for a more complete description of the low-frequency dynamics. 

Overall, the study shows that coherent structures in elliptical jets can be described using symmetry-consistent linear PSE modes, provided that multiple unstable PSE branches are combined. This shows that these coherent structures are largely driven by similar convective amplification mechanisms as those present in round jets.

%%%%%%%%%%%%%%%%%%%%%%%%%%%%%%%%%%%%%%%%%%%%%%
% ACKNOWLEDGEMENTS
%%%%%%%%%%%%%%%%%%%%%%%%%%%%%%%%%%%%%%%%%%%%%%

\noindent{\bf Acknowledgements\bf{.}} The authors would like to thank the technical staff of the \emph{Prométée platform}, especially Dr. Anton Lebedev and Eng. Damien Eysseric, as well as Dr. Barbara G. Hasparyk, for their valuable support during the experimental campaign.
F. R. do Amaral and M. P. Avanci are particularly grateful to Joselito and Lindomar (H\&R) for their unwavering support. \\

\noindent{\bf Funding\bf{.}} The authors would like to thank AIRBUS for its financial and technical support.
This work was supported by the \emph{Association Nationale de la Recherche et de la Technologie} (ANRT) through the \emph{Convention Industrielle de Formation par la Recherche} (CIFRE) grant No. 2019/1829. \\

\noindent{\bf Author contributions.} F. R. do Amaral performed the TR-PIV experiments and carried out the SPOD analysis of the database. M. P. Avanci conducted the 3D-PSE calculations. F. R. do Amaral and M. P. Avanci contributed equally to this work. J.-C. Robinet and P. Jordan contributed to the interpretation and discussion of the results. J. Huber and G. Pont contributed to the conception of the study.  \\

\noindent{\bf  Author ORCID\bf{.}} F. R. do Amaral , \url{https://orcid.org/0000-0003-1158-3216}; M. P. Avanci, \url{https://orcid.org/0000-0003-1586-6475}; J.-C. Robinet, \url{https://orcid.org/0000-0002-3529-6003}; P. Jordan, \url{https://orcid.org/0000-0001-8576-5587}; J. Huber, \url{https://orcid.org/0009-0003-3741-6181}; G. Pont \url{https://orcid.org/0009-0008-5009-3539}. \\

\noindent{\bf Declaration of Interests\bf{.}} The authors report no conflict of interest. \\

%%%%%%%%%%%%%%%%%%%%%%%%%%%%%%%%%%%%%%%%%%%%%%
\appendix
%%%%%%%%%%%%%%%%%%%%%%%%%%%%%%%%%%%%%%%%%%%%%%

%%%%%%%%%%%%%%%%%%%%%%%%%%%%%%%%%%%%%%%%%%%%%%
\section{D$_2$-decomposition}
\label{app:D2-decomposition}

Following the approach of~\citet{yeung2025spectral}, the D$_2$ symmetry is enforced by first estimating the symmetric-symmetric mean in the four quadrants, i.e.
\begin{equation}
    \overline{\boldsymbol{q}}_\mathit{SS}(x,y,z) = \frac{1}{4}
    \begin{bmatrix}
    \overline{u}(x,y,z) + \overline{u}(x,-y,z) + \overline{u}(x,y,-z) + \overline{u}(x,-y,-z) \\
    \overline{v}(x,y,z) - \overline{v}(x,-y,z) + \overline{v}(x,y,-z) - \overline{v}(x,-y,-z) \\
    \overline{w}(x,y,z) + \overline{w}(x,-y,z) - \overline{w}(x,y,-z) - \overline{w}(x,-y,-z)
    \end{bmatrix} \mbox{,}
    \label{eq:qSS}
\end{equation}
\noindent where $(x,y,z)$ denote the Cartesian coordinates, and $(u,v,w)$ the corresponding velocity components (see the picture and sketch in Figure~\ref{fig:picture_piv}), $t$ is time and overbars indicate time-averaged quantities.
The sign changes in the Cartesian coordinates indicate which quadrant of the domain is taken into account to extract the D$_2$ modes.

Once $\overline{\boldsymbol{q}}_\mathit{SS}$ is evaluated, the velocity fluctuation components are obtained as 
\begin{equation}
    \boldsymbol{q}^\prime(x,y,z,t) = \boldsymbol{q}(x,y,z,t) - \overline{\boldsymbol{q}}_\mathit{SS}(x,y,z) \mbox{,}
    \label{eq:Reynolds_decomposition}
\end{equation}
\noindent where 
\begin{equation}
    \boldsymbol{q}^\prime = {\boldsymbol{q}_\mathit{SS}}^\prime + {\boldsymbol{q}_\mathit{SA}}^\prime + {\boldsymbol{q}_\mathit{AS}}^\prime + {\boldsymbol{q}_\mathit{AA}}^\prime
    \label{eq:D2_fluctuations}
\end{equation}
\noindent as per~\eqref{eq:D2_decomposition}, and the four symmetry components are given by
\begin{subequations}
    \begin{align}
        {\boldsymbol{q}_\mathit{SS}}^\prime(x,y,z,t) = \frac{1}{4}
        \begin{bmatrix}
        u'(x,y,z,t) + u'(x,-y,z,t) + u'(x,y,-z,t) + u'(x,-y,-z,t) \\
        v'(x,y,z,t) - v'(x,-y,z,t) + v'(x,y,-z,t) - v'(x,-y,-z,t) \\
        w'(x,y,z,t) + w'(x,-y,z,t) - w'(x,y,-z,t) - w'(x,-y,-z,t)
        \end{bmatrix} \mbox{,} \\
        {\boldsymbol{q}_\mathit{SA}}^\prime(x,y,z,t) = \frac{1}{4}
        \begin{bmatrix}
        u'(x,y,z,t) + u'(x,-y,z,t) - u'(x,y,-z,t) - u'(x,-y,-z,t) \\
        v'(x,y,z,t) - v'(x,-y,z,t) - v'(x,y,-z,t) + v'(x,-y,-z,t) \\
        w'(x,y,z,t) + w'(x,-y,z,t) + w'(x,y,-z,t) + w'(x,-y,-z,t)
        \end{bmatrix} \mbox{,} \\
        {\boldsymbol{q}_\mathit{AS}}^\prime(x,y,z,t) = \frac{1}{4}
        \begin{bmatrix}
        u'(x,y,z,t) - u'(x,-y,z,t) + u'(x,y,-z,t) - u'(x,-y,-z,t) \\
        v'(x,y,z,t) + v'(x,-y,z,t) - v'(x,y,-z,t) - v'(x,-y,-z,t) \\
        w'(x,y,z,t) - w'(x,-y,z,t) - w'(x,y,-z,t) + w'(x,-y,-z,t)
        \end{bmatrix} \mbox{,} \\
        {\boldsymbol{q}_\mathit{AA}}^\prime(x,y,z,t) =
        \frac{1}{4} \begin{bmatrix}
        u'(x,y,z,t) - u'(x,-y,z,t) - u'(x,y,-z,t) + u'(x,-y,-z,t) \\
        v'(x,y,z,t) + v'(x,-y,z,t) + v'(x,y,-z,t) + v'(x,-y,-z,t) \\
        w'(x,y,z,t) - w'(x,-y,z,t) + w'(x,y,-z,t) - w'(x,-y,-z,t)
        \end{bmatrix} \mbox{.}
    \end{align}
\end{subequations}

%%%%%%%%%%%%%%%%%%%%%%%%%%%%%%%%%%%%%%%%%%%%%%
\section{SPOD convergence}
\label{app:SPOD_convergence}

In this appendix, a convergence analysis of the SPOD computations at $M_j = 0.4$ is performed using two subsets of the dataset: the full time series ($N_s = 151,356$ snapshots) and two subsets containing half of the snapshots ($N_s = 75,768$).
These subsets correspond to $N_b = 2367$ and $1183$ blocks, respectively, with the block size fixed at $N_{\mathit{fft}} = 128$ and a 50\% overlap, as described in \S~\ref{sec:experimental_database}.
Similar convergence behavior was observed for $M_j = 0.7$, which is not addressed here for brevity.

This split-record approach provides a practical assessment of the statistical robustness of the SPOD estimates when substantially longer experimental records are not available, and has been adopted in previous SPOD studies \citep{schmidt2019efficient,lesshafft2019resolvent,abreu2021spanwise,blanco2022improved,chiu2023vorticity}.

\begin{figure}
    \centering
    \includegraphics[width=\textwidth]{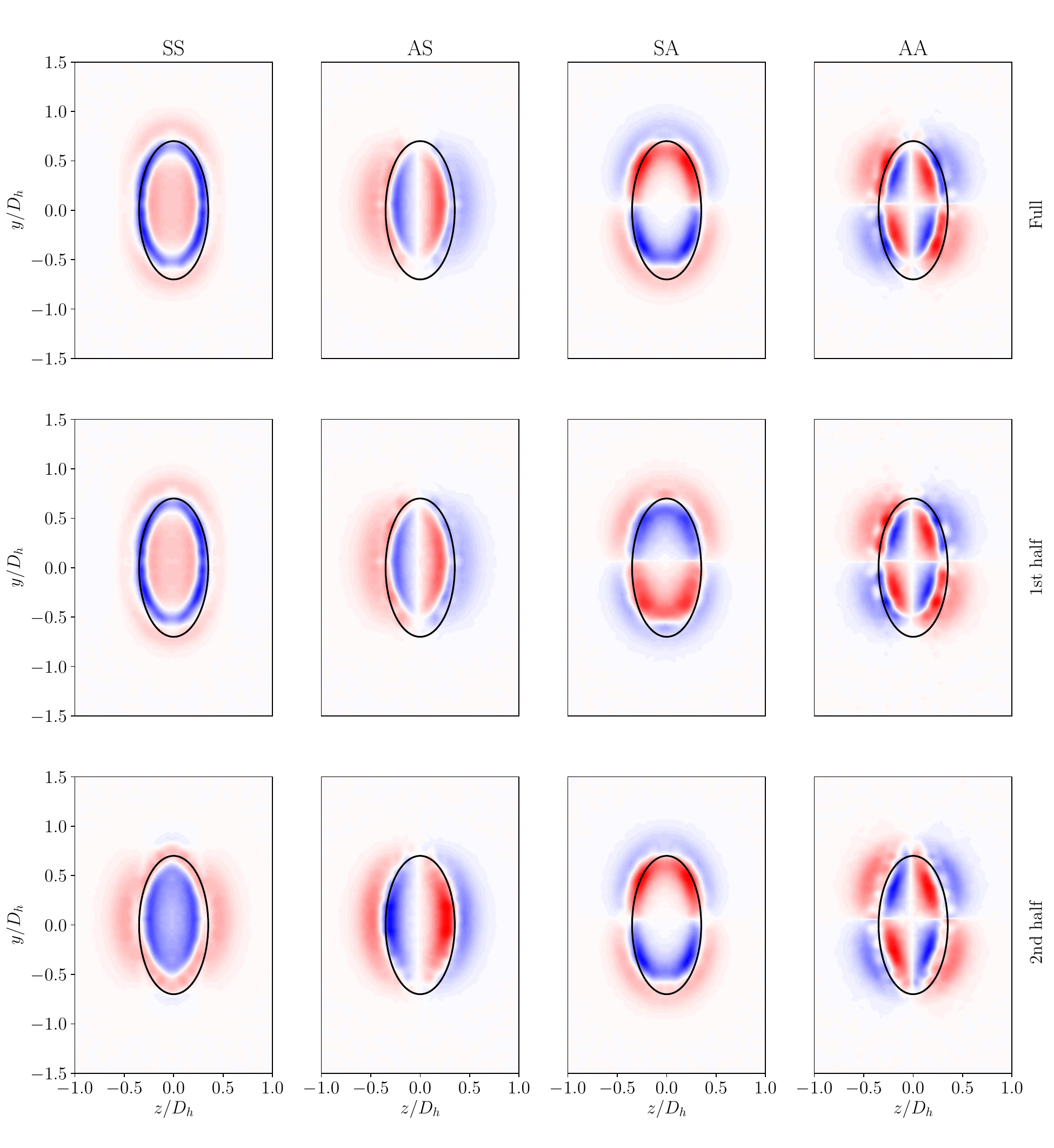}
    \caption{Real part of the leading SPOD mode for $(\mathit{St},~x/D_h) \approx (0.8,~1)$ and $M_j = 0.4$. Contours show the streamwise velocity fluctuation component, from blue (negative) to red (positive). The elliptical curve denotes the nozzle lip. Frames, from left to right: SS, AS, SA, and AA symmetry components. Top row: full time series. Middle row: first half of the time series. Bottom row: second half of the time series.}
    \label{fig:SPODmodes_convergence_xD1_St075}
\end{figure}

\begin{figure}
    \centering
    \includegraphics[width=\textwidth]{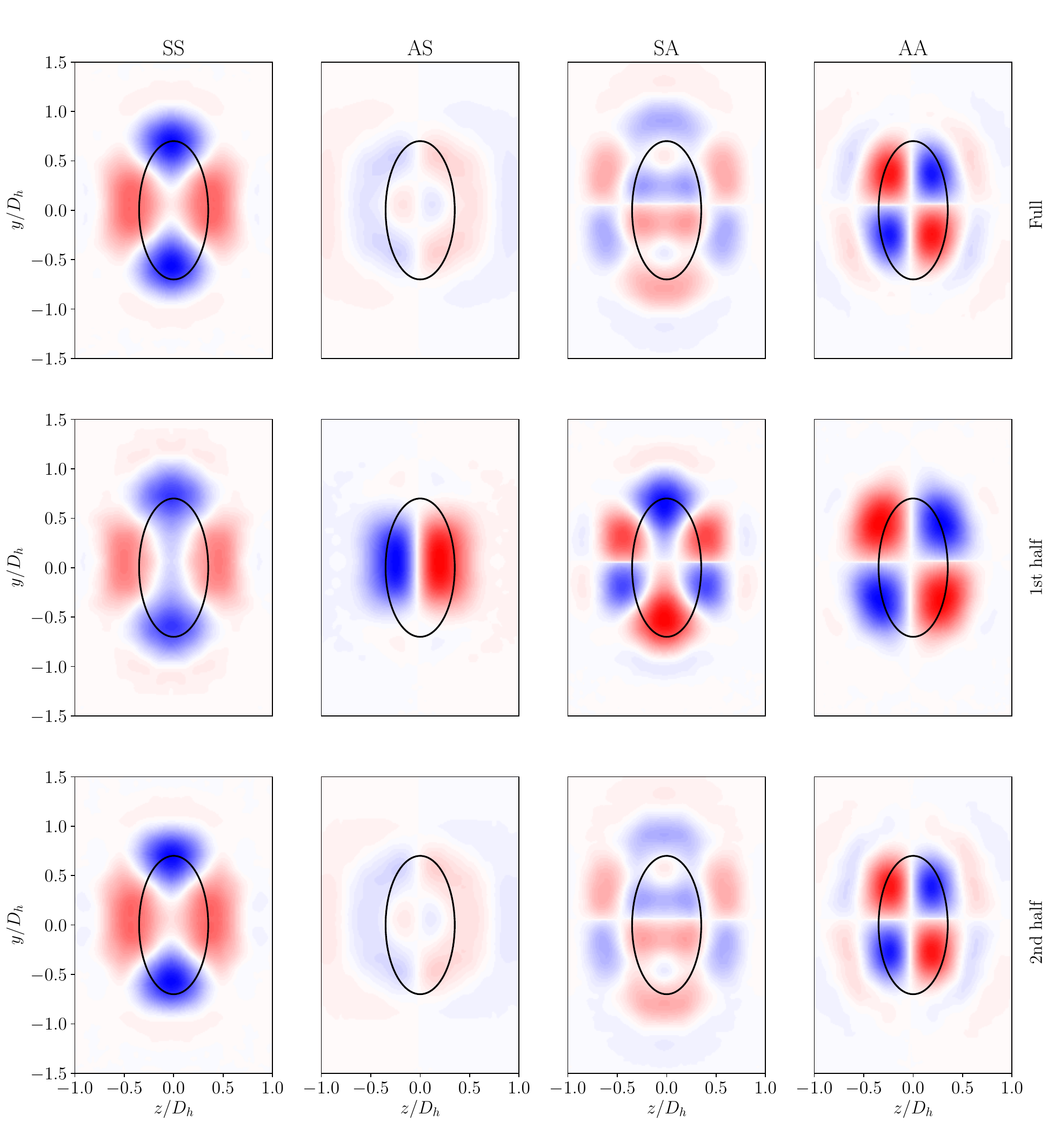}
    \caption{Real part of the leading SPOD mode for $(\mathit{St},~x/D_h) \approx (0.2,~6)$ and $M_j = 0.4$. See the comments in the caption of Figure~\ref{fig:SPODmodes_convergence_xD1_St075}.}
    \label{fig:SPODmodes_convergence_xD6_St02}
\end{figure}

We begin with a qualitative analysis, showing the real parts of the leading SPOD eigenfunctions (associated with the streamwise velocity component) in figures~\ref{fig:SPODmodes_convergence_xD1_St075} and~\ref{fig:SPODmodes_convergence_xD6_St02}, for $(\mathit{St},~x/D_h) \approx (0.8,~1)$ and $(\mathit{St},~x/D_h) \approx (0.2,~6)$, respectively.
In both cases, all symmetry components exhibit qualitatively consistent structures across the different subsets of the time series.
The agreement between the subsets also provides a check of statistical stationarity over the duration of the measurements.
More importantly for the present analysis, the persistence of the modal structures when the number of snapshots is substantially reduced indicates that the corresponding SPOD estimates are robust to the particular subset of the available data.
Thus, the comparison is used here as an empirical assessment of statistical convergence rather than as a test of stationarity alone.

Next, a quantitative measure of convergence is provided by the alignment metric defined in equation~\eqref{eq:beta}.
Here, the alignment is computed between SPOD modes obtained from the full dataset and those from each half-subset, rather than comparing SPOD to PSE modes as done in the main text.
A value of $\beta$ close to unity indicates that the spatial organisation of a given SPOD mode is insensitive to the reduction of the dataset, providing a quantitative measure of the robustness of the corresponding modal estimate.

\begin{figure}
    \centering
    \includegraphics[width=\textwidth]{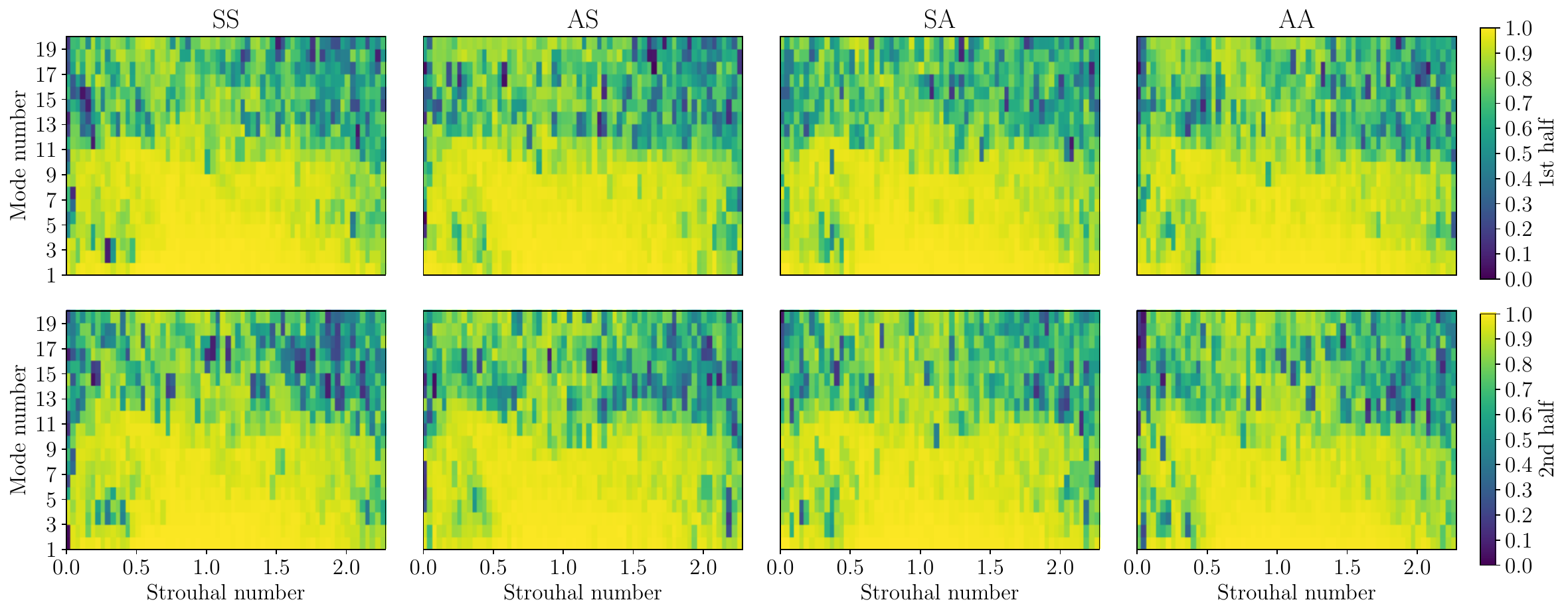}
    \caption{SPOD convergence maps for $M_j = 0.4$ and $x/D_h = 1$. Frames, from left to right: SS, SA, AS, and AA symmetry components.}
    \label{fig:SPOD_convergence_xD1}
\end{figure}

Figure~\ref{fig:SPOD_convergence_xD1} presents convergence maps as a function of Strouhal number for $x/D_h = 1$, showing the first 20 SPOD modes.
These maps are constructed following the methodology outlined in~\citet{kaplan2021nozzle}, and include all four symmetry components.
As observed, the modes are generally well converged—particularly the leading modes in the frequency range $0.5 \leq \mathit{St} \leq 1.5$.
The convergence level progressively decreases with increasing mode order, reflecting the greater sensitivity of the higher-order modes to the finite number of statistically independent realisations available for the SPOD estimation.
This behaviour is consistent with previous SPOD convergence studies, which have also reported a stronger sensitivity of higher-order modes to the available sample size \citep{lesshafft2019resolvent,chiu2023vorticity}.

\begin{figure}
    \centering
    \includegraphics[width=\textwidth]{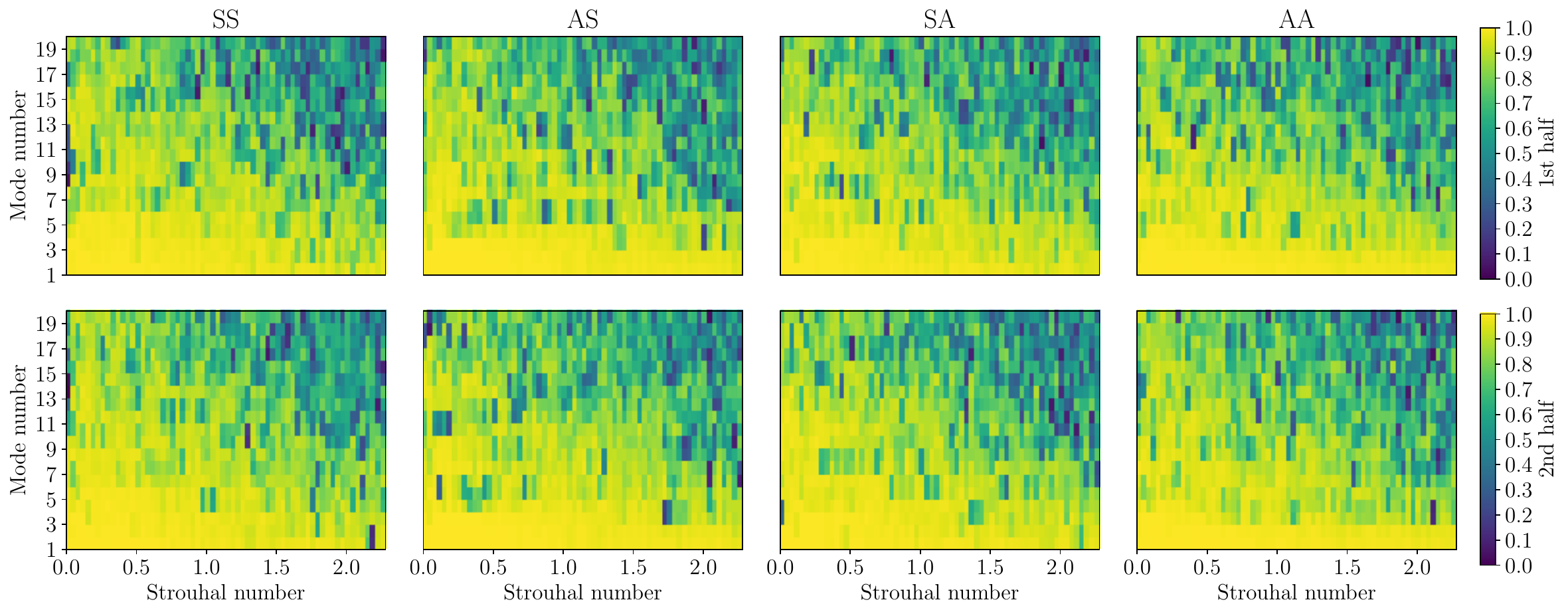}
    \caption{SPOD convergence maps for $M_j = 0.4$ and $x/D_h = 6$. See the comments in the caption of Figure~\ref{fig:SPOD_convergence_xD1}.}
    \label{fig:SPOD_convergence_xD6}
\end{figure}

Further downstream, figure~\ref{fig:SPOD_convergence_xD6} displays convergence maps at $x/D_h = 6$, constructed using the same approach as in figure~\ref{fig:SPOD_convergence_xD1}.
At this location, the modes show best convergence at lower frequencies, consistent with the trends observed in the alignment between SPOD and PSE modes discussed in \S~\ref{sec:psestructures}.

\begin{figure}
    \centering
    \includegraphics[width=\textwidth]{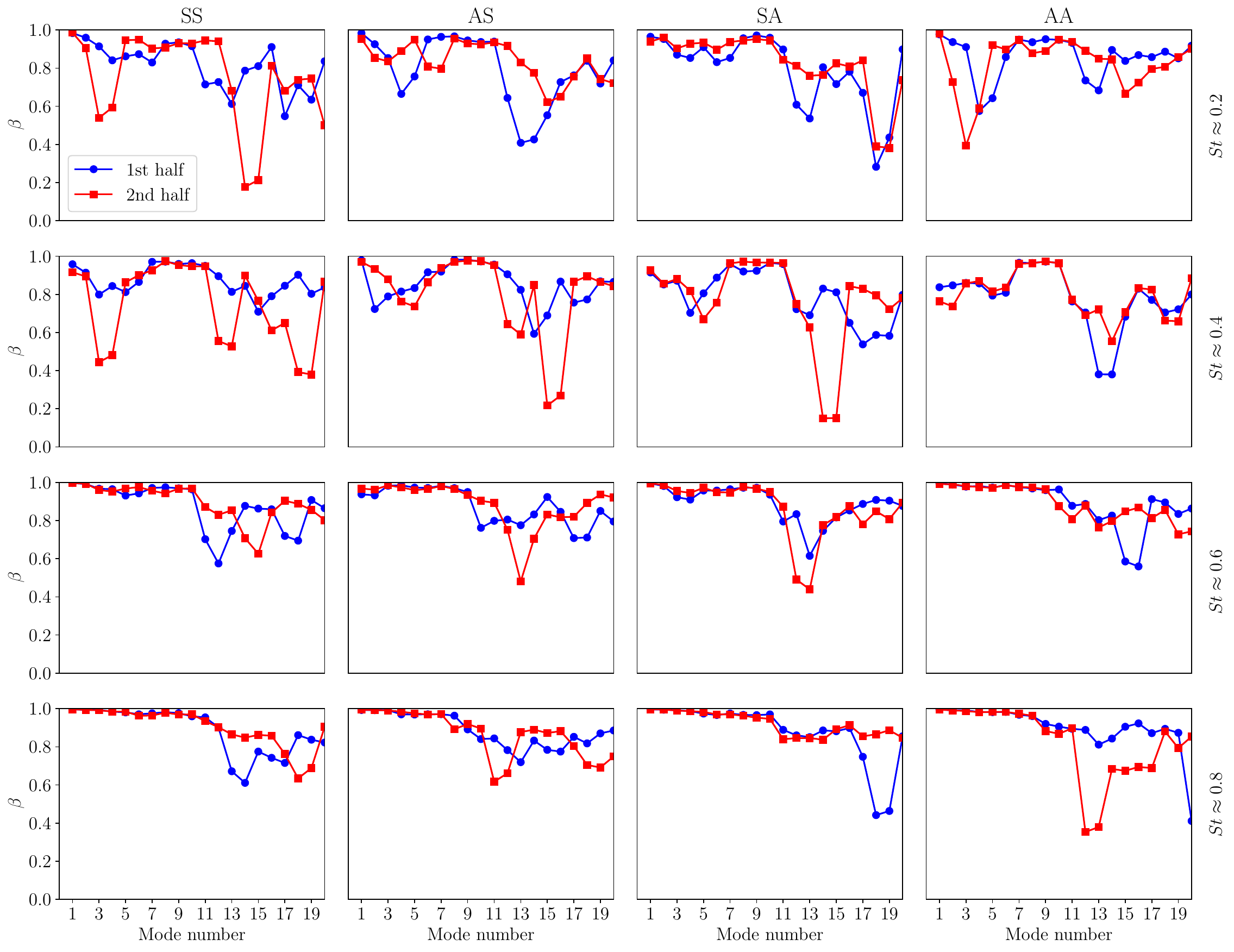}
    \caption{SPOD convergence maps for $M_j = 0.4$ at $x/D_h = 1$. Frames, from left to right: SS, SA, AS, and AA symmetry components. Frames, from top to bottom: $\mathit{St} \approx$ 0.2, 0.4, 0.6, and 0.8.}
    \label{fig:SPOD_convergence_xD1_St}
\end{figure}

\begin{figure}
    \centering
    \includegraphics[width=\textwidth]{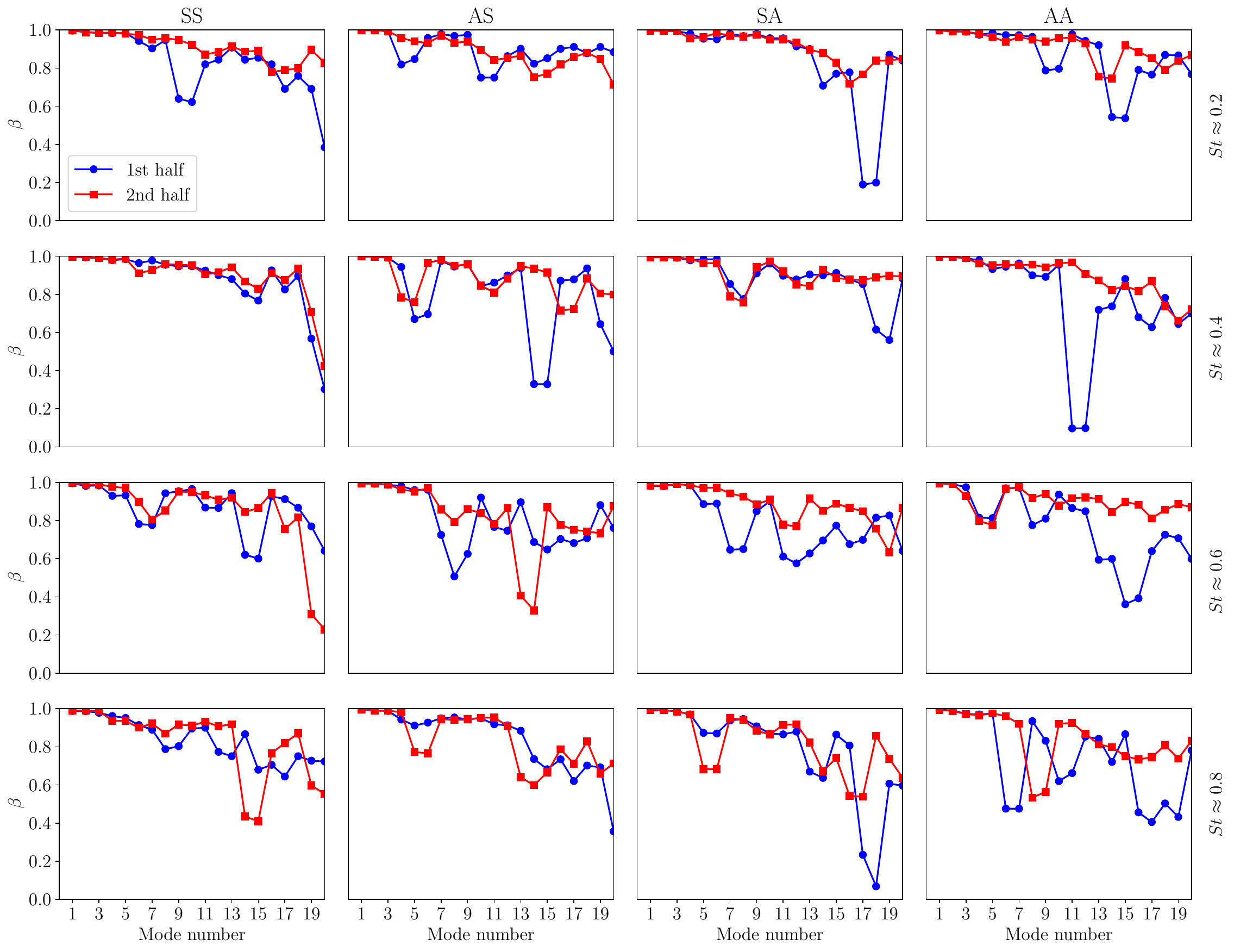}
    \caption{SPOD convergence maps for $M_j = 0.4$ at $x/D_h = 6$. See the comments in the caption of Figure~~\ref{fig:SPOD_convergence_xD1_St}.}
    \label{fig:SPOD_convergence_xD6_St}
\end{figure}

We now examine convergence at selected Strouhal numbers, namely $\mathit{St} \approx 0.2$, 0.4, 0.6, and 0.8, as shown in Figures~\ref{fig:SPOD_convergence_xD1_St} and~\ref{fig:SPOD_convergence_xD6_St}, following the approach of~\citet{lesshafft2019resolvent}.
These maps further highlight the spatial variation in convergence behavior: near the nozzle exit (e.g., $x/D_h = 1$), higher frequencies (above $\mathit{St} \approx 0.8$) tend to exhibit better convergence, as illustrated in Figure~\ref{fig:SPOD_convergence_xD1_St}.
Conversely, farther downstream (e.g., $x/D_h = 6$), the modes show stronger convergence at lower frequencies (around and below $\mathit{St} \approx 0.4$), consistent with the trends observed previously.

For comparison, the following figures repeat the convergence study for $M_j = 0.4$, now using the smaller database with $N_s = 27{,}000$ snapshots~\citep{amaral2023elliptical,amaral2024elliptical}.
Overall, the convergence deteriorates relative to the larger dataset, with high convergence values observed only for the first two or three leading modes.
Despite this reduction in convergence levels, the trends for the $N_s = 27{,}000$ snapshots database follow the same general pattern as those observed with $N_s = 151{,}356$.
Similar convergence behavior was found for the $M_j = 0.7$ database, which also used $N_s = 27{,}000$ snapshots; this case is omitted here for brevity.

\begin{figure}
    \centering
    \includegraphics[width=\textwidth]{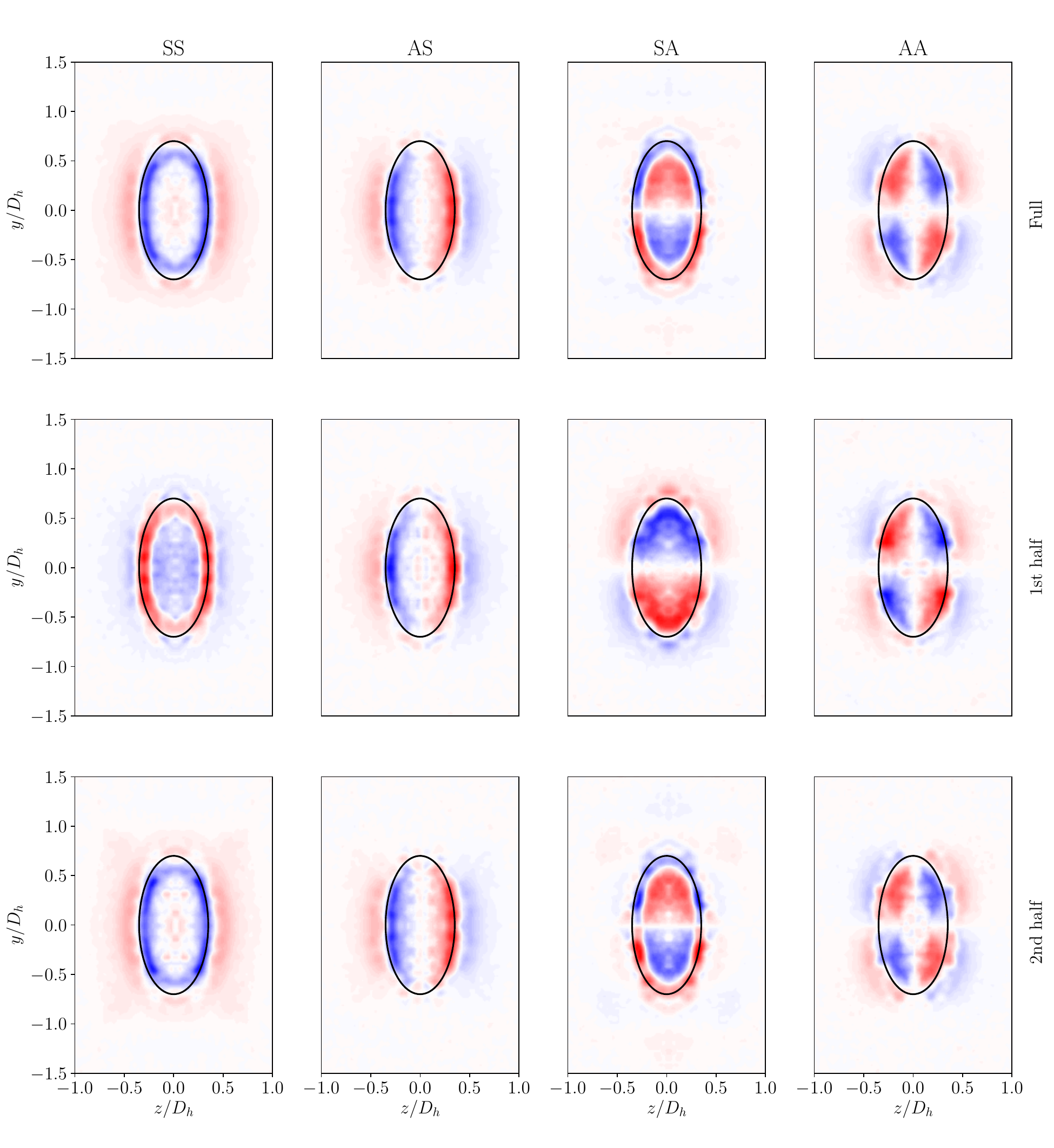}
    \caption{Real part of the leading SPOD mode for $(\mathit{St},~x/D_h) \approx (0.8,~1)$, $M_j = 0.4$, and $N_s = 27{,}000$. See the comments in the caption of Figure~\ref{fig:SPODmodes_convergence_xD1_St075}.}

    \label{fig:SPODmodes_convergence_xD1_St075_old}
\end{figure}

\begin{figure}
    \centering
    \includegraphics[width=\textwidth]{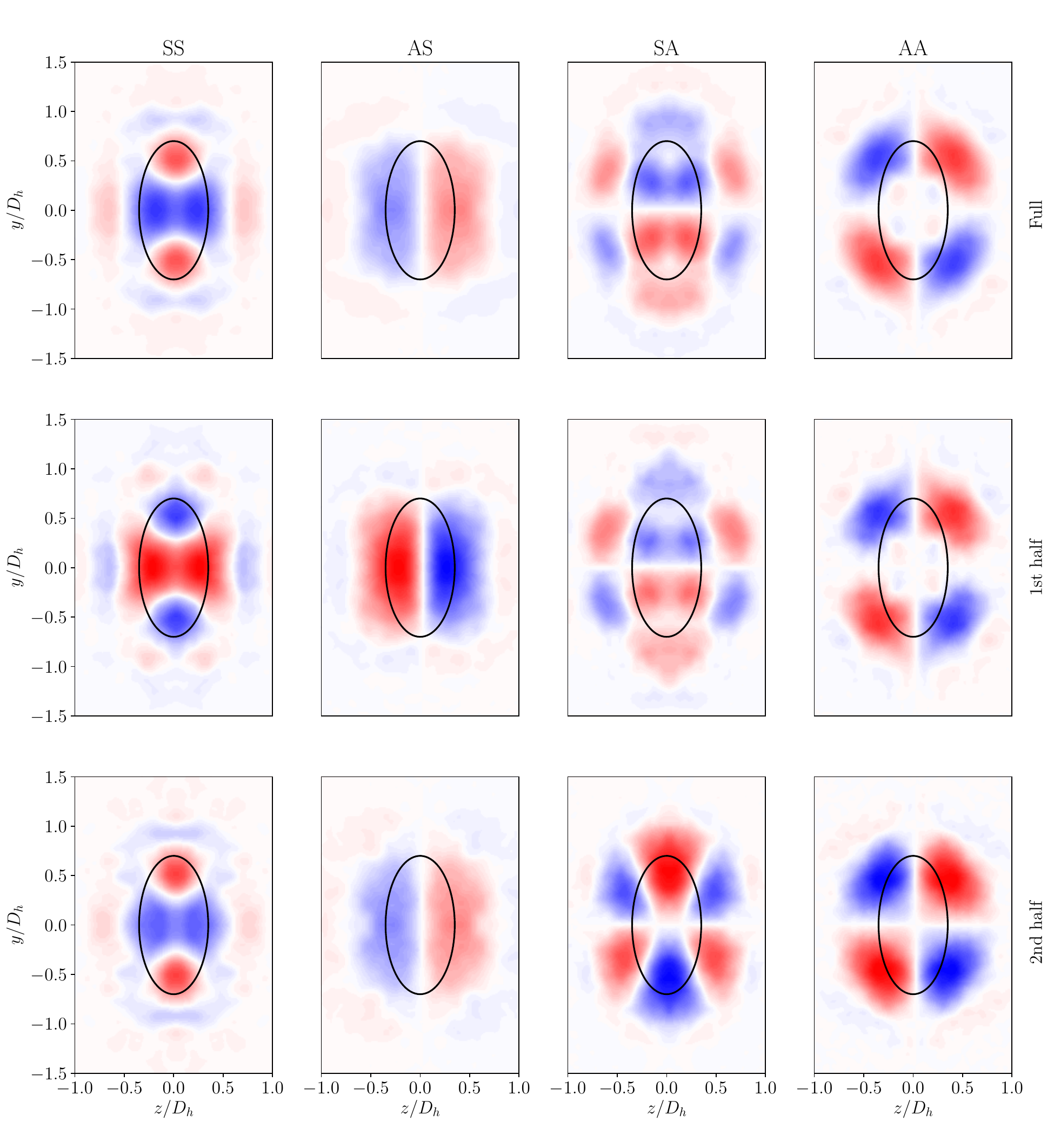}
    \caption{Real part of the leading SPOD mode for $(\mathit{St},~x/D_h) \approx (0.2,~6)$, $M_j = 0.4$, and $N_s = 27{,}000$. See the comments in the caption of Figure~\ref{fig:SPODmodes_convergence_xD1_St075}.}
    \label{fig:SPODmodes_convergence_xD6_St02_old}
\end{figure}

\begin{figure}
    \centering
    \includegraphics[width=\textwidth]{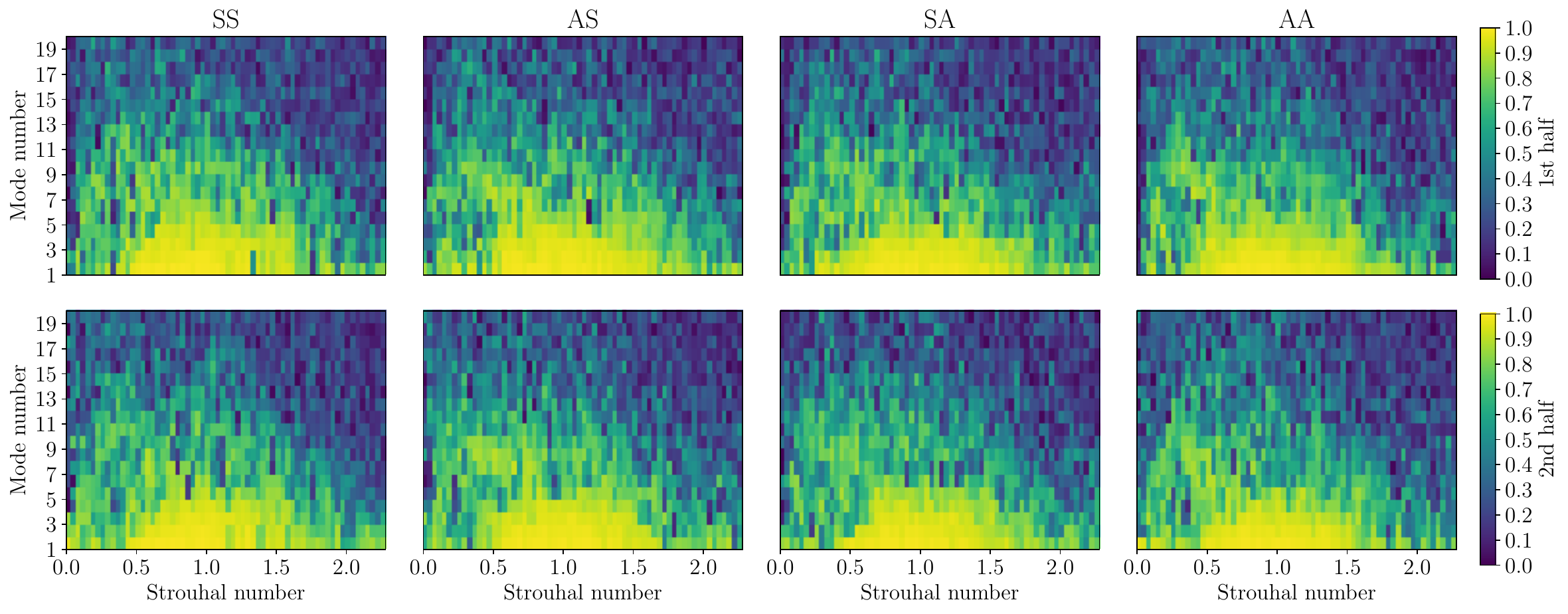}
    \caption{SPOD convergence maps for $M_j = 0.4$, $x/D_h = 1$, and $N_s = 27{,}000$. See the caption of Figure~\ref{fig:SPOD_convergence_xD1} for details.}
    \label{fig:SPOD_convergence_xD1_old}
\end{figure}

\begin{figure}
    \centering
    \includegraphics[width=\textwidth]{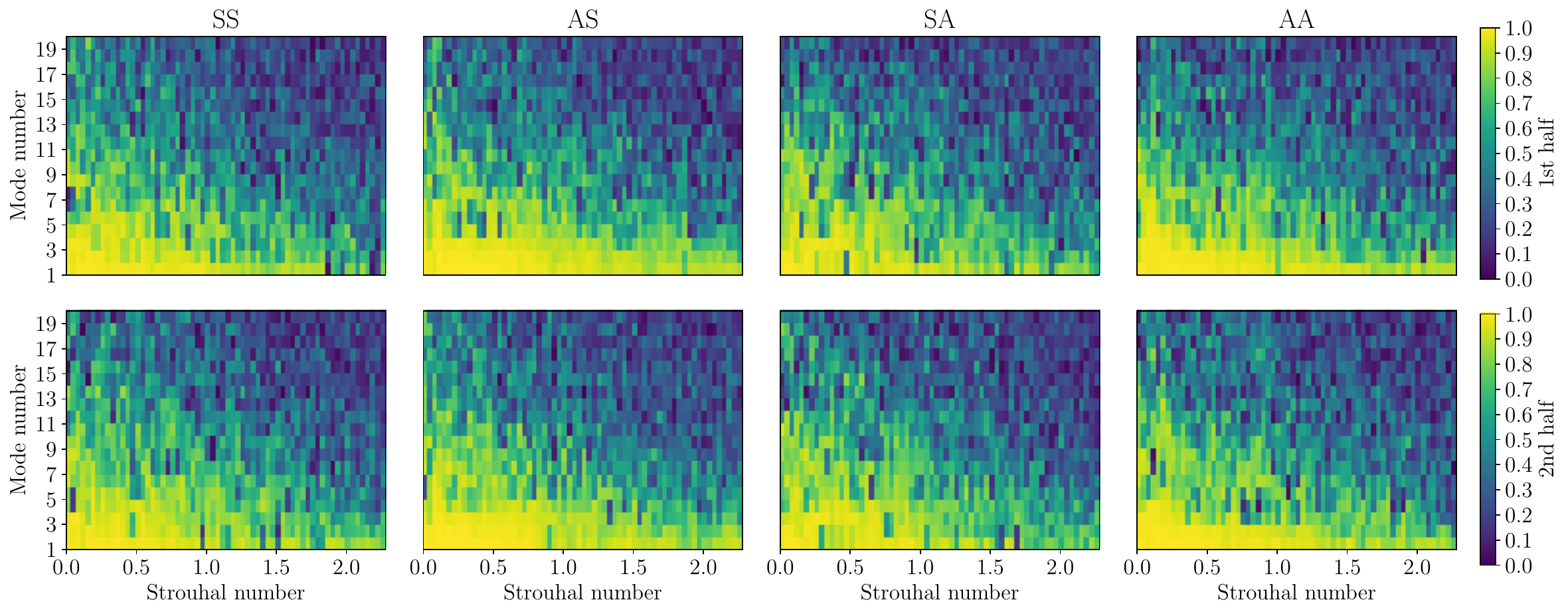}
    \caption{SPOD convergence maps for $M_j = 0.4$, $x/D_h = 6$, and $N_s = 27{,}000$. See the caption of Figure~\ref{fig:SPOD_convergence_xD1} for details.}
    \label{fig:SPOD_convergence_xD6_old}
\end{figure}

\begin{figure}
    \centering
    \includegraphics[width=\textwidth]{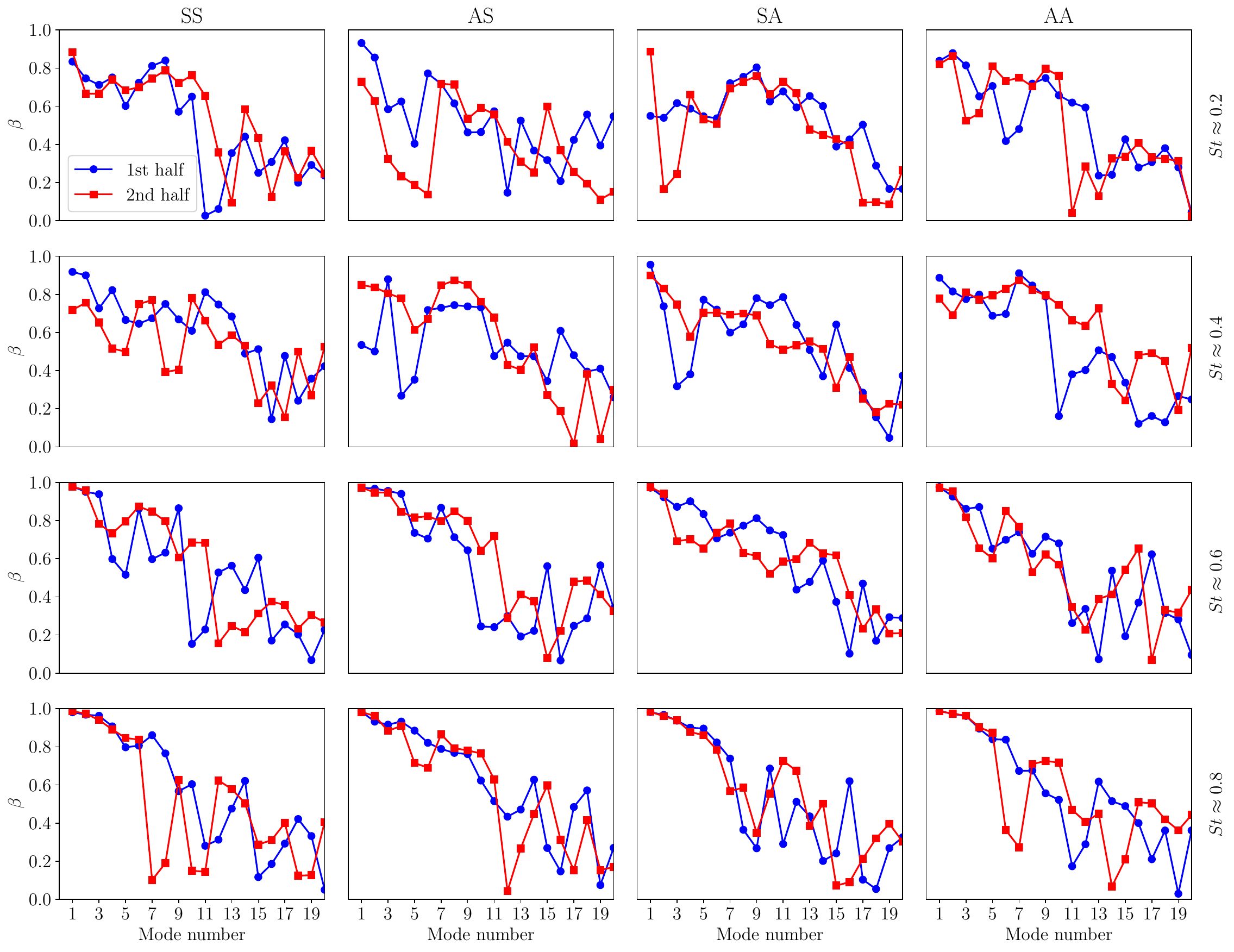}
    \caption{SPOD convergence maps for $M_j = 0.4$, $x/D_h = 1$, and $N_s = 27{,}000$. See the caption of Figure~\ref{fig:SPOD_convergence_xD1_St} for details.}
    \label{fig:SPOD_convergence_xD1_St_old}
\end{figure}

\begin{figure}
    \centering
    \includegraphics[width=\textwidth]{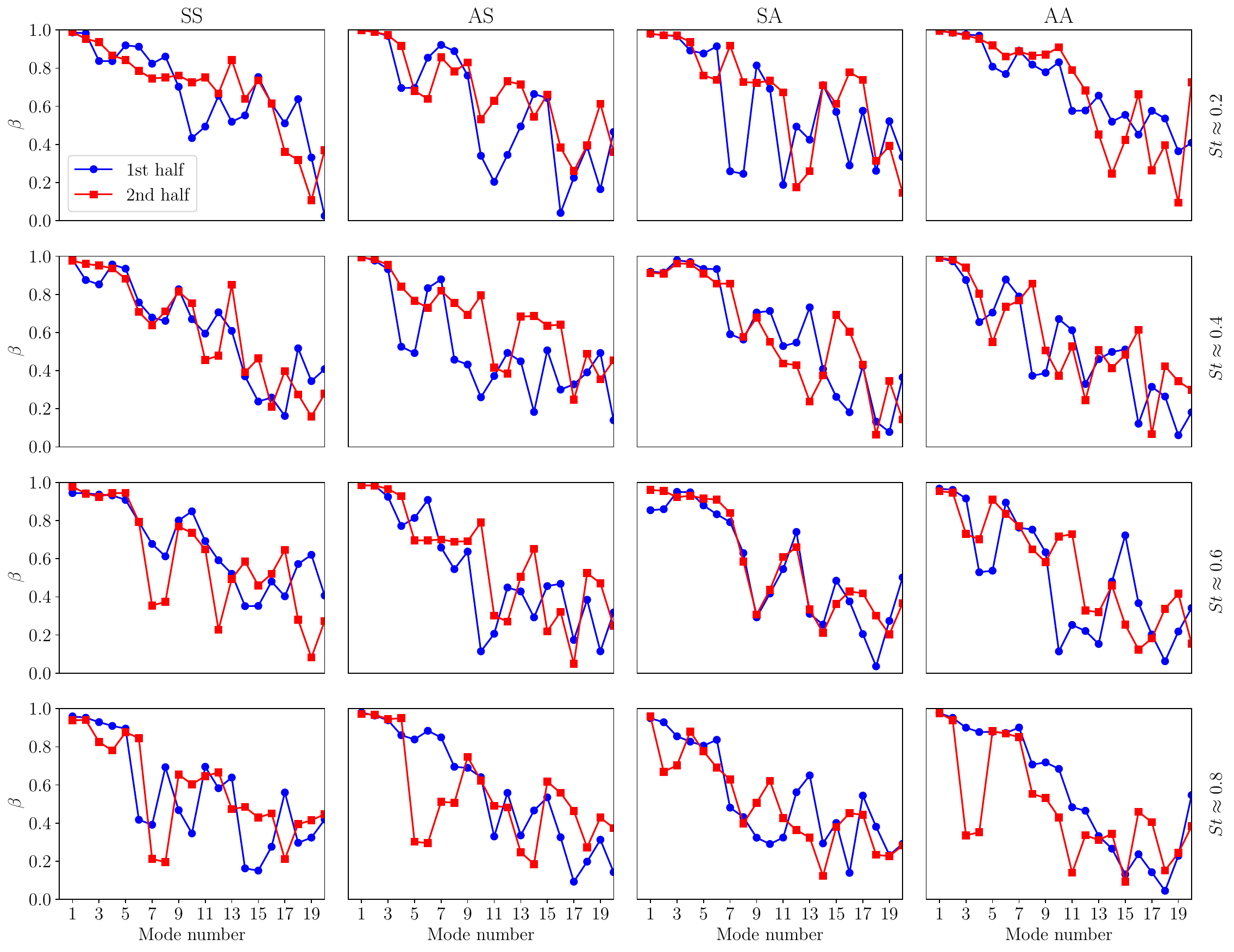}
    \caption{SPOD convergence maps for $M_j = 0.4$, $x/D_h = 6$, and $N_s = 27{,}000$. See the caption of Figure~\ref{fig:SPOD_convergence_xD1_St} for details.}
    \label{fig:SPOD_convergence_xD6_St_old}
\end{figure}

These results further support the focus of the main analysis on the leading SPOD modes.
While higher-order modes can contain non-negligible spectral energy, their reduced statistical robustness makes their detailed spatial organisation increasingly sensitive to the finite sample size.
Restricting the comparison with the 3D-PSE predictions to the leading modes therefore provides a more reproducible and physically robust assessment of the dominant coherent structures.

%%%%%%%%%%%%%%%%%%%%%%%%%%%%%%%%%%%%%%%%%%%%%%
\section{Compressible operators}
\label{app:compressible_operators}

%%%%%%%%%%%%%%%%%%%%%%%%%%%%%%%%%%%%%%%%%%%%%%
\subsection{3D-PSE}
\label{app:3dpse_operators}

Beginning with the compressible Navier--Stokes equations, we can derive the operators for the 3D-PSE process~\eqref{eq:pseprob}.

\begin{equation}
    \begin{aligned}
        \begin{pmatrix}
        %%%%%%%%%%%%%%%%%%%%%%%%% M1 %%%%%%%%%%%%%%%%%%%%%%%%%
        \mathcal{M}_{11} & 0 & 0 & 0 & 0 \\
        %%%%%%%%%%%%%%%%%%%%%%%%% M2 %%%%%%%%%%%%%%%%%%%%%%%%%
        0 & \mathcal{M}_{22} & 0 & 0 & 0 \\
        %%%%%%%%%%%%%%%%%%%%%%%%% M3 %%%%%%%%%%%%%%%%%%%%%%%%%
        0 & 0 & \mathcal{M}_{33} & \mathcal{M}_{34} & \mathcal{M}_{35} \\
        %%%%%%%%%%%%%%%%%%%%%%%%% M4 %%%%%%%%%%%%%%%%%%%%%%%%%
        0 & 0 & \mathcal{M}_{43} & \mathcal{M}_{44} & 0 \\ 
        %%%%%%%%%%%%%%%%%%%%%%%%% M5 %%%%%%%%%%%%%%%%%%%%%%%%%
        0 & 0 & \mathcal{M}_{53} & 0 & \mathcal{M}_{55} \\ 
        \end{pmatrix}
        %%%%%%%%%%%%%%%%%%%%%%%%%%%%%%%%%%%%%%%%%%%%%%%%%%%%%%
        \begin{pmatrix}
        \hat{w_x} \\
        \hat{v_x} \\
        \hat{u_x} \\
        \hat{\rho_x} \\
        \hat{\tau_x}
        \end{pmatrix}
        %%%%%%%%%%%%%%%%%%%%%%%%%%%%%%%%%%%%%%%%%%%%%%%%%%%%%%
        =
        %%%%%%%%%%%%%%%%%%%%%%%%%%%%%%%%%%%%%%%%%%%%%%%%%%%%%%
        \begin{pmatrix}
        %%%%%%%%%%%%%%%%%%%%%%%%% N1 %%%%%%%%%%%%%%%%%%%%%%%%%
        \mathcal{N}_{11} & \mathcal{N}_{12} & \mathcal{N}_{13} & \mathcal{N}_{14} & \mathcal{N}_{15} \\
        %%%%%%%%%%%%%%%%%%%%%%%%% N2 %%%%%%%%%%%%%%%%%%%%%%%%%
        \mathcal{N}_{21} & \mathcal{N}_{22} & \mathcal{N}_{23} & \mathcal{N}_{24} & \mathcal{N}_{25} \\
        %%%%%%%%%%%%%%%%%%%%%%%%% N3 %%%%%%%%%%%%%%%%%%%%%%%%%
        \mathcal{N}_{31} & \mathcal{N}_{32} & \mathcal{N}_{33} & \mathcal{N}_{34} & \mathcal{N}_{35} \\
        %%%%%%%%%%%%%%%%%%%%%%%%% N4 %%%%%%%%%%%%%%%%%%%%%%%%%
        \mathcal{N}_{41} & \mathcal{N}_{42} & \mathcal{N}_{43} & \mathcal{N}_{44} & 0 \\
        %%%%%%%%%%%%%%%%%%%%%%%%% N5 %%%%%%%%%%%%%%%%%%%%%%%%%
        \mathcal{N}_{51} & \mathcal{N}_{52} & \mathcal{N}_{53} & \mathcal{N}_{54} & \mathcal{N}_{55} 
        \end{pmatrix}
        %%%%%%%%%%%%%%%%%%%%%%%%%%%%%%%%%%%%%%%%%%%%%%%%%%%%%%
        \begin{pmatrix}
        \hat{w_x} \\
        \hat{v_x} \\
        \hat{u_x} \\
        \hat{\rho_x} \\
        \hat{\tau_x}
        \end{pmatrix} \mbox{.}
        %%%%%%%%%%%%%%%%%%%%%%%%%%%%%%%%%%%%%%%%%%%%%%%%%%%%%%
    \end{aligned}
    \label{eq:compse}
\end{equation}

The full operators in~\eqref{eq:compse} are
\begin{subequations}
\begin{align}
        \mathcal{M}_{11} &= \mathcal{M}_{22} = \mathcal{M}_{33} = \mathcal{M}_{55} = \rho_0 U_0 \,, \\
        \mathcal{M}_{34} &= \frac{1}{\gamma Ma^2}T_0 \,, \\
        \mathcal{M}_{35} &= \frac{1}{\gamma Ma^2}\rho_0 \,, \\
        \mathcal{M}_{43} &= \rho_0 \,, \\
        \mathcal{M}_{44} &= U_0 \,, \\
        \mathcal{M}_{53} &= \frac{\gamma-1}{\gamma}\rho_0 T_0 \,, \\
\end{align}
\end{subequations}
\noindent and
\begin{subequations}
\begin{align}
        \mathcal{N}_{11} &= i\omega\rho_0 - \rho_0 W_0\mathcal{D}_z - \rho_0\mathcal{D}_z W_0 - \rho_0 V_0\mathcal{D}_y - i\alpha\rho_0 U_0 \,, \\
        \mathcal{N}_{12} &= - \rho_0\mathcal{D}_y W_0 \,, \\
        \mathcal{N}_{13} &= -\rho_0 W_{x0} \,, \\
        \mathcal{N}_{14} &= -\frac{1}{\gamma Ma^2}(\mathcal{D}_z T_0 + T_0 \mathcal{D}_z) - W_0 \mathcal{D}_z W_0 - V_0 \mathcal{D}_y W_0 - U_0 W_{x0} \,, \\
        \mathcal{N}_{15} &= -\frac{1}{\gamma Ma^2}(\mathcal{D}_z \rho_0 + \rho_0 \mathcal{D}_z) \,, \\
\end{align}
\end{subequations}
\noindent and
\begin{subequations}
\begin{align}
        \mathcal{N}_{21} &= - \rho_0 \mathcal{D}_z V_0 \,, \\
        \mathcal{N}_{22} &= i\omega\rho_0 - \rho_0 W_0 \mathcal{D}_z - \rho_0 \mathcal{D}_y V_0 - \rho_0 V_0 \mathcal{D}_y - i\alpha \rho_0 U_0 \,, \\
        \mathcal{N}_{23} &= -\rho_0 V_{x0} \,, \\
        \mathcal{N}_{24} &= -\frac{1}{\gamma Ma^2}(\mathcal{D}_y T_0 + T_0 \mathcal{D}_y) - W_0 \mathcal{D}_z V_0 - V_0 \mathcal{D}_y V_0 - U_0 V_{x0} \,, \\
        \mathcal{N}_{25} &= -\frac{1}{\gamma Ma^2}(\mathcal{D}_y \rho_0 + \rho_0 \mathcal{D}_y) \,, \\
\end{align}
\end{subequations}
\noindent and
\begin{subequations}
\begin{align}
        \mathcal{N}_{31} &= -\rho_0 \mathcal{D}_z U_0 \,, \\
        \mathcal{N}_{32} &= -\rho_0 \mathcal{D}_y U_0 \,, \\
        \mathcal{N}_{33} &= i\omega\rho_0 - \rho_0 W_0 \mathcal{D}_z - \rho_0 U_{x0} - \rho_0 V_0 \mathcal{D}_y - i\alpha \rho_0 U_0 \,, \\
        \mathcal{N}_{34} &= -\frac{1}{\gamma Ma^2}(i\alpha T_0 + T_{x0}) - W_0 \mathcal{D}_z U_0 - V_0 \mathcal{D}_y U_0 - U_0 U_{x0} \,, \\
        \mathcal{N}_{35} &= -\frac{1}{\gamma Ma^2}(i\alpha \rho_0 + \rho_{x0})  \,, \\
\end{align}
\end{subequations}
\noindent and
\begin{subequations}
\begin{align}
        \mathcal{N}_{41} &= -\mathcal{D}_z \rho_0 - \rho_0 \mathcal{D}_z \,, \\
        \mathcal{N}_{42} &= -\mathcal{D}_y \rho_0 - \rho_0 \mathcal{D}_y \,, \\
        \mathcal{N}_{43} &= -i\alpha \rho_0 - \rho_{x0} \,, \\
        \mathcal{N}_{44} &= i\omega - \mathcal{D}_z W_0 - \mathcal{D}_y V_0 - W_{x0} - i\alpha U_0 - W_0 \mathcal{D}_z - V_0 \mathcal{D}_y \,, \\
\end{align}
\end{subequations}
\noindent and
\begin{subequations}
\begin{align}
        \mathcal{N}_{51} &= - \rho_0 \mathcal{D}_z T_0 - \frac{\gamma-1}{\gamma} \rho_0 T_0 \mathcal{D}_z \,, \\
        \mathcal{N}_{52} &= - \rho_0 \mathcal{D}_y T_0 - \frac{\gamma-1}{\gamma} \rho_0 T_0 \mathcal{D}_y \,, \\
        \mathcal{N}_{53} &= - \rho_0 T_{x0} - \frac{\gamma-1}{\gamma} i\alpha \rho_0 T_0 \,, \\
        \mathcal{N}_{54} &= - W_0 \mathcal{D}_z T_0 - V_0 \mathcal{D}_y T_0 - U_0 T_{x0} - \frac{\gamma-1}{\gamma} T_0 \mathcal{D}_z W_0 - \frac{\gamma-1}{\gamma} T_0 \mathcal{D}_y V_0 - \frac{\gamma-1}{\gamma} T_0 U_{x0} \,, \\
        \mathcal{N}_{55} &= i\omega \rho_0 - i\alpha \rho_0 U_0 - \rho_0 W_0 \mathcal{D}_z - \rho_0 V_0 \mathcal{D}_y - \frac{\gamma-1}{\gamma} \rho_0 \mathcal{D}_z W_0 - \frac{\gamma-1}{\gamma} \rho_0 \mathcal{D}_y V_0 - \frac{\gamma-1}{\gamma} \rho_0 U_{x0} \,.
\end{align}
\end{subequations}

%%%%%%%%%%%%%%%%%%%%%%%%%%%%%%%%%%%%%%%%%%%%%%
\subsection{2D-Local}
\label{app:2dlocal_operators}

Beginning with the compressible Navier--Stokes equations, we can derive the operators for the local eigenvalue problem~\eqref{eq:egv} as
\begin{equation}
    \begin{aligned}
        \begin{pmatrix}
        %%%%%%%%%%%%%%%%%%%%%%%%% A1 %%%%%%%%%%%%%%%%%%%%%%%%%
        \mathcal{A}_{11} & 0 & 0 & \mathcal{A}_{14} & \mathcal{A}_{15} \\
        %%%%%%%%%%%%%%%%%%%%%%%%% A2 %%%%%%%%%%%%%%%%%%%%%%%%%
        0 & \mathcal{A}_{22} & 0 & \mathcal{A}_{24} & \mathcal{A}_{25} \\
        %%%%%%%%%%%%%%%%%%%%%%%%% A3 %%%%%%%%%%%%%%%%%%%%%%%%%
        \mathcal{A}_{31} & \mathcal{A}_{32} & \mathcal{A}_{33} & 0 & 0 \\
        %%%%%%%%%%%%%%%%%%%%%%%%% A4 %%%%%%%%%%%%%%%%%%%%%%%%%
        \mathcal{A}_{41} & \mathcal{A}_{42} & 0 & \mathcal{A}_{44} & 0 \\ 
        %%%%%%%%%%%%%%%%%%%%%%%%% A5 %%%%%%%%%%%%%%%%%%%%%%%%%
        \mathcal{A}_{51} & \mathcal{A}_{52} & 0 & 0 & \mathcal{A}_{55} \\
        \end{pmatrix}
        %%%%%%%%%%%%%%%%%%%%%%%%%%%%%%%%%%%%%%%%%%%%%%%%%%%%%%
        \begin{pmatrix}
        \tilde{w} \\
        \tilde{v} \\
        \tilde{u} \\
        \tilde{\rho} \\
        \tilde{\tau}
        \end{pmatrix}
        %%%%%%%%%%%%%%%%%%%%%%%%%%%%%%%%%%%%%%%%%%%%%%%%%%%%%%
        =\alpha
        %%%%%%%%%%%%%%%%%%%%%%%%%%%%%%%%%%%%%%%%%%%%%%%%%%%%%%
        \begin{pmatrix}
        %%%%%%%%%%%%%%%%%%%%%%%%% B1 %%%%%%%%%%%%%%%%%%%%%%%%%
        \mathcal{B}_{11} & 0 & 0 & 0 & 0 \\
        %%%%%%%%%%%%%%%%%%%%%%%%% B2 %%%%%%%%%%%%%%%%%%%%%%%%%
        0 & \mathcal{B}_{22} & 0 & 0 & 0 \\
        %%%%%%%%%%%%%%%%%%%%%%%%% B3 %%%%%%%%%%%%%%%%%%%%%%%%%
        \mathcal{B}_{31} & \mathcal{B}_{32} & \mathcal{B}_{33} & \mathcal{B}_{34} & \mathcal{B}_{35} \\
        %%%%%%%%%%%%%%%%%%%%%%%%% AB4 %%%%%%%%%%%%%%%%%%%%%%%%%
        0 & 0 & \mathcal{B}_{43} & \mathcal{B}_{44} & 0 \\ 
        %%%%%%%%%%%%%%%%%%%%%%%%% B5 %%%%%%%%%%%%%%%%%%%%%%%%%
        0 & 0 & \mathcal{B}_{53} & 0 & \mathcal{B}_{55} \\
        \end{pmatrix}
        %%%%%%%%%%%%%%%%%%%%%%%%%%%%%%%%%%%%%%%%%%%%%%%%%%%%%%
        \begin{pmatrix}
        \tilde{w} \\
        \tilde{v} \\
        \tilde{u} \\
        \tilde{\rho} \\
        \tilde{\tau} 
        \end{pmatrix} \mbox{.}
        %%%%%%%%%%%%%%%%%%%%%%%%%%%%%%%%%%%%%%%%%%%%%%%%%%%%%%
    \end{aligned}
    \label{eq:comlocal}
\end{equation}

The full operators in~\eqref{eq:comlocal} are
\begin{subequations}
\begin{align}
    \mathcal{A}_{11} &= -i\omega\rho_0 \,, \\
    \mathcal{A}_{14} &= \frac{1}{\gamma Ma^2}(\mathcal{D}_z T_0 + T_0\mathcal{D}_z) \,, \\
    \mathcal{A}_{15} &= \frac{1}{\gamma Ma^2}(\mathcal{D}_z \rho_0 + \rho_0\mathcal{D}_z) \,, \\
\end{align}
\end{subequations}
\noindent and
\begin{subequations}
\begin{align}
    \mathcal{A}_{22} &= - i\omega\rho_0 \,, \\
    \mathcal{A}_{24} &= \frac{1}{\gamma Ma^2}(\mathcal{D}_y T_0 + T_0\mathcal{D}_y) \,, \\
    \mathcal{A}_{25} &= \frac{1}{\gamma Ma^2}(\mathcal{D}_y \rho_0 + \rho_0\mathcal{D}_y) \,, \\
\end{align}
\end{subequations}
\noindent and
\begin{subequations}
\begin{align}
    \mathcal{A}_{31} &= \rho_0\mathcal{D}_z U_0 \,, \\
    \mathcal{A}_{32} &= \rho_0\mathcal{D}_y U_0 \,, \\
    \mathcal{A}_{33} &= - i\omega\rho_0 \,, \\
\end{align}
\end{subequations}
\noindent and
\begin{subequations}
\begin{align}
    \mathcal{A}_{41} &= \mathcal{D}_z \rho_0 + \rho_0\mathcal{D}_z \,, \\
    \mathcal{A}_{42} &= \mathcal{D}_y \rho_0 + \rho_0\mathcal{D}_y \,, \\
    \mathcal{A}_{44} &= - i\omega \,, \\
\end{align}
\end{subequations}
\noindent and
\begin{subequations}
\begin{align}
    \mathcal{A}_{51} &= \rho_0\mathcal{D}_z T_0 + \frac{\gamma-1}{\gamma}\rho_0 T_0 \mathcal{D}_z \,, \\
    \mathcal{A}_{52} &= \rho_0\mathcal{D}_y T_0 + \frac{\gamma-1}{\gamma}\rho_0 T_0 \mathcal{D}_y \,, \\
    \mathcal{A}_{55} &= - i\omega\rho_0 \,,
\end{align}
\end{subequations}
\noindent and
\begin{subequations}
\begin{align}
    \mathcal{B}_{11} &= -i \rho_0 U_0 \,, \\
    \mathcal{B}_{22} &= -i \rho_0 U_0 \,, \\
    \mathcal{B}_{33} &= -i \rho_0 U_0 \,, \\
    \mathcal{B}_{34} &= -\frac{i}{\gamma Ma^2}T_0 \,, \\
    \mathcal{B}_{35} &= -\frac{i}{\gamma Ma^2}\rho_0 \,, \\
    \mathcal{B}_{38} &= -\frac{1}{Re} \,, \\
\end{align}
\end{subequations}
\noindent and
\begin{subequations}
\begin{align}
    \mathcal{B}_{43} &= -i\rho_0 \,, \\
    \mathcal{B}_{44} &= -iU_0 \,, \\
\end{align}
\end{subequations}
\noindent and
\begin{subequations}
\begin{align}
    \mathcal{B}_{53} &= -i\frac{\gamma-1}{\gamma}\rho_0T_0 \,, \\
    \mathcal{B}_{55} &= -i\rho_0 U_0 \,.
\end{align}
\end{subequations}

%%%%%%%%%%%%%%%%%%%%%%%%%%%%%%%%%%%%%%%%%%%%%%
\section{TR-PIV time series length effect}
\label{app:time_series_length}

\begin{figure}
    \centering
    \begin{subfigure}[b]{0.95\textwidth}
        \centering
        \includegraphics[width=\textwidth]{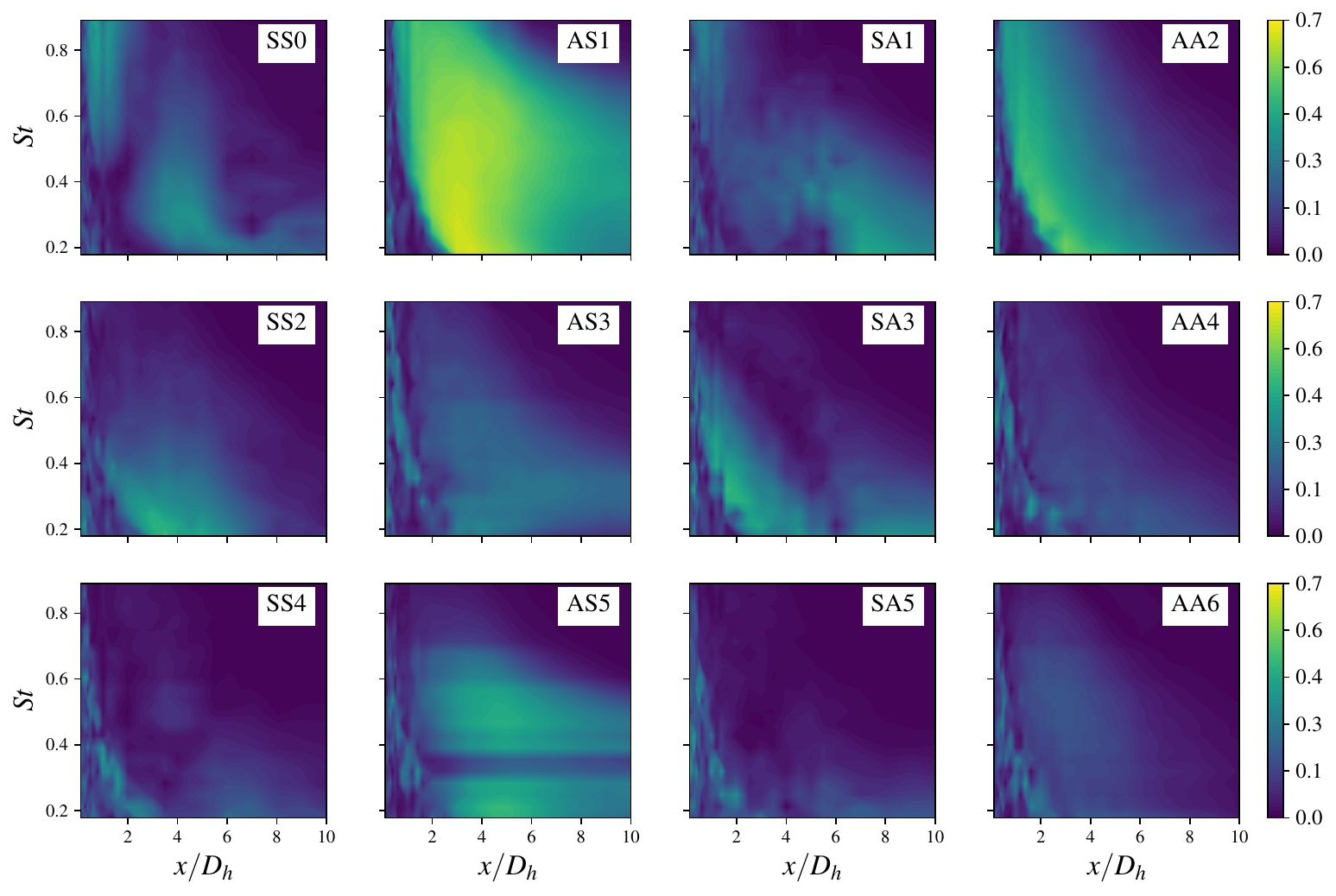}
        \caption{SPOD basis computed using $T = 2.16$ seconds of acquisition time.}
    \end{subfigure}
    \begin{subfigure}[b]{0.95\textwidth}
        \centering
        \includegraphics[width=\textwidth]{Figures/All_beta_maps_new.pdf}
        \caption{SPOD basis computed using $T = 12.12$ seconds of acquisition time.}
    \end{subfigure}
    \caption{Alignment coefficient $\beta$ maps computed using SPOD bases obtained from different acquisition times.}
\label{fig:beta_map_comp}
\end{figure}

Figure~\ref{fig:beta_map_comp} shows a comparison of the alignment coefficients $\beta$ computed using SPOD bases obtained from acquisition times of $T = 2.16$ seconds and $T = 12.12$ seconds, for $M_j = 0.4$.
Only minor differences are observed overall, except for mode AS5, which exhibits improved alignment when using the more converged basis.

These results indicate that although the $T = 12.12$ seconds base is more converged (see Appendix~\ref{app:SPOD_convergence}), the SPOD basis obtained from $T = 2.16$ seconds of data is already sufficient to provide meaningful comparisons with the PSE results.

\FloatBarrier

%%%%%%%%%%%%%%%%%%%
% REFERENCES
%%%%%%%%%%%%%%%%%%%

\bibliographystyle{unsrtnatabbrv}
\bibliography{references}

\end{document}